\documentclass[12pt]{article}
\usepackage{epsfig}
\usepackage{amsmath}
\usepackage{amssymb}
\usepackage{graphicx}
\usepackage{lscape}
\usepackage {subfigure}
\def\be{\begin{equation}}
\def\ee{\end{equation}}
\def\ba{\begin{array}}
\def\ea{\end{array}}
\def\beqn{\begin{eqnarray}}
\def\eeqn{\end{eqnarray}}

\def\bt{\begin{tabular}}
\def\et{\end{tabular}}
\def\bc{\begin{center}}
\def\ec{\end{center}}
\usepackage{epstopdf}
\usepackage{hyperref, cite}
\begin{document}
\title{New possibilities of hybrid texture of neutrino mass matrix}

\author{Madan Singh$^{*} $\\
\it Department of Physics, National Institute of Technology Kurukshetra,\\
\it Haryana,136119, India.\\
\it $^{*}$singhmadan179@gmail.com
}
\maketitle

\begin{abstract}
In this paper, we investigate the noval possibilities of hybrid textures comprising a vanishing minor (or element) and two equal elements (or cofactors) in light neutrino mass matrix $M_{\nu}$. Such type of texture  structures lead to sixty phenomenological cases each, out of which only fifty six are viable with texture containing a vanishing minor and an equality between the elements in $M_{\nu}$, while fifty are found to be viable with texture containing a vanishing element and an equality of cofactors in $M_{\nu}$  under the current experimental test at 3$\sigma$ confidence level. Detailed numerical analysis of all the possible cases have been presented. 

\end{abstract}

\maketitle
\section{Introduction}
During the last two decades, our knowledge regarding the neutrino sector has enriched to a great extent. Thanks to solar, atmospheric, reactor and accelerator based experiments which convincingly reveal that neutrinos have non-zero and non degenerate masses and can convert from one flavor to another.  While the developments over the past two decades have brought out a coherent picture of neutrino mixing, there are still several intriguing issues without which our understanding of neutrino physics remains incomplete. For instance, the present available data does not throw any light on the neutrino mass spectrum, which may be normal/inverted and may even be degenerate.  In addition, nature of neutrino mass whether Dirac or Majorana particle, determination of absolute neutrino mass,  leptonic CP violation and Dirac CP phase $\delta$ are still open issues.  Also the information regarding the lightest neutrino mass has to be sharpened further to pin point the specific possibility of neutrino mass spectrum.

 After the precise measurement of reactor mixing angle $\theta_{13}$ in T2K, MINOS, Double Chooz, Daya Bay and RENO experiments \cite{1,2,3,4,5}, five parameters in the neutrino sector have been well measured by neutrino
oscillation experiments. In general, there are nine parameters in the lightest neutrinos mass
matrix. The remaining four unknown parameters may be taken as the lightest neutrino mass, the
Dirac CP violating phase and two Majorana phases. The Dirac CP violating phase is expected to be measured in future
long baseline neutrino experiments, and the lightest mass can be determined from beta
decay and cosmological experiments. If neutrinoless double beta decay ($0\nu \beta \beta$) is detected,
a combination of the two Majorana phases can also be probed. Clearly,
the currently available data on neutrino masses and mixing are insufficient for an
unambiguous reconstruction of neutrino mass matrices.

In the lack of a convincing fermion flavor theory, several phenomenological ansatze have been proposed in the literature  such as some elements of neutrino mass matrix are considered to be zero or equal  \cite{6, 7, 8, 9, 10} or some co-factors of neutrino mass matrix to be either zero or equal \cite {8, 11, 12}.  The main motivation for invoking different mass matrix ansatze is to relate fermion masses and mixing angles
in a testable manner which reduces the number of free parameters in the neutrino mass matrix. In particular, mass matrices with zero textures (or cofactors) have been extensively studied \cite {6, 11} due to their connections to flavor symmetries. In addition, texture specific mass matrices with  one zero element (or minor) and an equality between two independent elements (or cofactors) have also been studied in the literature \cite{7,9,10,12}. Out of sixty possiblities, only fifty four are found to be compatible with the neutrino oscillation data \cite{10} for texture structures having one zero element and an equal matrix elements in the neutrino mass matrix  (also known as hybrid texture), while for texture with one vanishing minor and  an equal cofactors in the neutrino mass matrix ( also known as inverse hybrid texture) only fifty two cases are able to survive the data \cite{12}.

In the present paper, we propose the noval possibilities of hybrid textures where we assume one texture zero and an equality between the cofactors (referred as type X) or one zero minor and an equality between the elements (referred as type Y) in the Majorana neutrino mass matrix $M_{\nu}$. Such type of texture structures sets two conditions on the parameter space and hence reduces the number of free parameters to seven. Therefore the proposed texture structures are as predictive as texture two zeros and any other hybrid textures.
There are total sixty such possibilities in each case which have
been summarized in Table \ref{tab1}.

In Ref. \cite{8}, it is demonstrated that an equality between the elements of $M_{\nu}$ can be realized  through type-II seesaw mechanism \cite{13} while an equality between cofactors of $M_{\nu}$ can be generated
from type-I seesaw mechanism \cite{14}. The zeros element(or minor) in $M_{\nu}$ can be obtained using $Z_{n}$ flavor symmetry \cite{11, 15}. Therefore the viable cases of proposed hybrid texture  can be realized within the framework of seesaw mechanism.

In the present work, we have systematically, investigated all the of sixty possible cases belonging to type X and type Y structures, respectively. We have studied the implication of these textures  for Dirac CP violating phase ($\delta$) and two Majorana phases ($\rho, \sigma$).  We, also, calculate the effective Majorana mass and lowest neutrino mass for all viable hybrid textures belonging to type X and type Y structures. In addition, we present the correlation plots
between different parameters of the hybrid textures of neutrinos for $3\sigma$ allowed ranges
of the known parameters.

The layout of the paper is planned as follows: In Section 2, we shall discuss the methodology to obtain the constraint equations. Section 3 is devoted to numerical analysis. Section 4 will summarize our result.

\section{Methodology} 
Before proceeding further, we briefly underline the methodology relating the elements of the mass matrices to those of the mixing matrix. In the flavor basis, where the charged lepton mass matrix is diagonal, the Majorana neutrino mass matrix can be expressed as,
\begin{equation}\label{eq1}
 M_{\nu}=P_{l}UP_{\nu}M^{\textrm{diag}}P_{\nu}^{T}U^{T}P_{l}^{T},
\end{equation}
where $M^{\textrm{diag}}$ = diag($m_{1}$, $m_{2}$, $m_{3}$) is the diagonal matrix of neutrino masses and $U$ is the flavor mixing matrix, 
and 
\begin{equation}\label{eq2}
P_{\nu}=\left(
\begin{array}{ccc}
    e^{i\rho} & 0& 0 \\
  0 & e^{i\sigma}& 0\\
  0& 0& 1  \\
\end{array}
\right), \qquad \qquad P_{l}= \left(
\begin{array}{ccc}
    e^{i \phi_{e}}& 0& 0 \\
  0 & e^{i \phi_{\mu}} & 0\\
  0& 0& e^{i \phi_{\tau}} \\
\end{array}
\right);
\end{equation}
where $P_{\nu}$ is diagonal phase matrix containing Majorana neutrinos $\rho, \sigma$. $P_{l}$ is unobservable phase matrix and depends on phase convention. 
Eq. \ref{eq1} can be re-written as
\begin{equation}\label{eq3}
M_{\nu}=\left(
\begin{array}{ccc}
   M_{ee}& M_{e\mu}& M_{e\tau} \\
  M_{e\mu} & M_{\mu \mu} & M_{\mu \tau}\\
  M_{e\tau}& M_{\mu \tau}& M_{\tau \tau} \\
  \end{array}
  \right)=P_{l}U\left(
\begin{array}{ccc}
    \lambda_{1}& 0& 0 \\
  0 & \lambda_{2} & 0\\
  0& 0& \lambda_{3} \\
\end{array}
\right)U^{T}P_{l}^{T},
\end{equation}
where
$\lambda_{1} = m_{1} e^{2i\rho},\lambda_{2} = m_{2} e^{2i\sigma} ,\lambda_{3} = m_{3}.$
For the present analysis, we consider the following parameterization of $U$ \cite{16}:
\begin{equation}\label{eq4}
U=\left(
\begin{array}{ccc}
 c_{12}c_{13}& s_{12}c_{13}& s_{13} \\
-c_{12}s_{23}s_{13}-s_{12}c_{23}e^{-i\delta} & -s_{12}s_{23}s_{13}+c_{12}c_{23}e^{-i\delta} & s_{23}c_{13}\\
 -c_{12}c_{23}s_{13}+s_{12}s_{23}e^{-i\delta}& -s_{12}c_{23}s_{13}-c_{12}s_{23}e^{-i\delta}& c_{23}c_{13} \\
\end{array}
 \right),
\end{equation}
where, $c_{ij} = \cos \theta_{ij}$, $s_{ij}= \sin \theta_{ij}$. Here, $U$ is a 3 $\times$ 3 unitary matrix consisting of three flavor mixing angles ($\theta_{12}$, $\theta_{23}$, $\theta_{13}$) and one Dirac CP-violating phase $\delta$.

For the illustration of type X and Y structures, we consider a case $A_{1}$, satisfying following conditions  
\begin{equation}\label{eq5}
C_{11}=M_{\mu \mu}M_{\tau \tau}-M_{\mu \tau}M_{\mu \tau}=0,
\end{equation}
and
\begin{equation}\label{eq6}
M_{e\mu}-M_{e\tau}=0,
\end{equation}
for type X, while in case of type Y , it contains  
\begin{equation}\label{eq7}
M_{ee} = 0,
\end{equation}
and
\begin{equation}\label{eq8}
C_{12}-C_{13}=0,
\end{equation}
or
\begin{equation}\label{eq9}
(-1). (M_{e\tau}M_{\tau \tau}-M_{\mu \tau}M_{e \tau})-(M_{e \mu}M_{\mu \tau}-M_{\mu \mu}M_{e \tau}) = 0,
\end{equation}
where $C_{ij}$ denotes cofactor corresponding to $i^{th}$ row and $j^{th}$ column.  
Then $A_{1}$ can be denoted in a matrix form as

\begin{equation}\label{eq10}
\left(
\begin{array}{ccc}
    0& \Delta& \Delta \\
  \Delta & \times & \times\\
  \Delta& \times& \times \\
\end{array}
\right),
\end{equation}
where  $"\Delta"$ stands for  nonzero and equal elements (or cofactors), while "$0$" stands for
vanishing element (or minor) in neutrino mass matrix. The "$\times$" stands for
arbitrary elements.\\ 

\subsection{One Vanishing minor with Two Equal Elements of $M_{\nu}$}
Using Eq. \ref{eq1}, any element $M_{pq}$ in the neutrino mass matrix can be expressed in terms of mixing matrix elements as
\begin{equation}\label{eq11}
M_{pq}=e^{i(\phi_{p}+\phi_{q})}\sum_{i=1,2,3}U _{p i}U_{qi} \lambda_{i}, 
\end{equation}
where $p, q$ run over e, $\mu$ and $\tau$, and $e^{i(\phi_{p}+\phi_{q})}$ is phase factor. 

The existence of a zero minor in the Majorana neutrino 
mass matrix implies
\begin{equation}\label{eq12}
M_{pq}M_{rs}-M_{tu}M_{vw}=0
\end{equation}
The above condition yields a complex equation as
\begin{eqnarray}\label{eq13}
 \sum_{i,j =1,2,3} ( e^{i(\phi_{p}+\phi_{q}+\phi_{r}+\phi_{s})}U_{pi}U_{qi}U_{rj}U_{sj}-e^{i(\phi_{t}+\phi_{u}+\phi_{v}+\phi_{w})} U_{ti}U_{ui}U_{vj}U_{wj}) \lambda_{i} \lambda_{j}=0,
 \end{eqnarray}
It is observed that for any cofactor there
is an inherent property as $\phi_{p}+\phi_{q}+\phi_{r}+\phi_{s} = \phi_{t}+\phi_{u}+\phi_{v}+\phi_{w}$. Thus we can extract this total phase factor from the bracket in Eq.\ref{eq13}. \\
 Hence Eq.\ref{eq13} can be rewritten as
 \begin{equation}\label{eq14}
X_{3}\lambda_{1}\lambda_{2}+X_{1}\lambda_{2}\lambda_{3}+X_{2}\lambda_{3}\lambda_{1}=0,
\end{equation}
 where
 \begin{equation}\label{eq15}
 X_{k}=(U_{pi}U_{qi}U_{rj}U_{sj}-U_{ti}U_{ui}U_{vj}U_{wj})+(i \leftrightarrow j),
 \end{equation}
with ($i, j, k$) as the cyclic permutation of (1, 2, 3).

On the other hand, the condition of two equal elements in $M_{\nu}$ yields following equation
 \begin{equation}\label{eq16}
 M_{ab}-M_{cd}=0.
 \end{equation}
Eq. \ref{eq16} yields a following complex equation  
\begin{equation}\label{eq17}
\sum_{i=1,2,3} (P_{1} U_{a i}U_{b i}-P_{2} U_{ci}U_{di}) \lambda_{i}=0,
\end{equation}
where $P_{1}=e^{i(\phi_{a}+\phi_{b})}$ and $P_{2}=e^{i(\phi_{c}+\phi_{d})}.$\\
or
\begin{equation}\label{eq18}
\sum_{i=1,2,3} (P U_{ai}U_{bi}- U_{ci}U_{di}) \lambda_{i}=0
\end{equation} 
where $P \equiv\frac{P_{1}}{P_{2}}=e^{i(a+b-c-d)}$ and $a, b, c, d$ run over $e$, $\mu$ and $\tau$.\\
Eq. \ref{eq18} can be rewritten as
\begin{equation}\label{eq19}
 Y_{1}\lambda_{1}+Y_{2}\lambda_{2}+ Y_{3}\lambda_{3}=0
 \end{equation}
 where $Y_{1}=(PU_{a1}U_{b1}-U_{c1}U_{d1}$),  $Y_{2}=(PU_{a2}U_{b2}-U_{c2}U_{d2}$), $Y_{3}=(PU_{a3}U_{b3}-U_{c3}U_{d3})$.\\
   Solving Eqs. \ref{eq14} and \ref{eq19} simultaneously lead to  the following complex mass ratio  in terms of $(\lambda_{13})_{\pm}$
 
 \begin{equation}\label{eq20}
 (\lambda_{13})_{+} =\frac{-(Y_{1}X_{1}-Y_{2}X_{2}+Y_{3}X_{3}+\sqrt{C})}{2Y_{1}X_{3}}, 
 \end{equation}
 and
 \begin{equation}\label{eq21}
  (\lambda_{13})_{-}  =\frac{-(Y_{1}X_{1}-Y_{2}X_{2}+Y_{3}X_{3}-\sqrt{C})}{2Y_{1}X_{3}}. 
  \end{equation}
Using Eqs. \ref{eq14}, \ref{eq20} and \ref{eq21}, we obtain the relations for complex mass ratio in terms of $(\lambda_{23})_{\pm}$    
  
 \begin{equation}\label{eq22}
 (\lambda_{23})_{+}= -\frac{X_{2}}{X_{3}}\times \frac{Y_{1}X_{1}-Y_{2}X_{2}+Y_{3}X_{3}+ \sqrt{C}}{-Y_{1}X_{1}-Y_{2}X_{2}+Y_{3}X_{3}+ \sqrt{C}},
 \end{equation}
  and
  \begin{equation}\label{eq23}
  (\lambda_{23})_{-} = -\frac{X_{2}}{X_{3}}\times \frac{Y_{1}X_{1}-Y_{2}X_{2}+Y_{3}X_{3}-\sqrt{C}}{-Y_{1}X_{1}-Y_{2}X_{2}+Y_{3}X_{3}-\sqrt{C}}, 
 \end{equation}
 where $C =(-Y_{1}X_{1}+Y_{2}X_{2}+Y_{3}X_{3})^{2}-4X_{2}X_{3}Y_{2}Y_{3}$, and $ (\lambda_{13})_{\pm}\equiv \bigg(\frac{\lambda_{1}}{\lambda_{3}}\bigg)_{\pm}$, $ (\lambda_{23})_{\pm}\equiv \bigg(\frac{\lambda_{2}}{\lambda_{3}}\bigg)_{\pm}$.  
 The magnitudes of the two neutrino mass ratios in Eqs. \ref{eq20}, \ref{eq21}, \ref{eq22} and \ref{eq23}, are given by $\xi_{\pm}=|(\lambda_{13})_{\pm}|$,      
 $\zeta_{\pm}=|(\lambda_{23})_{\pm}|$, while the Majorana CP-violating phases $\rho$ and $\sigma$ can be given as $\rho=\frac{1}{2} arg(\lambda_{13})_{\pm}, \sigma=\frac{1}{2} arg(\lambda_{23})_{\pm}$.\\

\subsection{One Vanishing element with Two Equal Cofactors of $M_{\nu}$} 
If one of the elements of $M_{\nu}$ is considered zero, [e.g. $M_{\alpha \beta} = 0$],  we obtain the following constraint equation
\begin{equation}\label{eq24}
\sum_{i=1,2,3}U_{\alpha i}U_{\beta i}\lambda_{i}=0,  
\end{equation}
or
\begin{equation}\label{eq25}
\lambda_{1} A_{1} +\lambda_{2} A_{2}+\lambda_{3} A_{3}=0,
\end{equation}
where $A_{1}= U_{\alpha 1}U_{\beta 1}, A_{2}= U_{\alpha 2}U_{\beta 2}$ and $A_{3}= U_{\alpha 3}U_{\beta 3}$.\\
The condition for two equal cofactors [e.g., $C_{mn}=C_{m^{'}n^{'}}$] in neutrino mass matrix implies
 \begin{equation}\label{eq26}
  (-1)^{m+n}(M_{ab}M_{cd}-M_{ef}M_{gh})-(-1)^{m^{'}+n^{'}}(M_{a^{'}b^{'}}M_{c^{'}d^{'}}-M_{e^{'}f^{'}}M_{g^{'}h^{'}})=0,
 \end{equation}
 or
  \begin{eqnarray}\label{eq27}
&&\sum_{i,j =1,2,3} \big \{ (-1)^{m+n}( Q_{3} U_{ai}U_{bi}U_{cj}U_{dj}- Q_{4} U_{ei}U_{fi}U_{gj}U_{hj})
\nonumber  \\ 
 &&~~~~~ -(-1)^{m^{'}+n^{'}}(Q_{5} U_{a^{'}i}U_{b^{'}i}U_{c^{'}j}U_{d^{'}j}- Q_{6} U_{e^{'}i}U_{f^{'}i}U_{g^{'}j}U_{h^{'}j}) \big \} \lambda_{i} \lambda_{j}=0,
  \end{eqnarray}
  where $Q_{3}=Q_{4}$ and $Q_{5}=Q_{6}$ due to inherent property of any cofactor. Thus we can write
  \begin{eqnarray}\label{eq28}
&&\sum_{i,j =1,2,3} \big \{ (-1)^{m+n}Q_{3}( U_{ai}U_{bi}U_{cj}U_{dj}-  U_{ei}U_{fi}U_{gj}U_{hj})
\nonumber  \\ 
 &&~~~~~ -(-1)^{m^{'}+n^{'}}Q_{5}( U_{a^{'}i}U_{b^{'}i}U_{c^{'}j}U_{d^{'}j}- U_{e^{'}i}U_{f^{'}i}U_{g^{'}j}U_{h^{'}j}) \big \} \lambda_{i} \lambda_{j}=0,
  \end{eqnarray}
   or 
  \begin{eqnarray}\label{eq29}
&&\sum_{i,j =1,2,3} \big \{ (-1)^{m+n}Q( U_{ai}U_{bi}U_{cj}U_{dj}-  U_{ei}U_{fi}U_{gj}U_{hj})
\nonumber  \\ 
 &&~~~~~ -(-1)^{m^{'}+n^{'}}( U_{a^{'}i}U_{b^{'}i}U_{c^{'}j}U_{d^{'}j}-  U_{e^{'}i}U_{f^{'}i}U_{g^{'}j}U_{h^{'}j}) \big \} \lambda_{i} \lambda_{j}=0,
  \end{eqnarray}
  where $Q \equiv \dfrac{Q_{3}}{Q_{5}}= e^{i(\phi_{a}+\phi_{b}+\phi_{c}+\phi_{d}-\phi_{a^{'}}-\phi_{b^{'}}-\phi_{c^{'}}-\phi_{d^{'}})}.$  \\
 Eq. \ref{eq29} can be rewritten as
\begin{equation}\label{eq30}
 \lambda_{1}\lambda_{2}B_{3}+\lambda_{2}\lambda_{3}B_{1}+\lambda_{3}\lambda_{1}B_{2}=0,
 \end{equation}
 where 
 
 \begin{eqnarray}\label{eq31}
 &&\quad\qquad  B_{k}=(-1)^{m+n}Q(U_{ai}U_{bi}U_{cj}U_{dj}-U_{ei}U_{fi}U_{gj}U_{hj}),
 \nonumber \\
 &&~~~~~~-(-1)^{m^{'}+n^{'}}(U_{a^{'}i}U_{b^{'}i}U_{c^{'}j}U_{d^{'}j}-U_{e^{'}i}U_{f^{'}i}U_{g^{'}j}U_{h^{'}j})+(i \leftrightarrow j),
 \end{eqnarray}
 
 with ($i, j, k$) a cyclic permutation of (1, 2, 3). 
 
Solving Eqs. \ref{eq25} and \ref{eq30} simultaneously  we obtain the analytical expressions of $(\lambda_{13})_{\pm}$
 \begin{small}
 \begin{eqnarray}\label{eq32}
 (\lambda_{13})_{+} =\frac{-(B_{1}A_{1}-B_{2}A_{2}+B_{3}A_{3}+\sqrt{D})}{2B_{1}A_{3}}, 
 \end{eqnarray}
 \end{small}
 and
 \begin{small}
  \begin{eqnarray}\label{eq33}
 (\lambda_{13})_{-}=\frac{-(B_{1}A_{1}-B_{2}A_{2}+B_{3}A_{3}-\sqrt{D})}{2B_{1}A_{3}}.
 \end{eqnarray}
 \end{small}

Using Eqs. \ref{eq30}, \ref{eq32} and \ref{eq33}, we get the relations for complex mass ratio in terms of $(\lambda_{23})_{\pm}$ 
  \begin{small}
 \begin{equation}\label{eq34}
 (\lambda_{23})_{+} = -\frac{B_{2}}{B_{3}}\times \frac{(B_{1}A_{1}-B_{2}A_{2}+B_{3}A_{3}+ \sqrt{D})}{(-B_{1}A_{1}-B_{2}A_{2}+B_{3}A_{3}+\sqrt{D})},
 \end{equation}
 \end{small}
  and
  \begin{small}
 \begin{equation}\label{eq35}
  (\lambda_{23})_{-} = -\frac{B_{2}}{B_{3}}\times \frac{(B_{1}A_{1}-B_{2}A_{2}+B_{3}A_{3}-\sqrt{D})}{(-B_{1}A_{1}-B_{2}A_{2}+B_{3}A_{3}-\sqrt{D})}, 
  \end{equation}
  \end{small}
 where, $D= (-B_{1}A_{1}+B_{2}A_{2}+B_{3}A_{3})^{2}-4A_{2}A_{3}B_{2}B_{3}$. 
 
 The magnitudes of the two neutrino mass ratios  are given by $\xi_{\pm}=|(\lambda_{13})_{\pm}|$,      
 $\zeta_{\pm}=|(\lambda_{23})_{\pm}|$, while the Majorana CP-violating phases $\rho$ and $\sigma$ can be given as $\rho=\frac{1}{2} arg(\lambda_{13})_{\pm}, \sigma=\frac{1}{2} arg(\lambda_{23})_{\pm}$.
 
 The solar and atmospheric mass squared differences ($\delta m^{2}, \Delta m^{2}$), where $\delta m^{2}$ corresponds to solar mass-squared difference and $\Delta m^{2}$ corresponds to atmospheric mass-squared difference, can be defined as \cite{9}
\begin{equation}\label{eq36}
 \delta m^{2}=(m_{2}^{2}-m_{1}^{2}),\;
 \end{equation}
 \begin{equation}\label{eq37}
  \Delta m^{2}=m_{3}^{2}-\frac{1}{2}(m_{1}^{2}+m_{2}^{2}).
 \end{equation}
  The sign of $\Delta m^{2}$ is still unknown:  $\Delta m^{2}>0$ or $\Delta m^{2}<0$ implies normal mass spectrum (NS) or inverted mass spectrum (IS). The lowest neutrino mass ($m_{0}$) is $m_{1}$ for NS and $m_{3}$ for IS. 
  The experimentally determined solar and atmospheric neutrino mass-squared differences can be related to  $\xi$ and $\zeta$ as
 \begin{equation}\label{eq38}
 R_{\nu} \equiv\frac{\delta m^{2}} {|\Delta m^{2}|} =\frac{2(\zeta ^{2}-\xi ^{2})}{\left |2-(\zeta^{2}+\xi ^{2})  \right |},
 \end{equation}
 and the three neutrino masses  can be determined using following relations 
 \begin{equation}\label{eq39}
 m_{3}=\sqrt{\frac{\delta m^{2}}{\zeta^{2}-\xi^{2}}},  \qquad m_{2}=m_{3} \zeta, \qquad m_{1}=m_{3} \xi.
 \end{equation}
 
 \begin{center}
 \begin{table}
\begin{footnotesize}
\noindent\makebox[\textwidth]{
\begin{tabular}{|c|c|c|c|c|}
 \hline
  Parameter& Best Fit & 1$\sigma$ & 2$\sigma$ & 3$\sigma$ \\
  \hline
   $\delta m^{2}$ $[10^{-5}eV^{2}]$ & $7.60$& $7.42$ - $7.79$ & $7.26$ - $7.99$ & $7.11$ - $8.18$ \\
   \hline
   $|\Delta m^{2}_{31}|$ $[10^{-3}eV^{2}]$ (NS) & $2.48$ & $2.41$ - $2.53$ & $2.35$ - $2.59$ & $2.30$ - $2.65$\\
   \hline
  $|\Delta m^{2}_{31}|$ $[10^{-3}eV^{2}]$ (IS) & $2.38$ & $2.32$ - $2.43$ & $2.26$ - $2.48$ & $2.20$ - $2.54$ \\
  \hline
  $\theta_{12}$ & $34.6^{\circ}$ & $33.6^{\circ}$ - $35.6^{\circ}$ & $32.7^{\circ}$ - $36.7^{\circ}$ & $31.8^{\circ}$ - $37.8^{\circ}$\\
  \hline
  $ \theta_{23}$ (NS) & $48.9^{\circ}$ &$41.7^{\circ}$ - $50.7^{\circ}$  & $40.0^{\circ}$ - $52.1^{\circ}$ & $38.8^{\circ}$ - $53.3^{\circ}$ \\
  \hline
  $\theta_{23}$ (IS)& $49.2^{\circ}$ & $46.9^{\circ}$ - $50.7^{\circ}$ & $41.3^{\circ}$ - $52.0^{\circ} $& $39.4^{\circ}$ - $53.1^{\circ}$ \\
  \hline
  $\theta_{13}$ (NS) & $8.6^{\circ}$ & $8.4^{\circ}$ - $8.9^{\circ}$ & $8.2^{\circ}$ - $9.1^{\circ}$& $7.9^{\circ}$ - $9.3^{\circ}$ \\
  \hline
  $\theta_{13}$ (IS) & $8.7^{\circ}$ & $8.5^{\circ}$ - $8.9^{\circ}$ & $8.2^{\circ}$ - $9.1^{\circ}$ & $8.0^{\circ}$ - $9.4^{\circ}$ \\
  \hline
  $\delta$ (NS) & $254^{\circ}$ & $182^{\circ}$ - $353^{\circ}$& $0^{\circ}$ - $360^{\circ}$ & $0^{\circ}$ - $360^{\circ}$ \\
  \hline
  $\delta$ (IS) &$266^{\circ}$& $210^{\circ}$ - $322^{\circ}$ & $0^{\circ}$ - $16^{\circ}$ $\oplus$ $ 155^{\circ}$ - $360^{\circ}$  & $0^{\circ}$ - $360^{\circ}$ \\
\hline
\end{tabular}}
\caption{\label{tab1} Current neutrino oscillation parameters from global fits at 1$\sigma$, 2$\sigma$ and 3$\sigma$ confidence level (CL) \cite{17}. NS(IS) refers to normal (inverted) neutrino mass spectrum.}
 \end{footnotesize}

\end{table}

\end{center}
 From the analysis, it is found  that cases belonging to type X (or type Y) exhibit the identical phenomenological implications and are related through permutation symmetry \cite{13,16}.  This corresponds to permutation of the 2-3 rows and 2-3 columns of $M_{\nu}$. The corresponding permutation matrix can be given by
\begin{equation}\label{eq40}
 P_{23} = \left(
\begin{array}{ccc}
    1& 0& 0 \\
  0 & 0 & 1\\
  0& 1& 0 \\
\end{array}
\right).
\end{equation}
With the help of permutation symmetry, one obtains the following relations among the neutrino oscillation parameters
\begin{equation}\label{eq41}
\theta_{12}^{l}=\theta_{12}^{m}, \ \
\theta_{23}^{l}=90^{\circ}-\theta_{23}^{m},\ \
\theta_{13}^{l}=\theta_{13}^{m}, \ \ \delta^{l}=\delta^{m} -180^{\circ},
\end{equation}
where $l$ and $m$ denote the cases related by 2-3 permutation. The following pairs among sixty possibilities of type X (or type Y) are related via permutation symmetry
 \\
($A_{1}, A_{1});\quad (A_{2}, A_{8});\quad (A_{3}, A_{7});\quad (A_{4}, A_{6});\\
\quad (A_{5}, A_{5});\quad (A_{9}, A_{10});\quad(B_{1}, C_{1});\quad(B_{2}, C_{7}); \\
\quad(B_{3}, C_{6}); \quad(B_{4},C_{5});\quad (B_{5}, C_{4}); \quad (B_{6}, C_{3})\\
\quad(B_{7}, C_{2}); \quad(B_{8}, C_{10});\quad (B_{9}, C_{9});\quad (B_{10}, C_{8}); \\
\quad(D_{1}, F_{2});\quad (D_{2}, F_{1});\quad (D_{3}, F_{4});\quad(D_{4}, F_{3}); \\
\quad(D_{5}, F_{5});\quad (D_{6}, F_{9});\quad (D_{7}, F_{8});\quad (D_{8}, F_{7});\\
\quad(D_{9}, F_{6});\quad (D_{10}, F_{10});\quad (E_{1}, E_{2});\quad (E_{3}, E_{4});\\
\quad (E_{5}, E_{5});\quad(E_{6}, E_{9});\quad (E_{7}, E_{8});\quad (E_{10}, E_{10}).$ \\
      Clearly we are left with only thirty two independent cases. It is worthwhile to mention that  $A_{1}, A_{5}, E_{5}$ and $ E_{10}$ are invariant under the permutations of 2- and 3-rows and columns.

 \section{Numerical analysis}
The effective Majorana mass relevant for neutrinoless double beta ($0\nu\beta\beta$) decay is given by
\begin{equation}\label{eq42}
|M|_{ee}=|m_{1}c_{12}^{2}c_{13}^{2}e^{2i\rho}+m_{2}s_{12}^{2}c_{13}^{2}e^{2i\sigma}+m_{3}s_{13}^{2}|.
\end{equation}
This effective mass is just the absolute value of
$M_{ee}$ component of the neutrino mass matrix.
 The observation of $0\nu\beta\beta$  would establish neutrinos to be Majorana particles.  For recent reviews see Refs. \cite{18, 19}.  Several running and forthcoming neutrinoless double decay experiments such as CUORICINO \cite{20}, CUORE \cite{21}, GERDA \cite{22}, MAJORANA \cite{23}, SuperNEMO \cite{24}, EXO \cite{25}, GENIUS \cite{26} target to achieve a sensitivity upto 0.01eV for $|M|_{ee}$.
\begin{figure}[h!]
\begin{center}
\subfigure[]{\includegraphics[width=0.40\columnwidth]{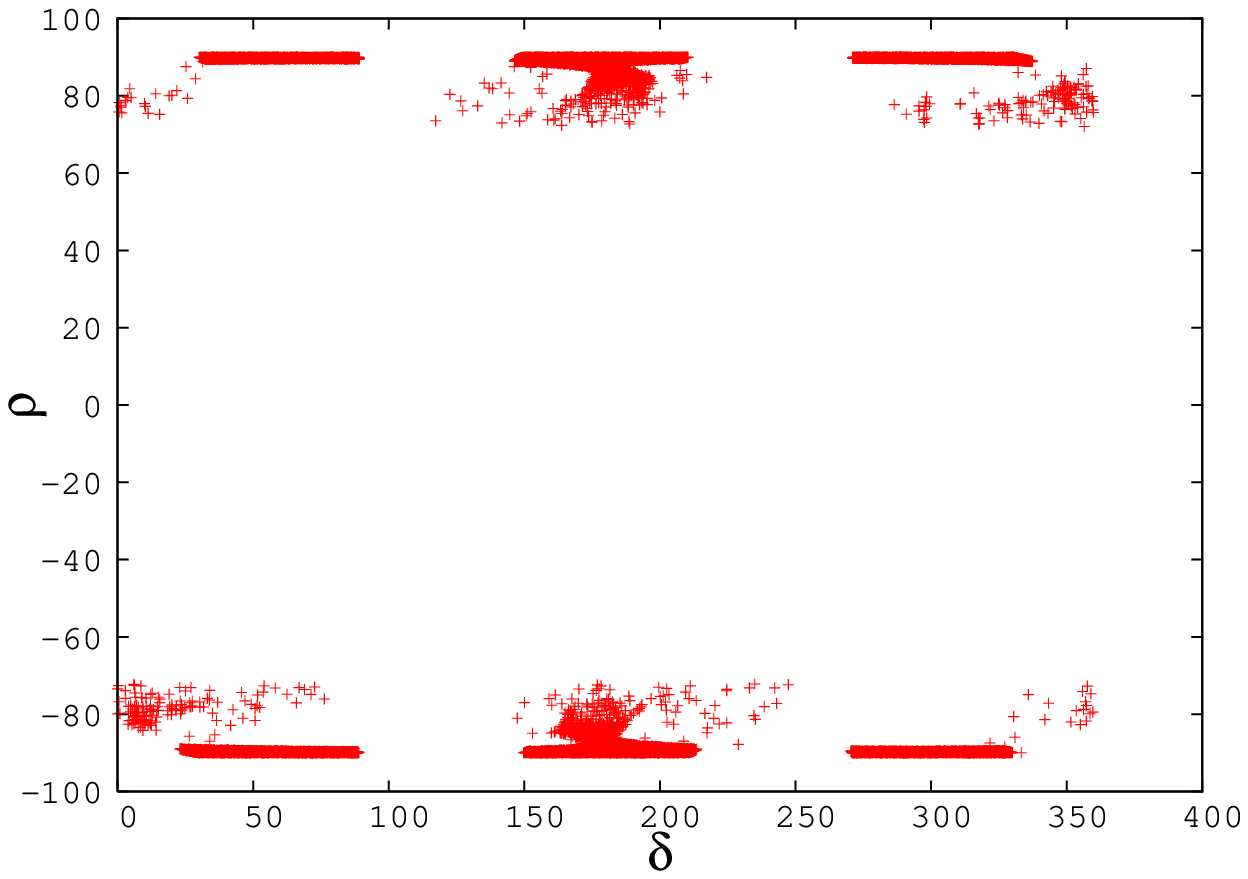}}
\subfigure[]{\includegraphics[width=0.40\columnwidth]{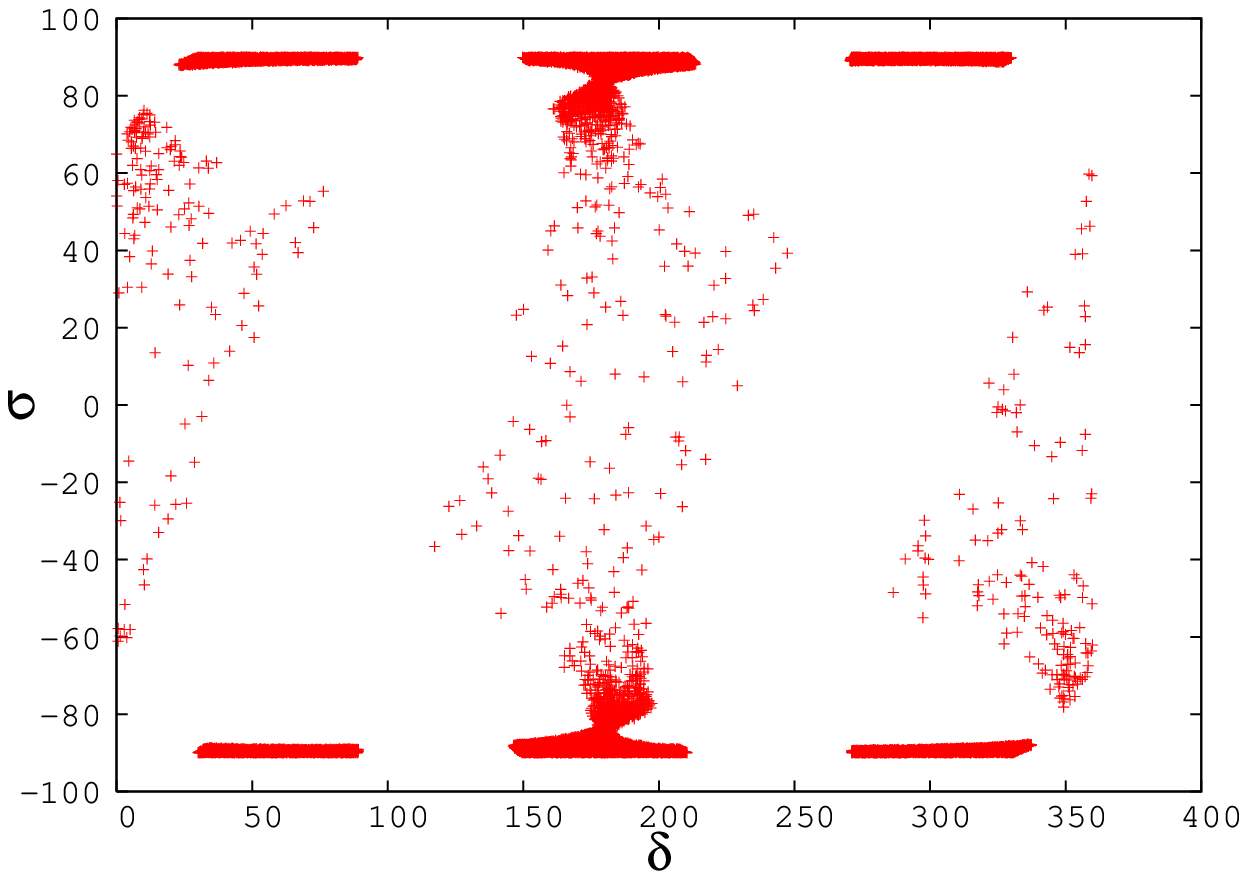}}
\subfigure[]{\includegraphics[width=0.40\columnwidth]{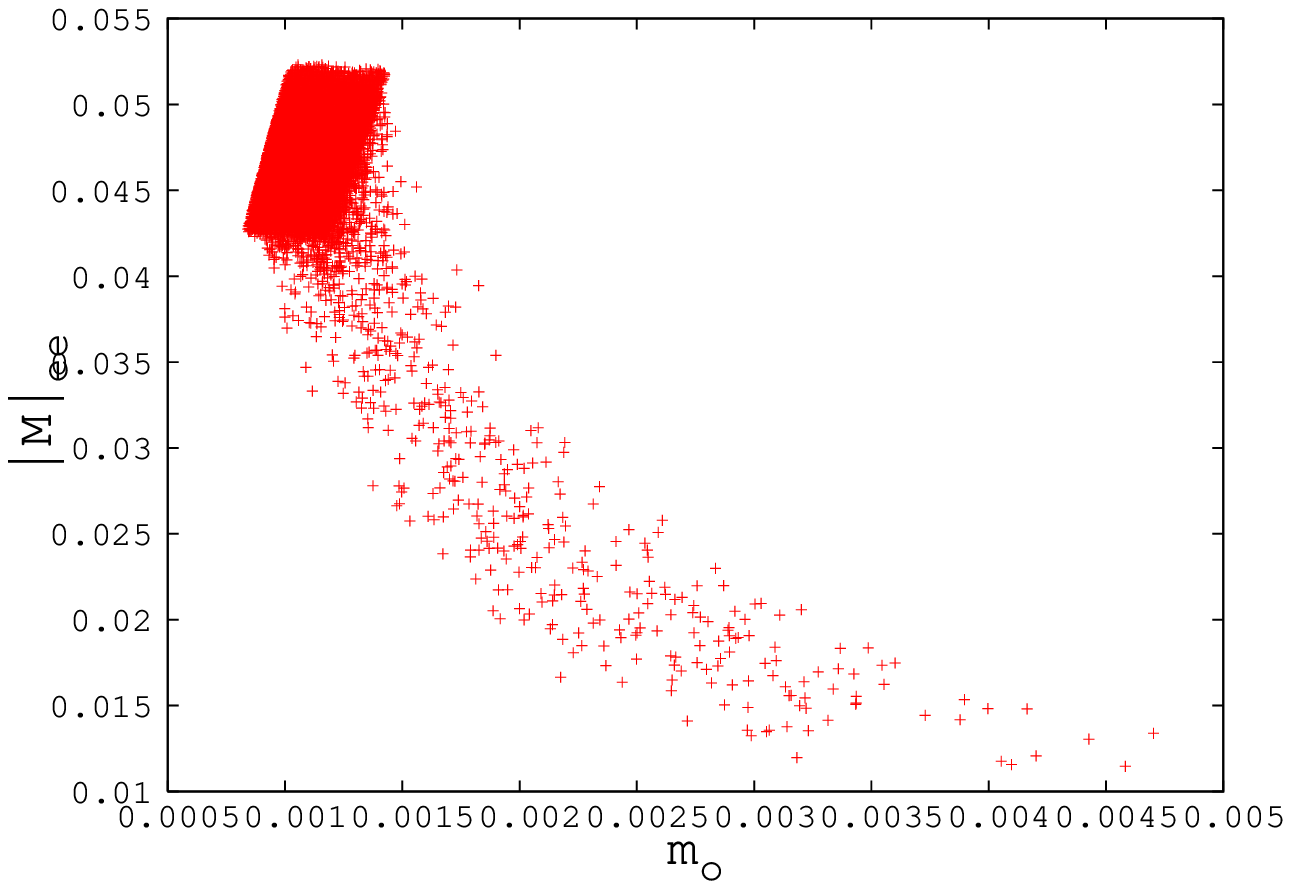}}
\caption{\label{fig1} Correlation plots for texture $A_{1}$ (IS) for type X at 3 $\sigma$ CL. The symbols have their usual meaning. The  $\delta, \rho, \sigma$ are measured in degrees, while $|M|_{ee}$ and $m_{0}$ are in eV units. }
\end{center}
\end{figure}
\begin{figure}[h!]
\begin{center}
\subfigure[]{\includegraphics[width=0.40\columnwidth]{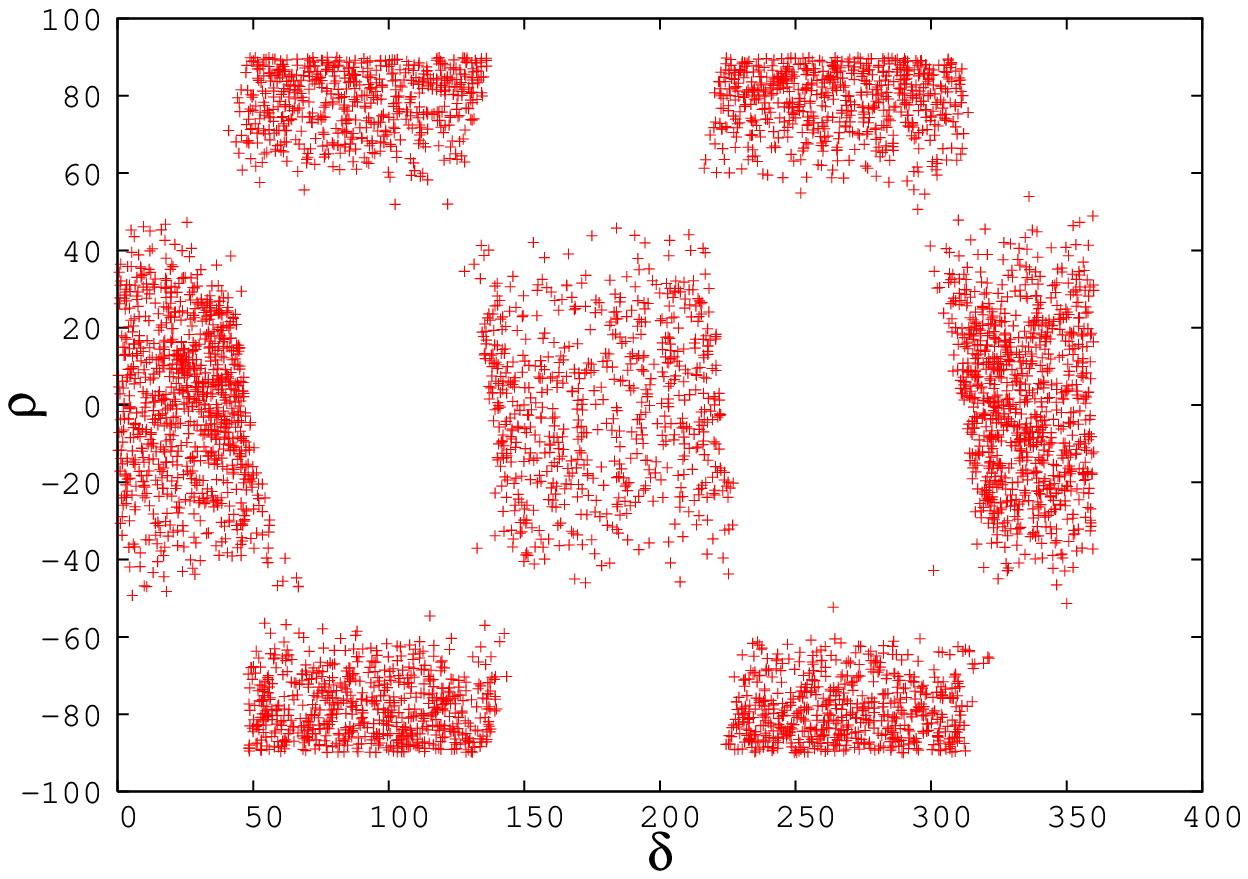}}
\subfigure[]{\includegraphics[width=0.40\columnwidth]{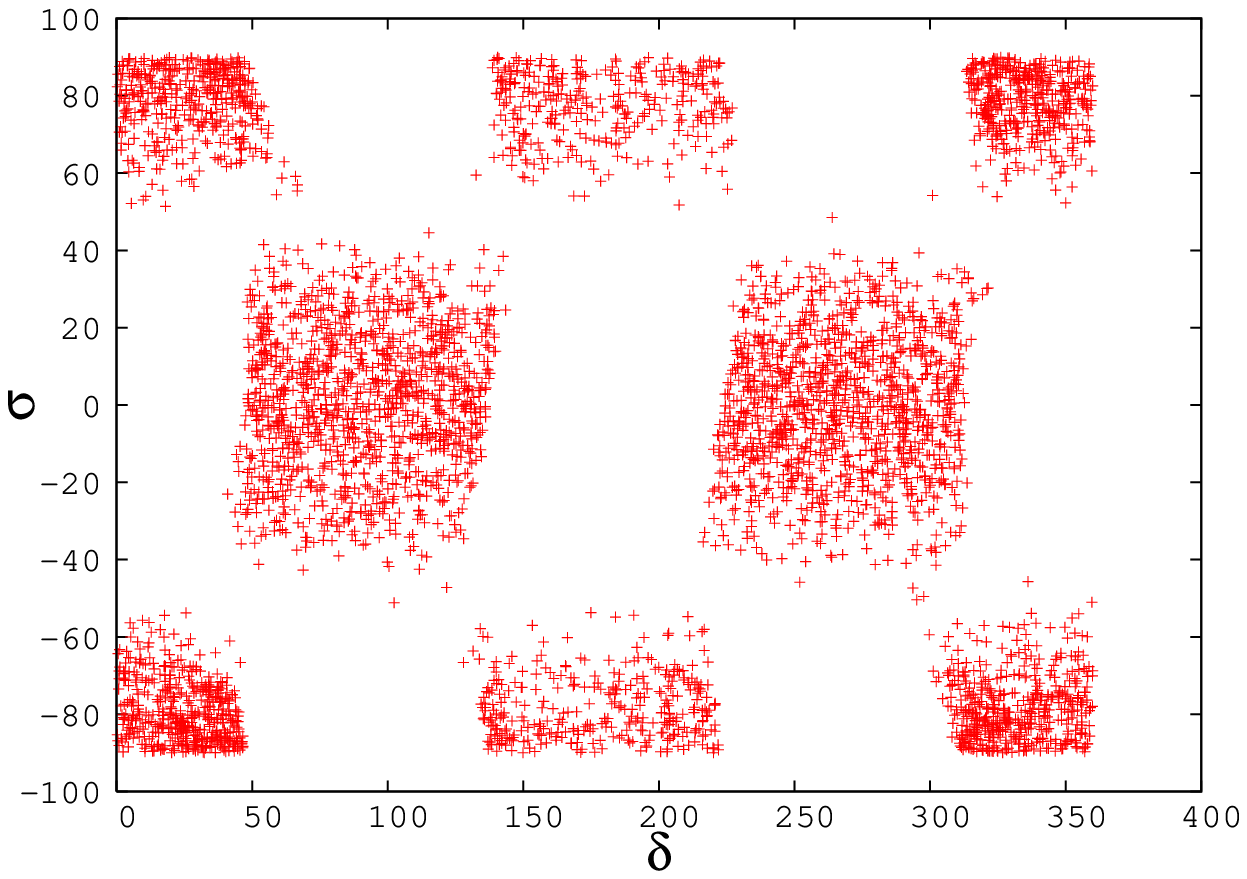}}
\subfigure[]{\includegraphics[width=0.40\columnwidth]{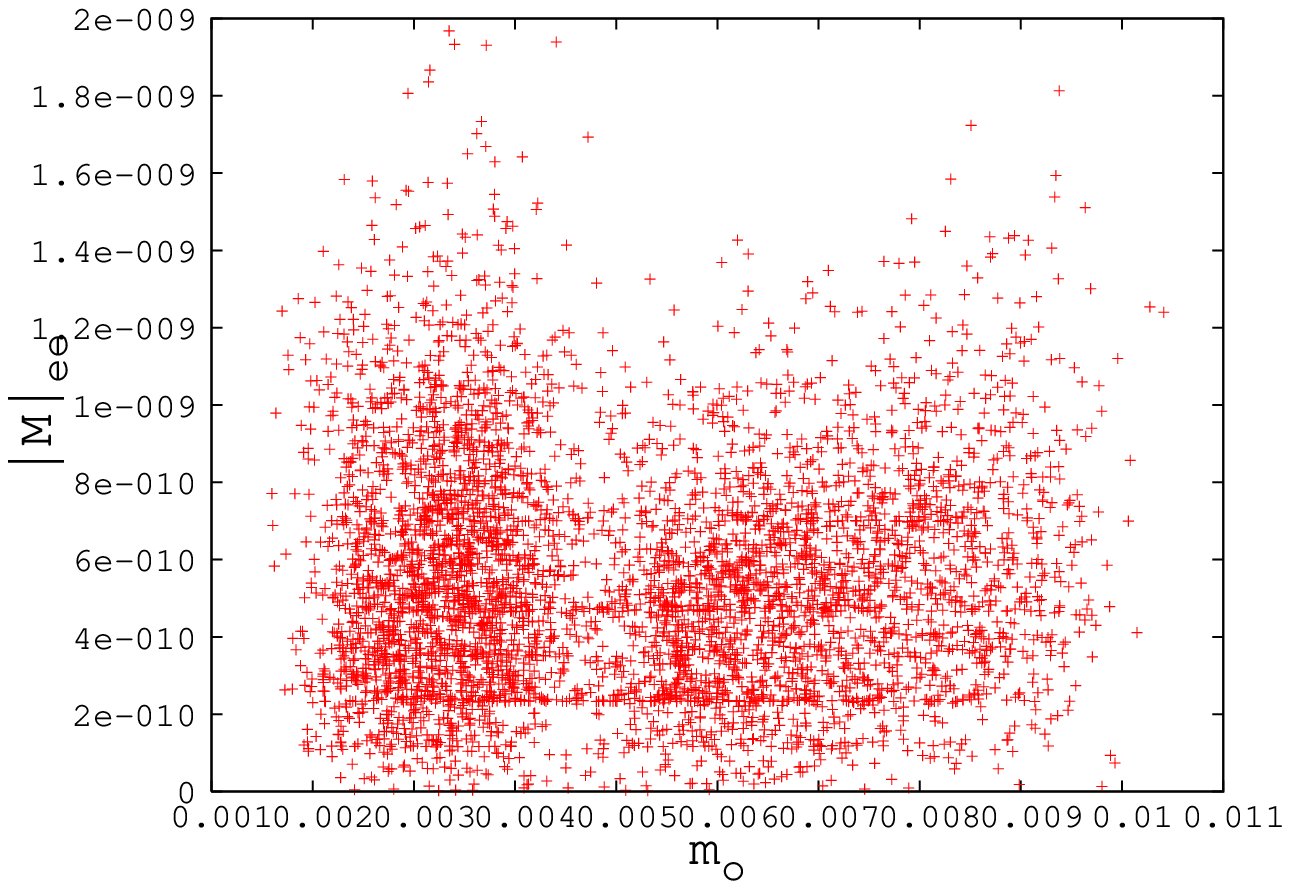}}
\caption{\label{fig2}Correlation plots for texture $A_{1}$ (NS) for type Y at 3 $\sigma$ CL. The symbols have their usual meaning. The  $\delta, \rho, \sigma$ are measured in degrees, while $|M|_{ee}$ and $m_{0}$ are in eV units.}
\end{center}
\end{figure}
In the present analysis,  we consider more conservative upper bound on $|M|_{ee}$ i.e. $|M|_{ee} < 0.5eV$  at 3$\sigma$ CL \cite{19}.  We span the parameter space of input neutrino oscillation parameters ($\theta_{12}$, $\theta_{23}$, $\theta_{13}$, $\Delta m^{2}$, $\Delta m^{2}$) lying in their $3\sigma$ ranges by randomly generating points of the order of $10^{7}$. Since the Dirac CP-violating phase $\delta$ is experimentally unconstrained at $3\sigma$ level, therefore, we vary $\delta$ within its full possible range [$0^\circ$, $360^{\circ}$]. Using Eq. \ref{eq38} and the experimental inputs on neutrino mixing angles and mass-squared differences, the parameter space of $\delta$, $\rho$, $\sigma$ and $|M|_{ee}$ and $m_{0}$ can be subsequently constrained.

\begin{figure}[h!]
\begin{center}
\subfigure[]{\includegraphics[width=0.40\columnwidth]{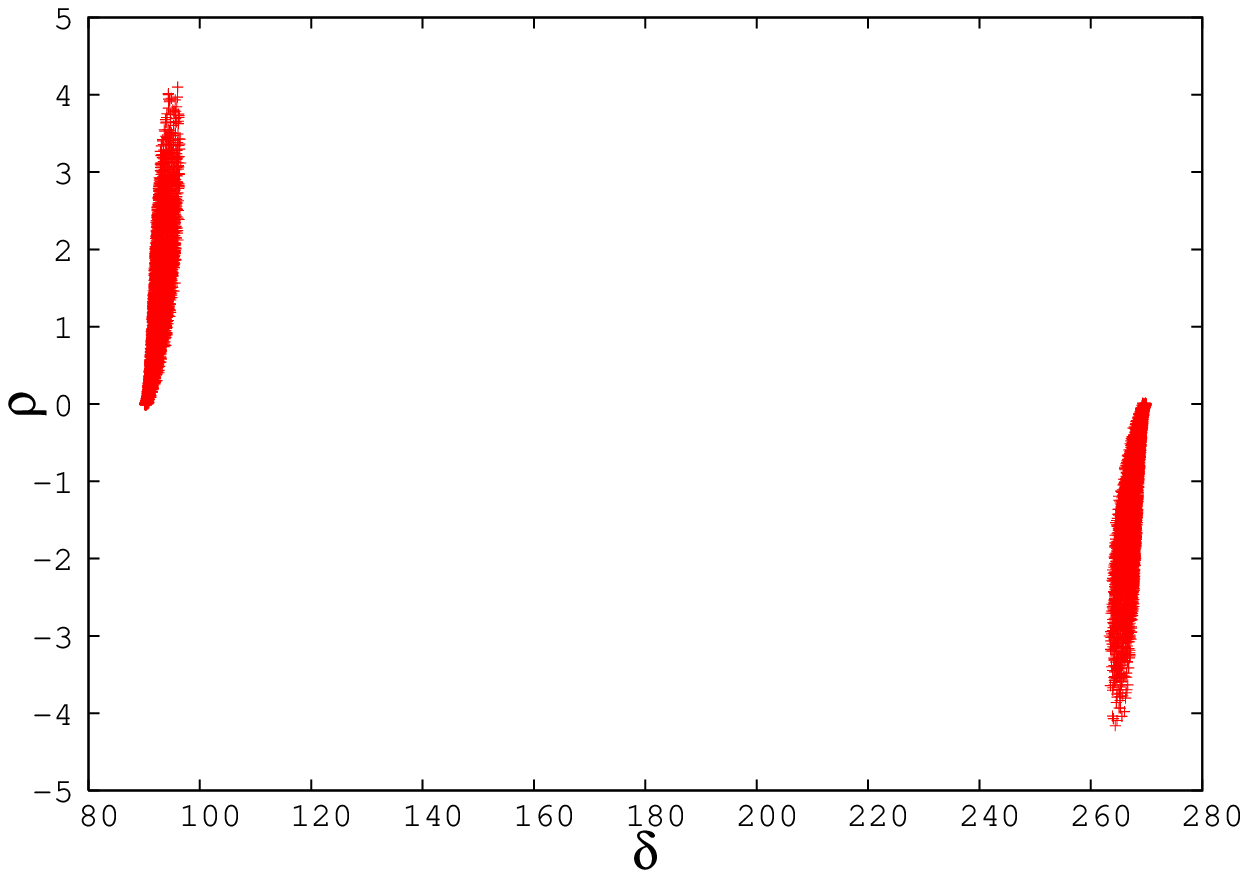}}
\subfigure[]{\includegraphics[width=0.40\columnwidth]{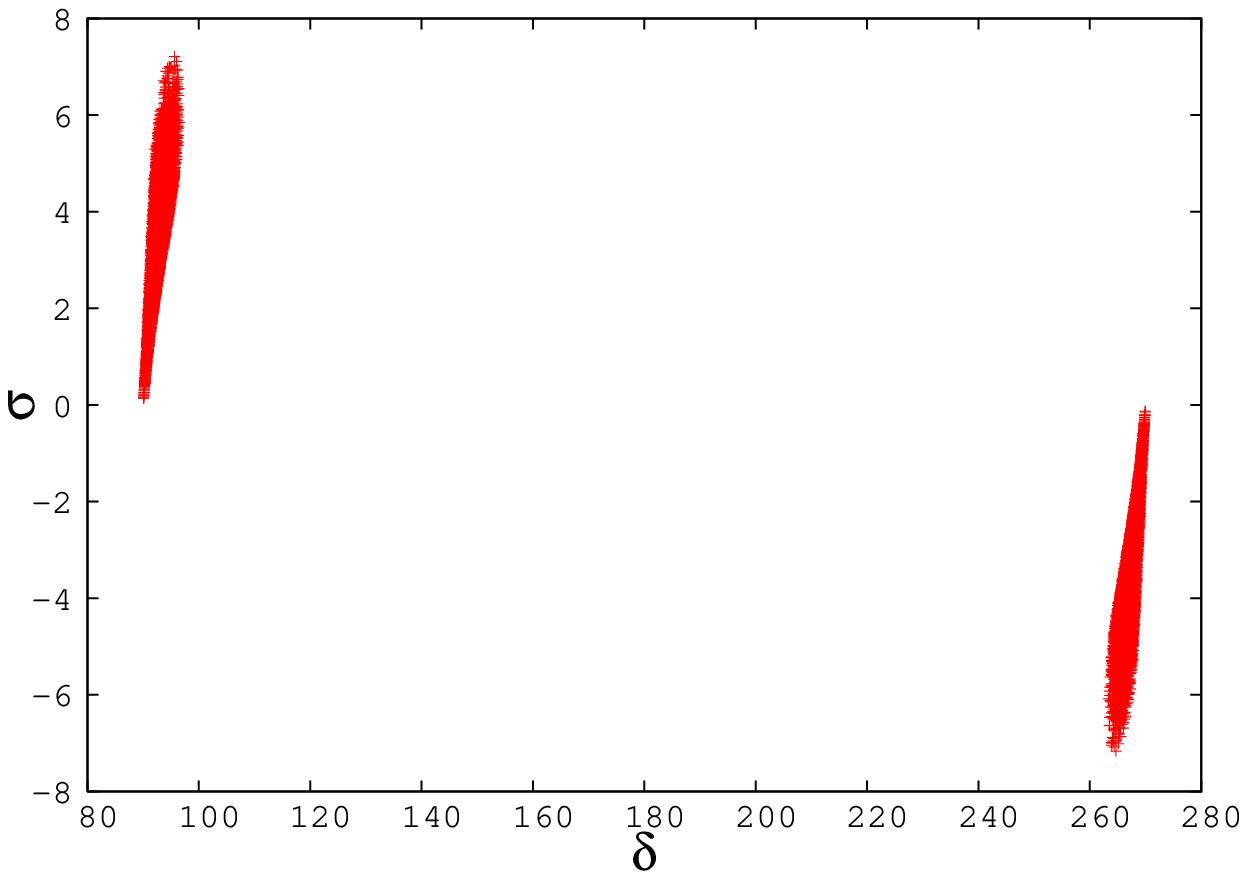}}
\subfigure[]{\includegraphics[width=0.40\columnwidth]{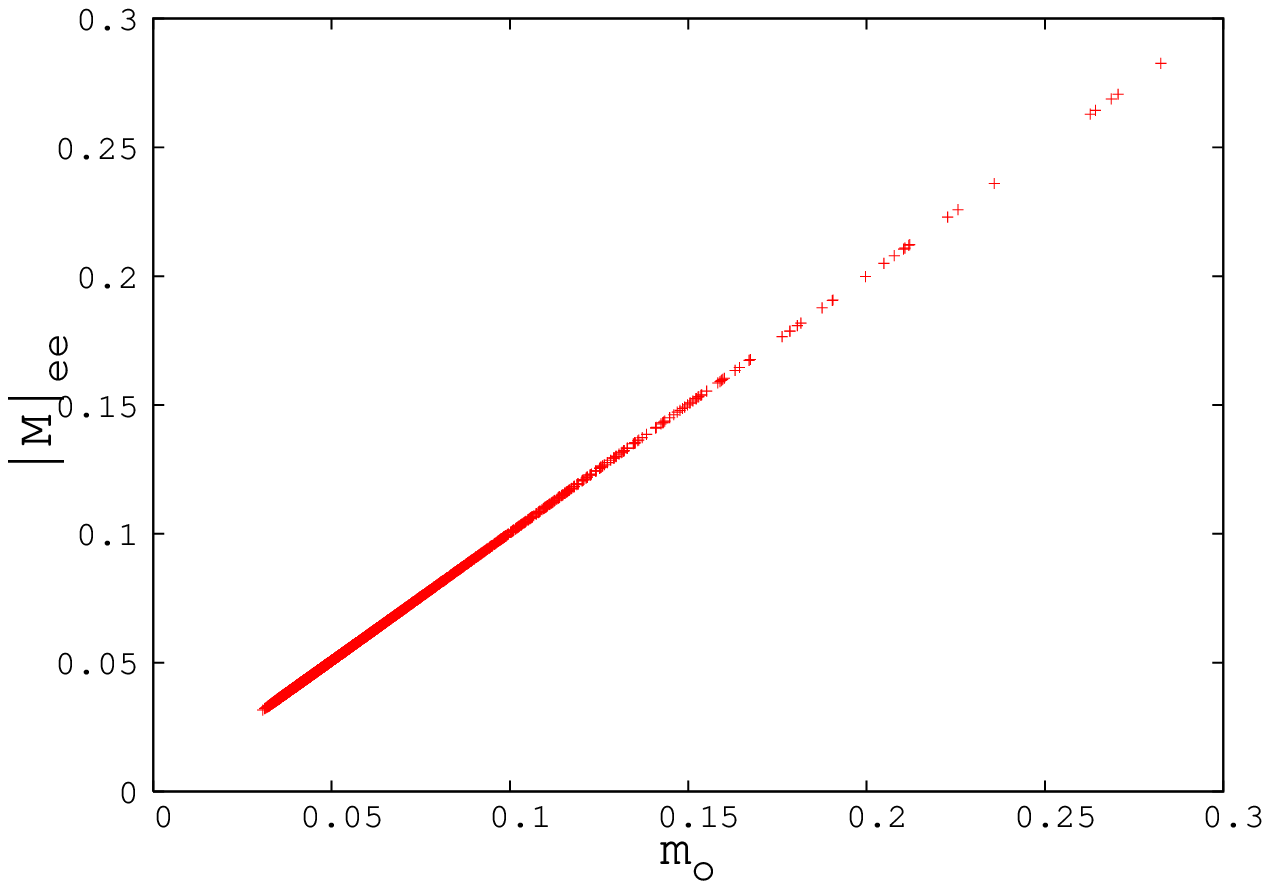}}
\caption{\label{fig3}Correlation plots for texture $B_{2}$ (NS) for type X at 3 $\sigma$ CL. The symbols have their usual meaning. The  $\delta, \rho, \sigma$ are measured in degrees, while $|M|_{ee}$ and $m_{0}$ are in eV units.}
\end{center}
\end{figure}

\begin{figure}[h!]
\begin{center}
\subfigure[]{\includegraphics[width=0.40\columnwidth]{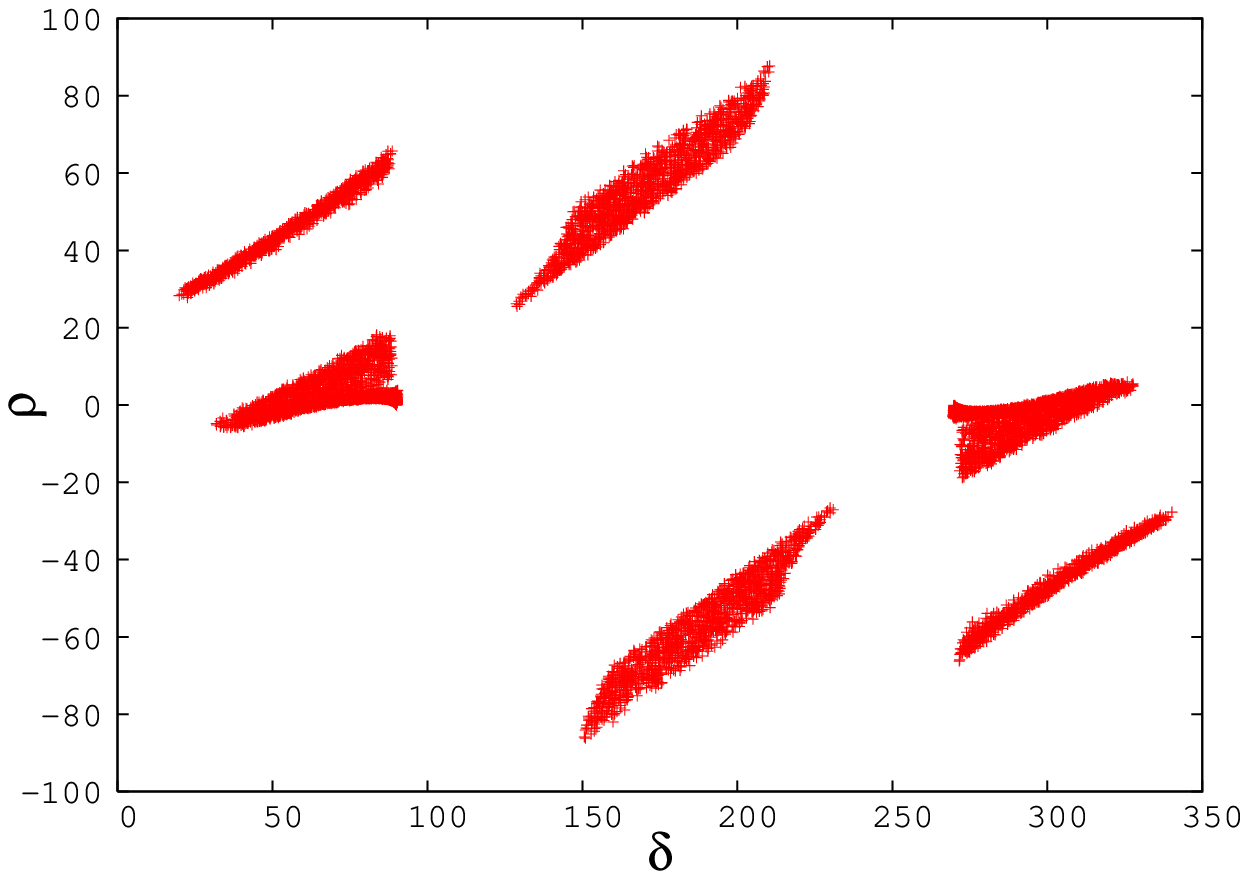}}
\subfigure[]{\includegraphics[width=0.40\columnwidth]{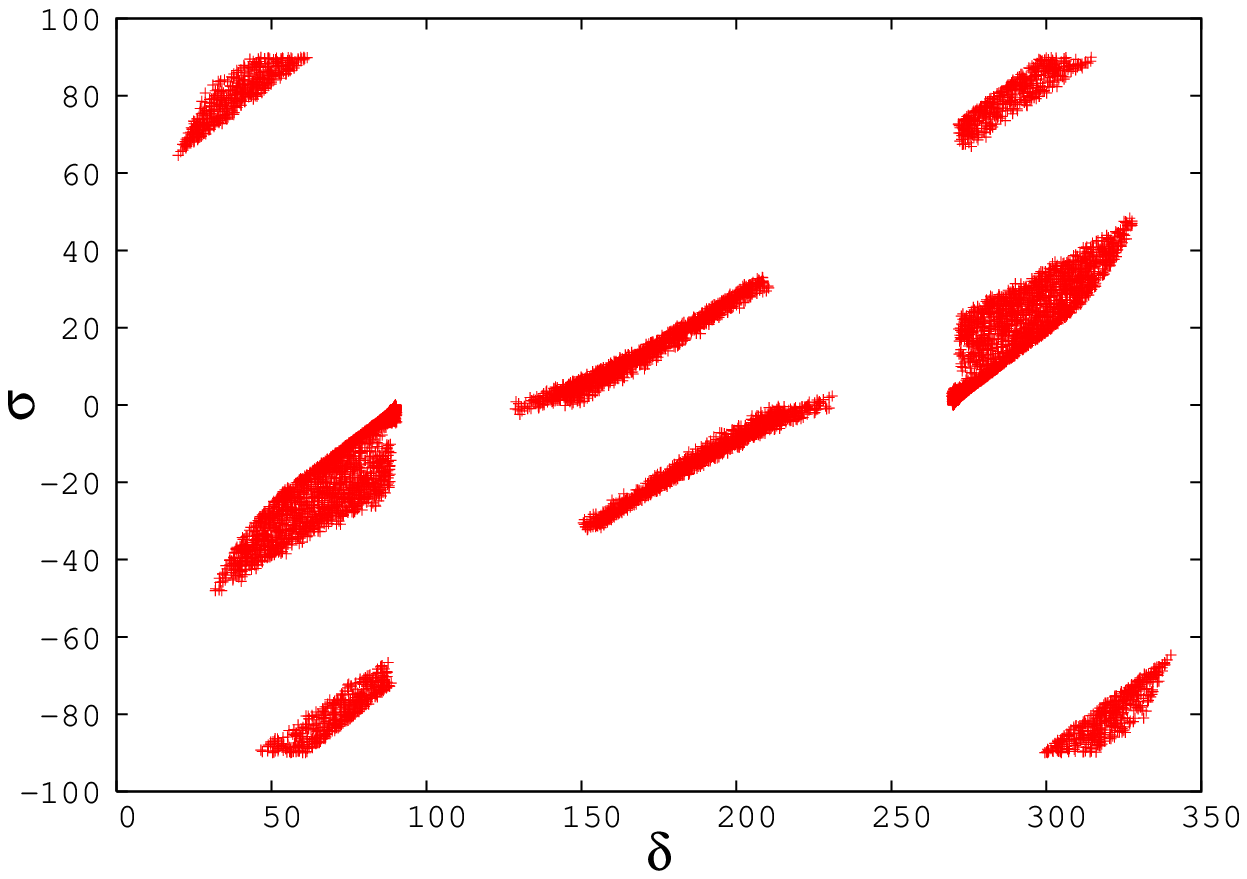}}
\subfigure[]{\includegraphics[width=0.40\columnwidth]{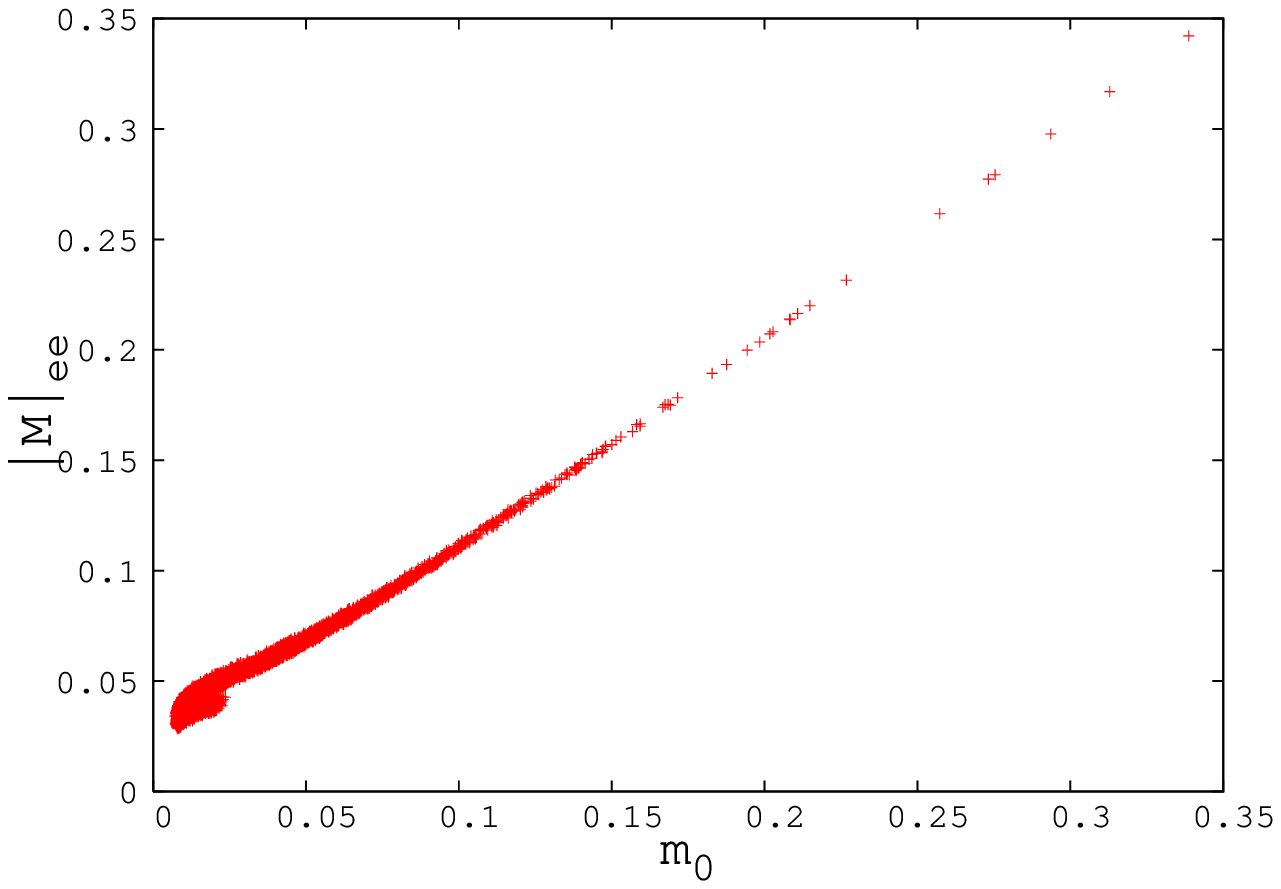}}
\caption{\label{fig4}Correlation plots for texture $B_{2}$ (IS) for type X at 3 $\sigma$ CL. The symbols have their usual meaning. The  $\delta, \rho, \sigma$ are measured in degrees, while $|M|_{ee}$ and $m_{0}$ are in eV units.}
\end{center}
\end{figure}

\begin{figure}[h!]
\begin{center}
\subfigure[]{\includegraphics[width=0.40\columnwidth]{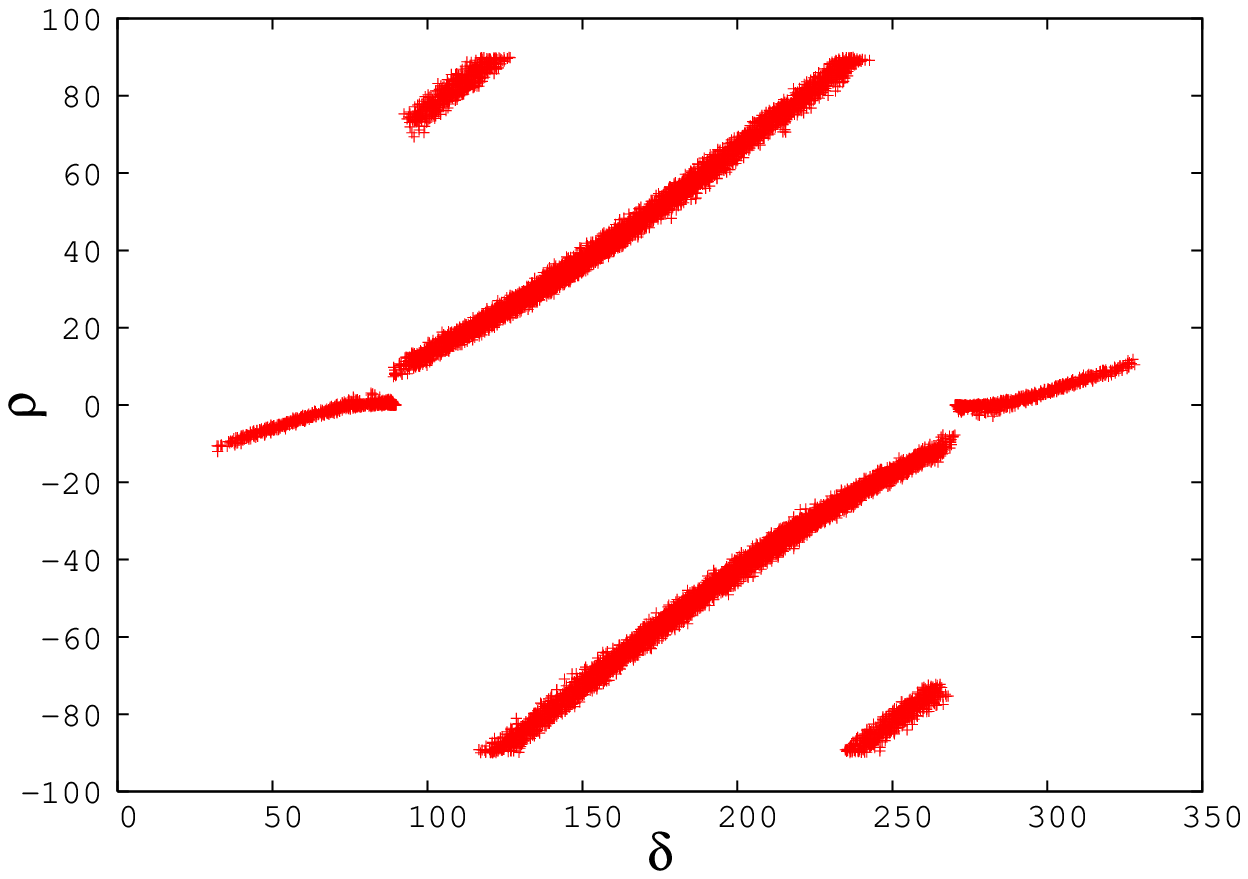}}
\subfigure[]{\includegraphics[width=0.40\columnwidth]{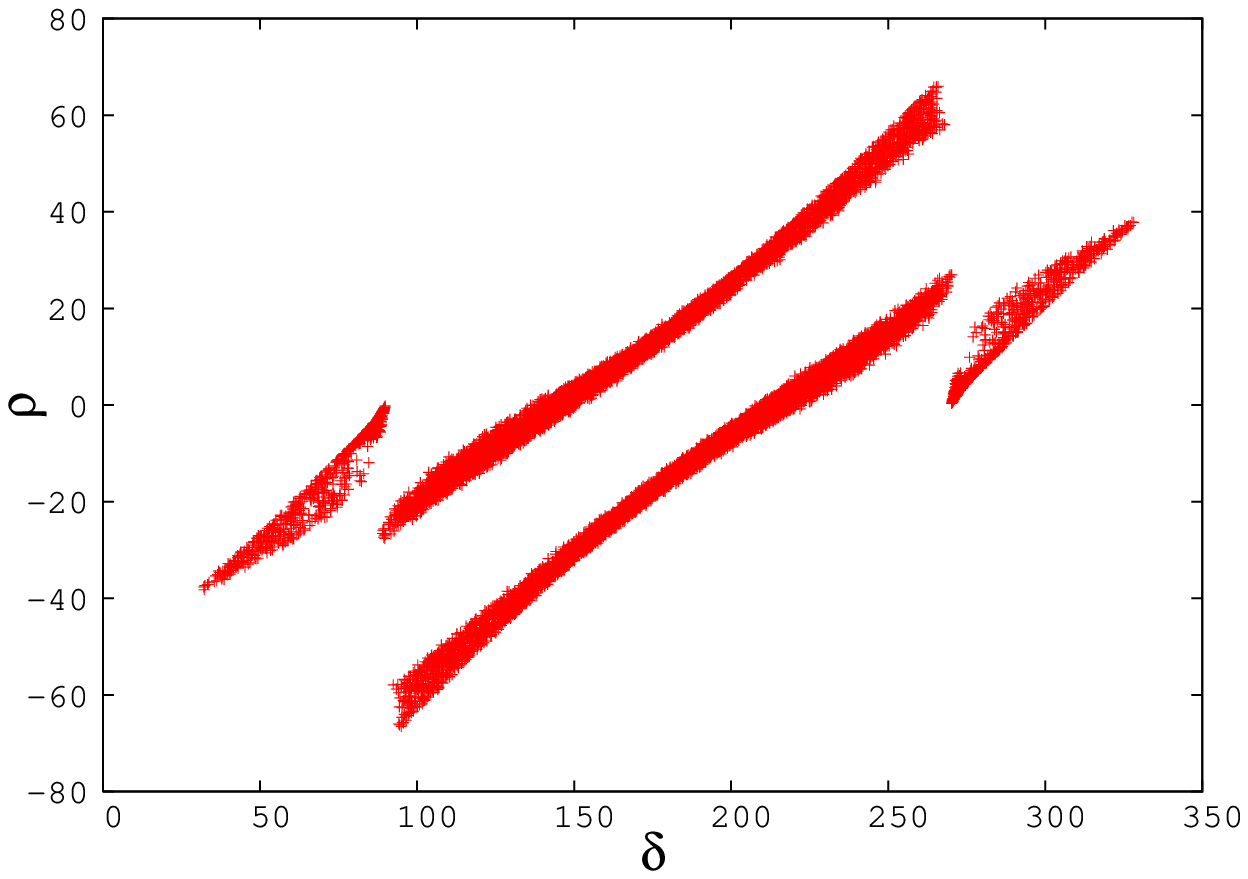}}
\subfigure[]{\includegraphics[width=0.40\columnwidth]{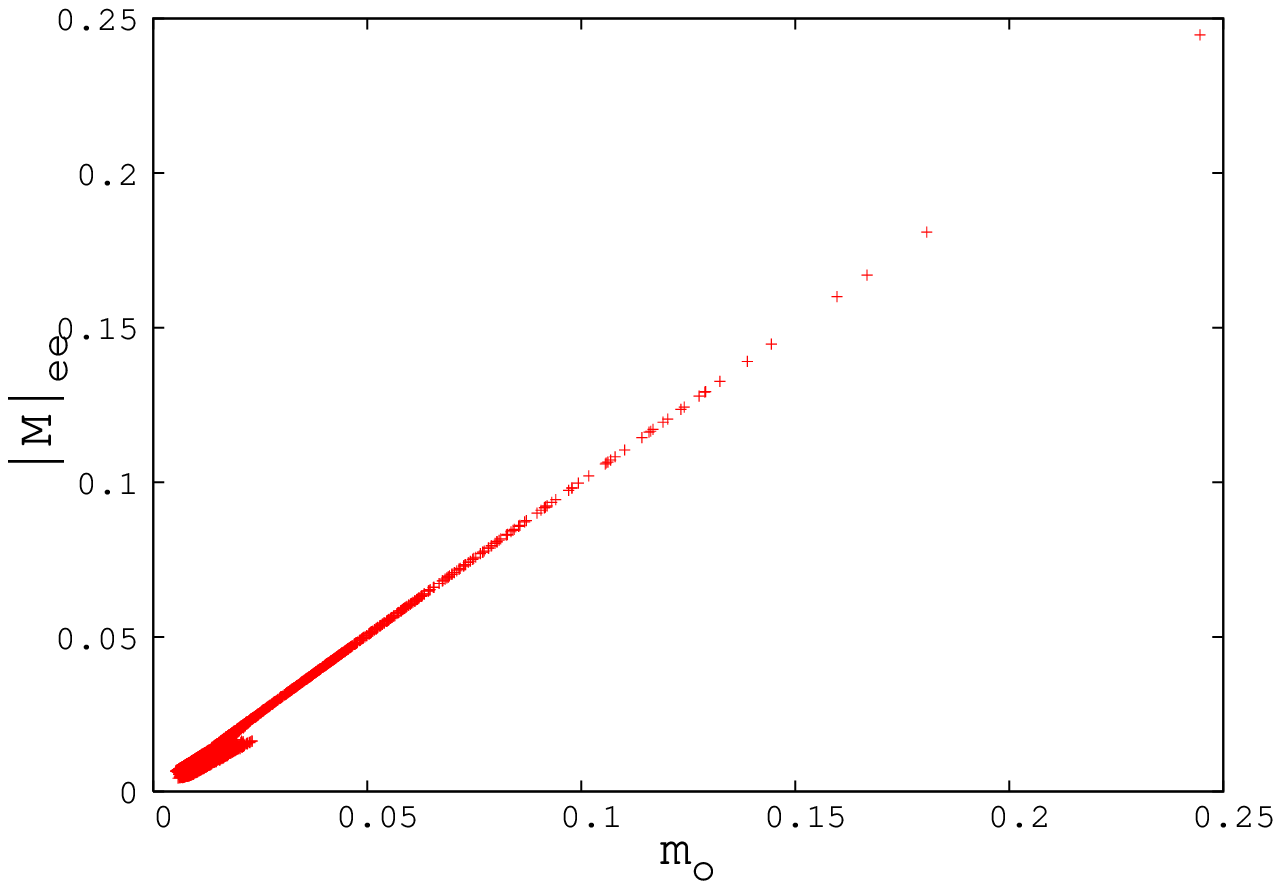}}
\caption{\label{fig5}Correlation plots for texture $B_{2}$ (NS) for type Y at 3 $\sigma$ CL. The symbols have their usual meaning. The  $\delta, \rho, \sigma$ are measured in degrees, while $|M|_{ee}$ and $m_{0}$ are in eV units.}
\end{center}
\end{figure}

\begin{figure}[h!]
\begin{center}
\subfigure[]{\includegraphics[width=0.40\columnwidth]{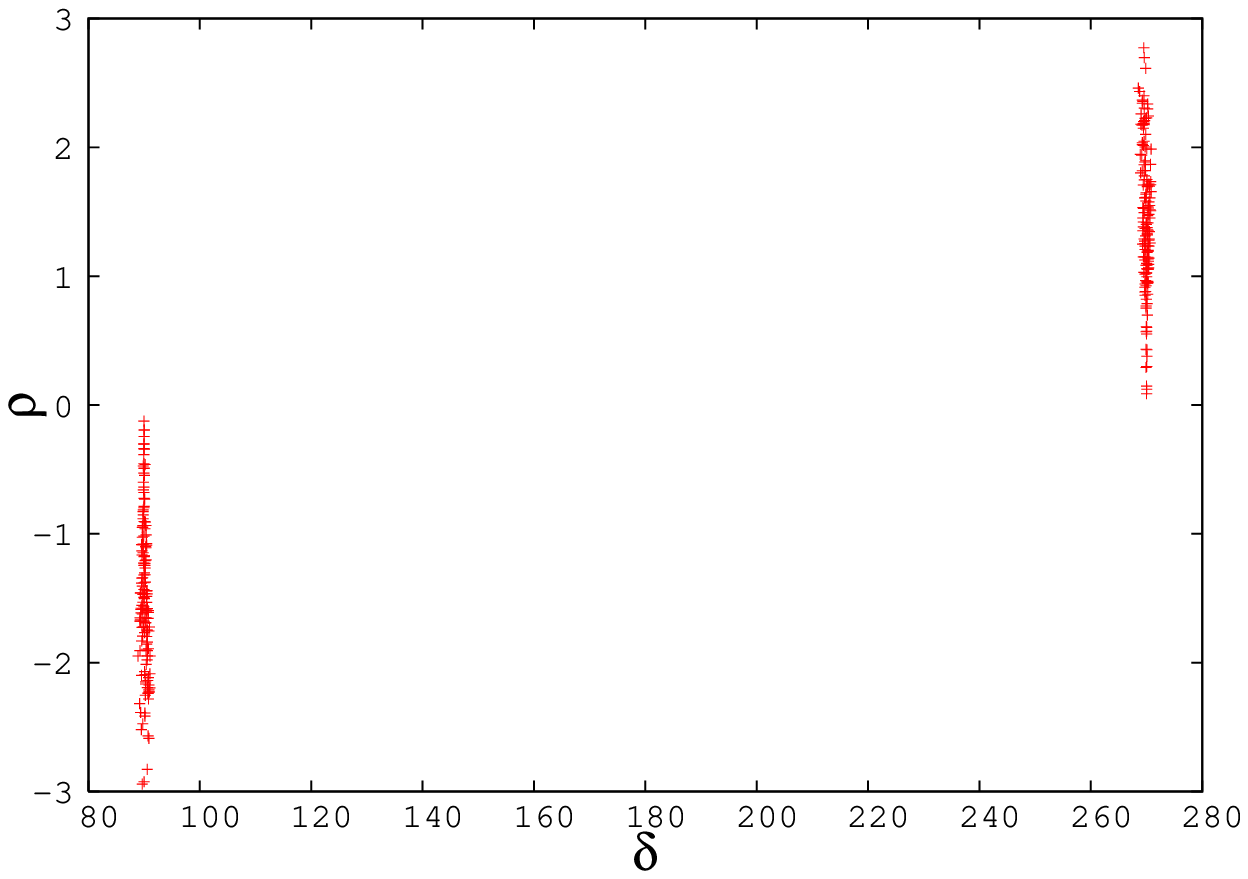}}
\subfigure[]{\includegraphics[width=0.40\columnwidth]{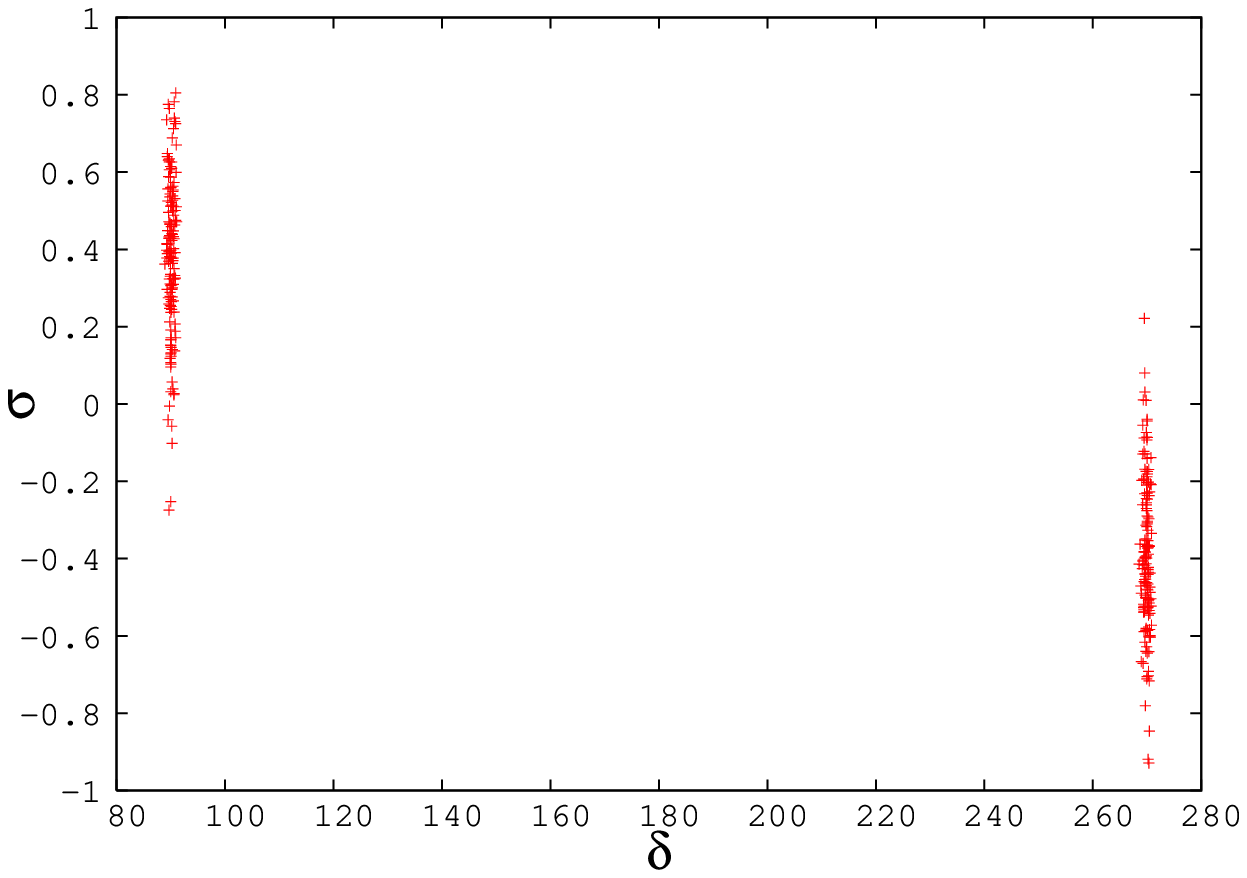}}
\subfigure[]{\includegraphics[width=0.40\columnwidth]{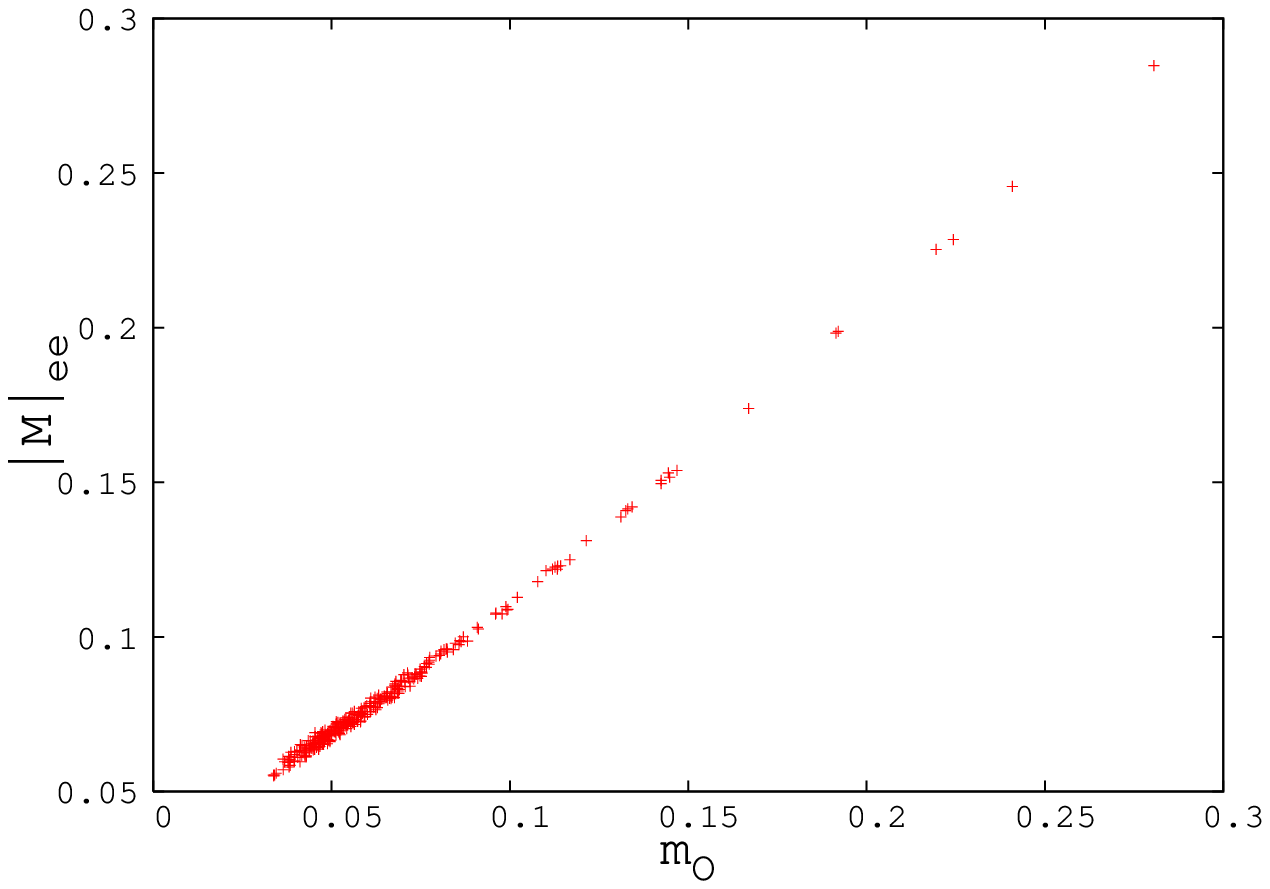}}
\caption{\label{fig6}Correlation plots for texture $B_{2}$ (IS) for type Y at 3 $\sigma$ CL. The symbols have their usual meaning. The  $\delta, \rho, \sigma$ are measured in degrees, while $|M|_{ee}$ and $m_{0}$ are in eV units.}
\end{center}
\end{figure}

\begin{figure}[h!]
\begin{center}
\subfigure[]{\includegraphics[width=0.40\columnwidth]{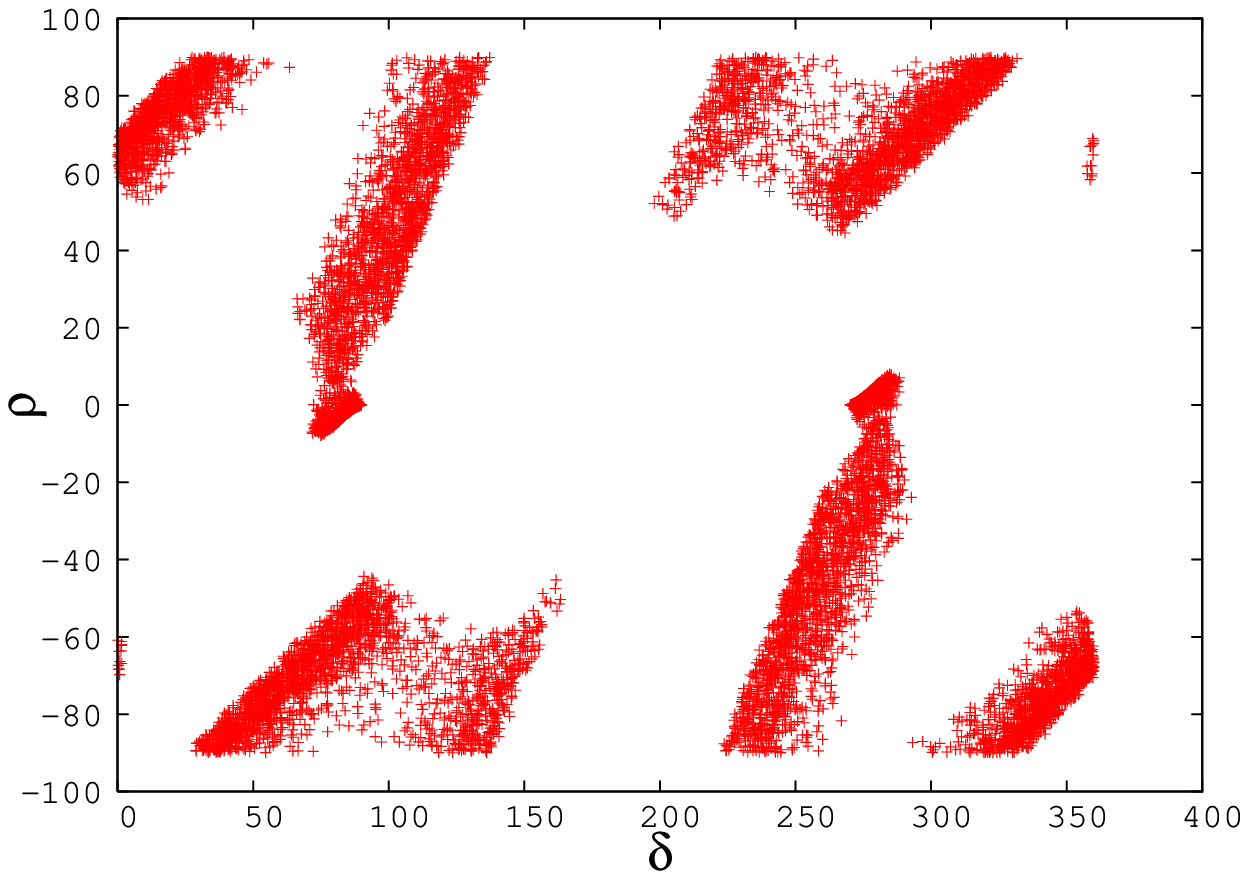}}
\subfigure[]{\includegraphics[width=0.40\columnwidth]{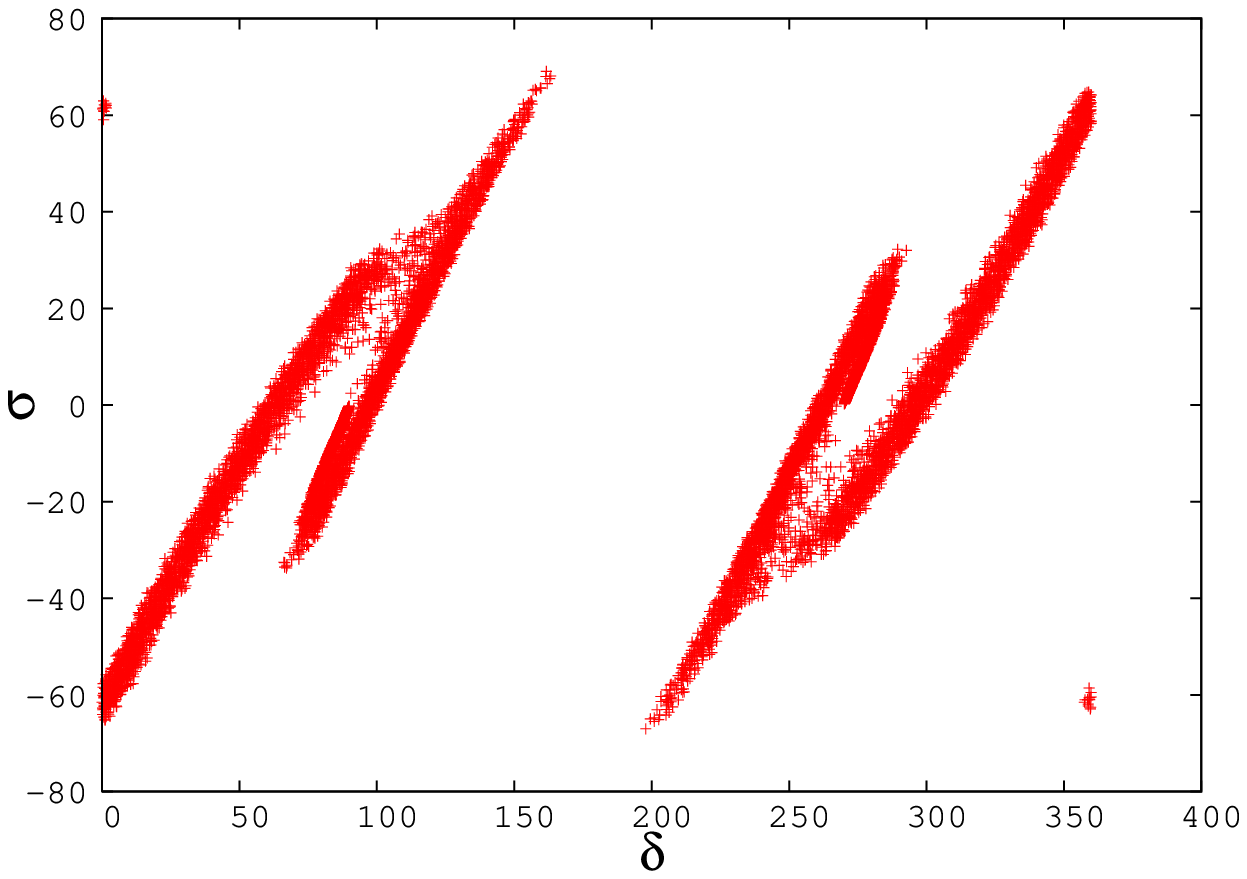}}
\subfigure[]{\includegraphics[width=0.40\columnwidth]{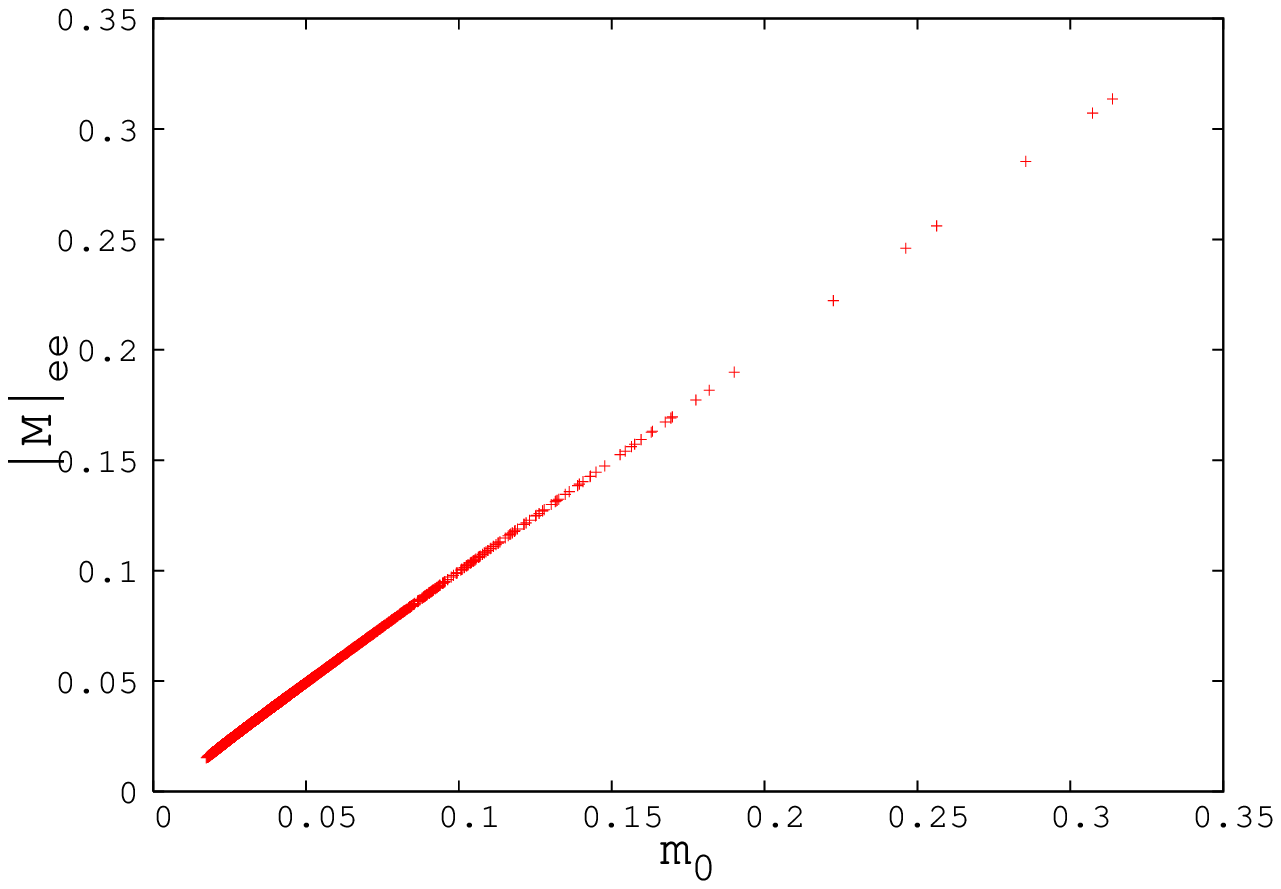}}
\caption{\label{fig7}Correlation plots for texture $D_{7}$ (NS) for type X at 3 $\sigma$ CL. The symbols have their usual meaning. The  $\delta, \rho, \sigma$ are measured in degrees, while $|M|_{ee}$ and $m_{0}$ are in eV units.}
\end{center}
\end{figure}

\begin{figure}[h!]
\begin{center}
\subfigure[]{\includegraphics[width=0.40\columnwidth]{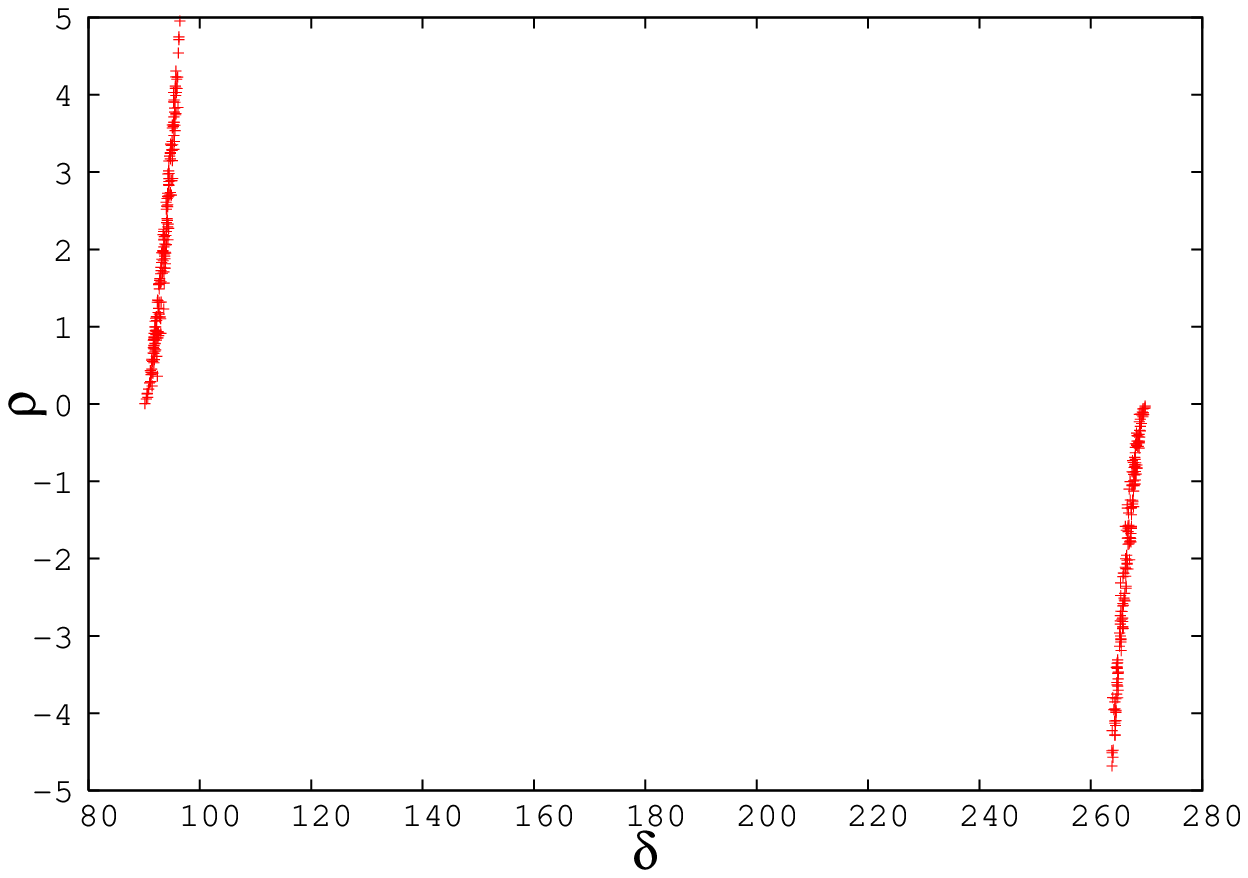}}
\subfigure[]{\includegraphics[width=0.40\columnwidth]{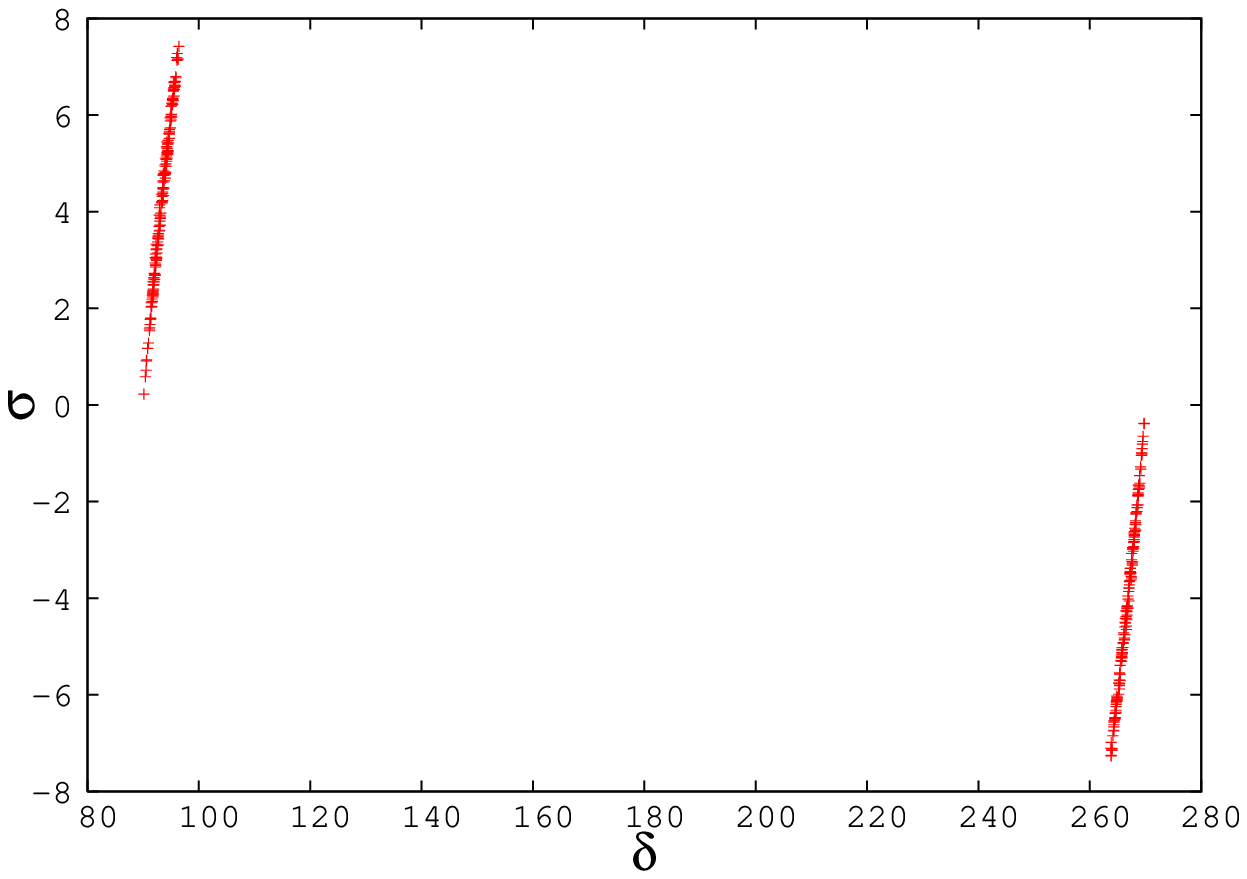}}
\subfigure[]{\includegraphics[width=0.40\columnwidth]{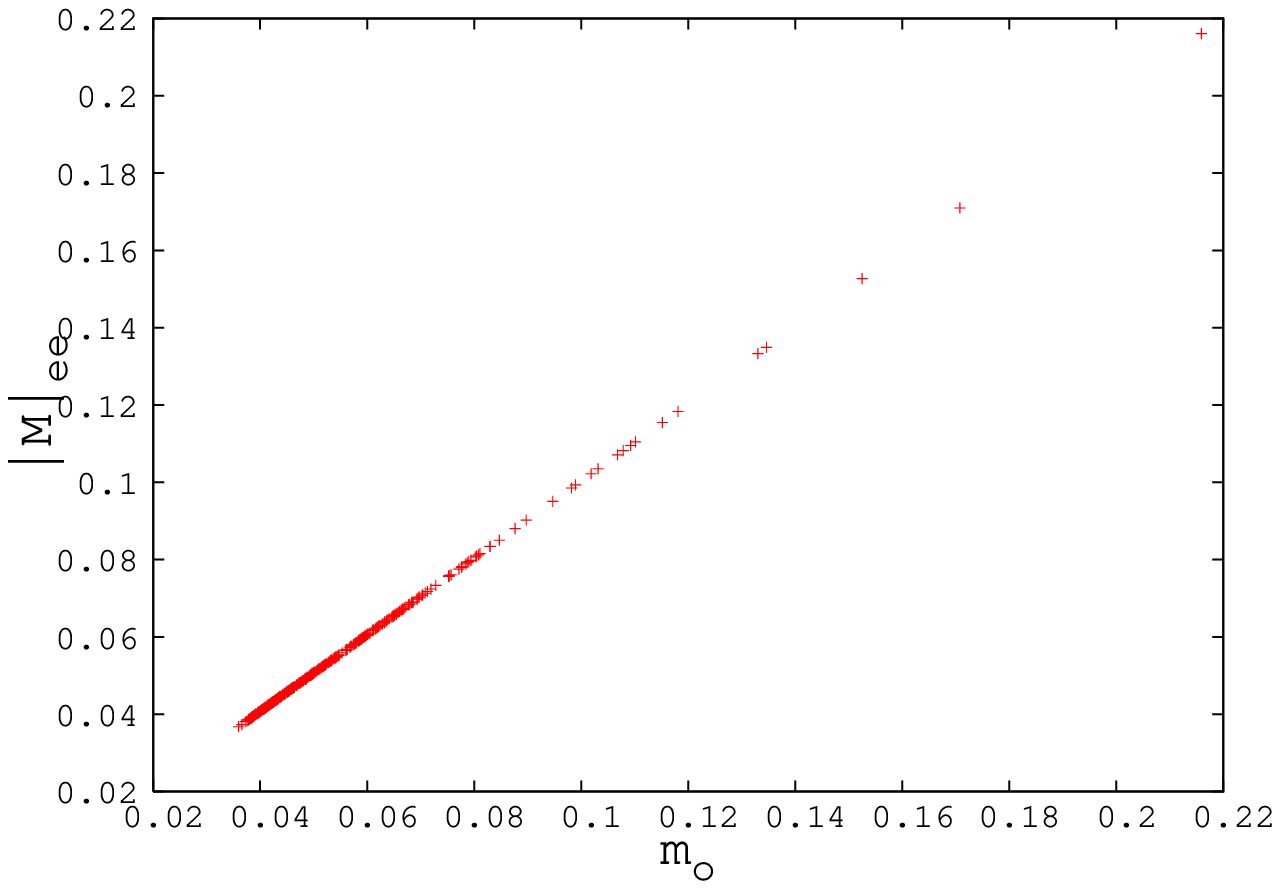}}
\caption{\label{fig8}Correlation plots for texture $D_{7}$ (IS) for type X at 3 $\sigma$ CL. The symbols have their usual meaning. The  $\delta, \rho, \sigma$ are measured in degrees, while $|M|_{ee}$ and $m_{0}$ are in eV units.}
\end{center}
\end{figure}

\begin{figure}[h!]
\begin{center}
\subfigure[]{\includegraphics[width=0.40\columnwidth]{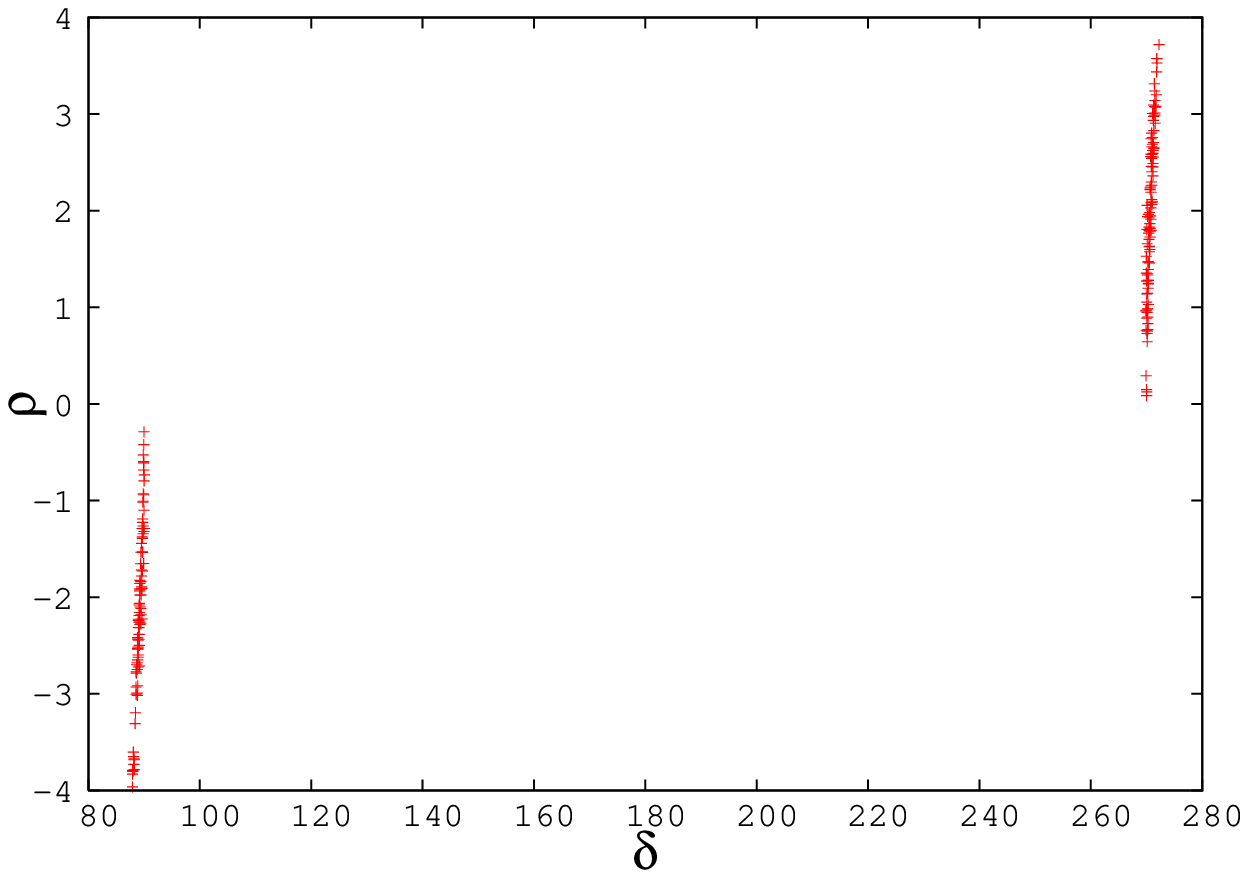}}
\subfigure[]{\includegraphics[width=0.40\columnwidth]{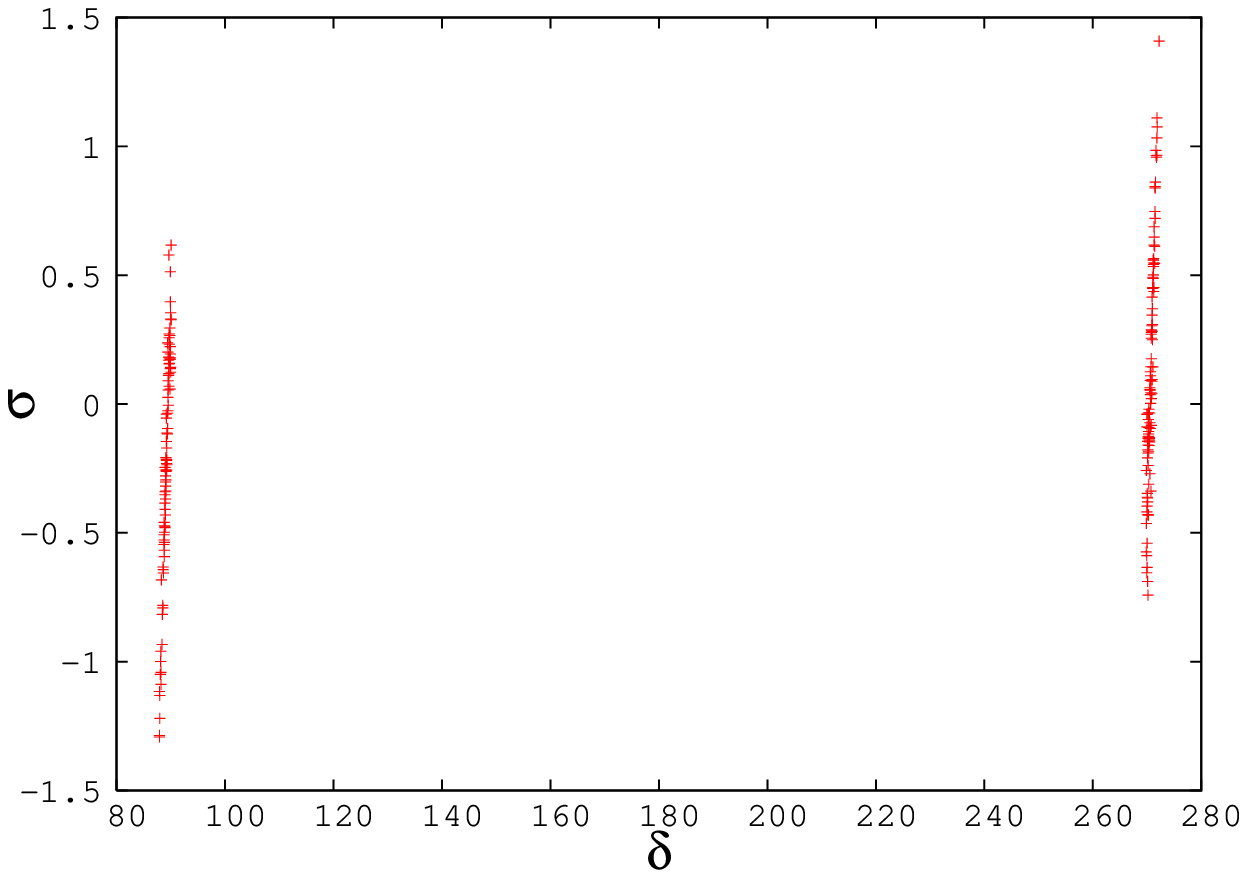}}
\subfigure[]{\includegraphics[width=0.40\columnwidth]{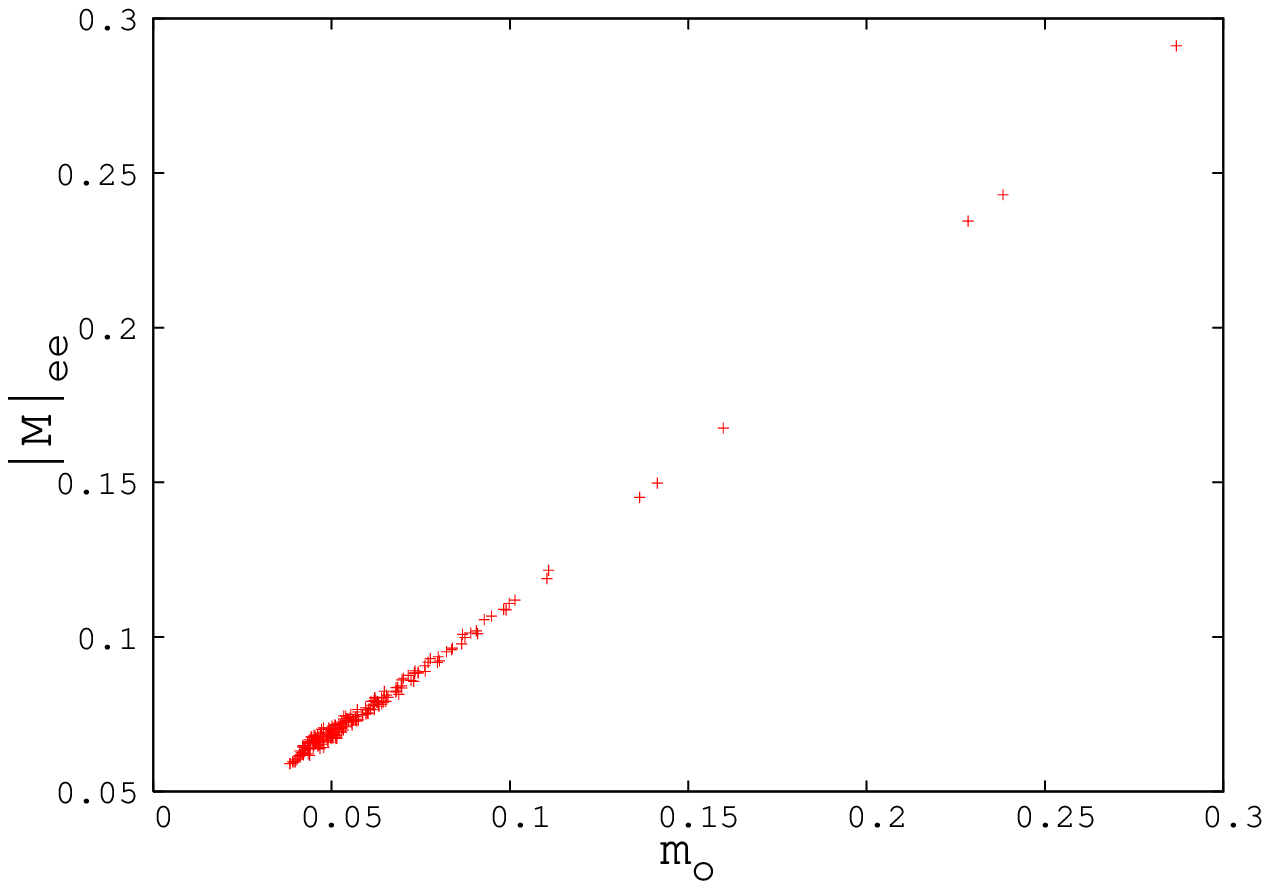}}
\caption{\label{fig9}Correlation plots for texture $D_{7}$ (NS) for type Y at 3 $\sigma$ CL. The symbols have their usual meaning. The  $\delta, \rho, \sigma$ are measured in degrees, while $|M|_{ee}$ and $m_{0}$ are in eV units.}
\end{center}
\end{figure}

\begin{figure}[h!]
\begin{center}
\subfigure[]{\includegraphics[width=0.40\columnwidth]{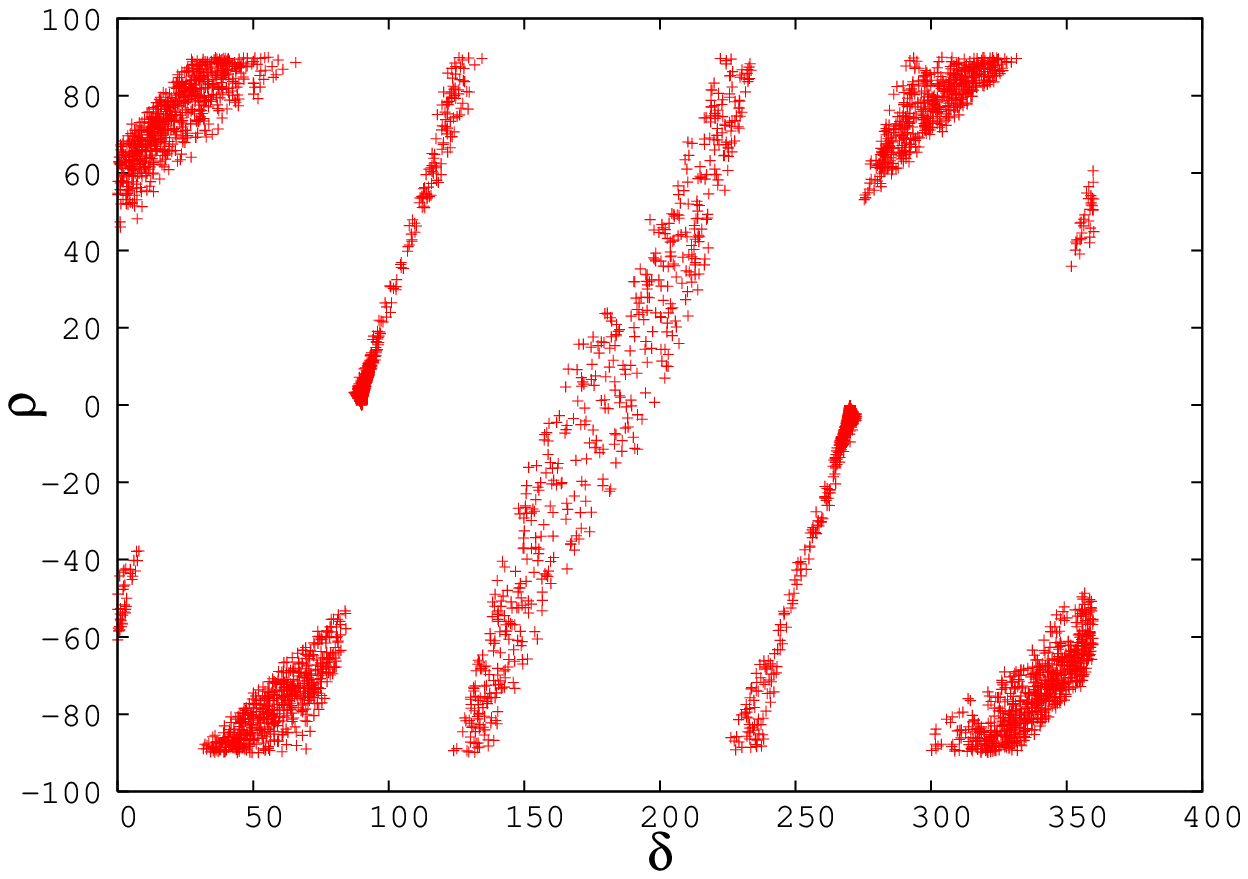}}
\subfigure[]{\includegraphics[width=0.40\columnwidth]{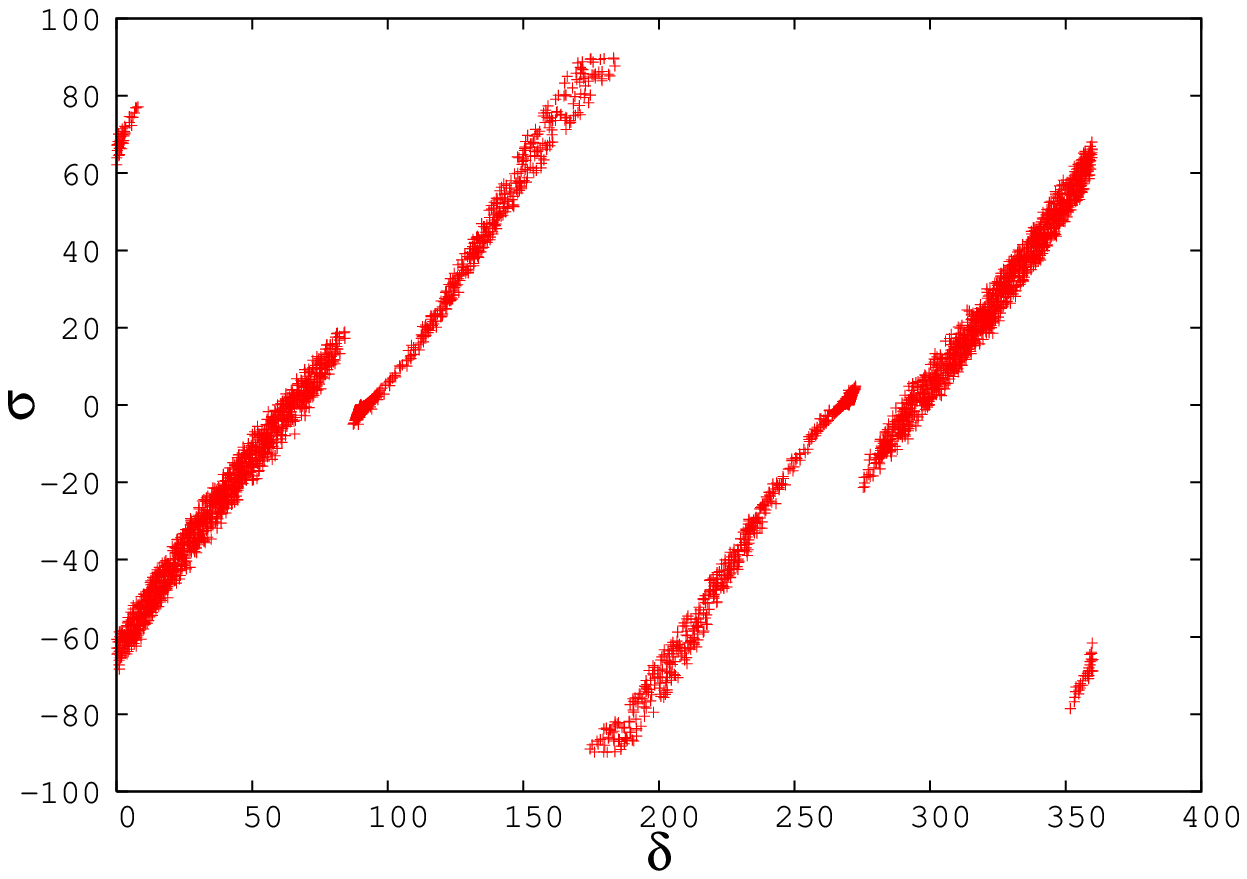}}
\subfigure[]{\includegraphics[width=0.40\columnwidth]{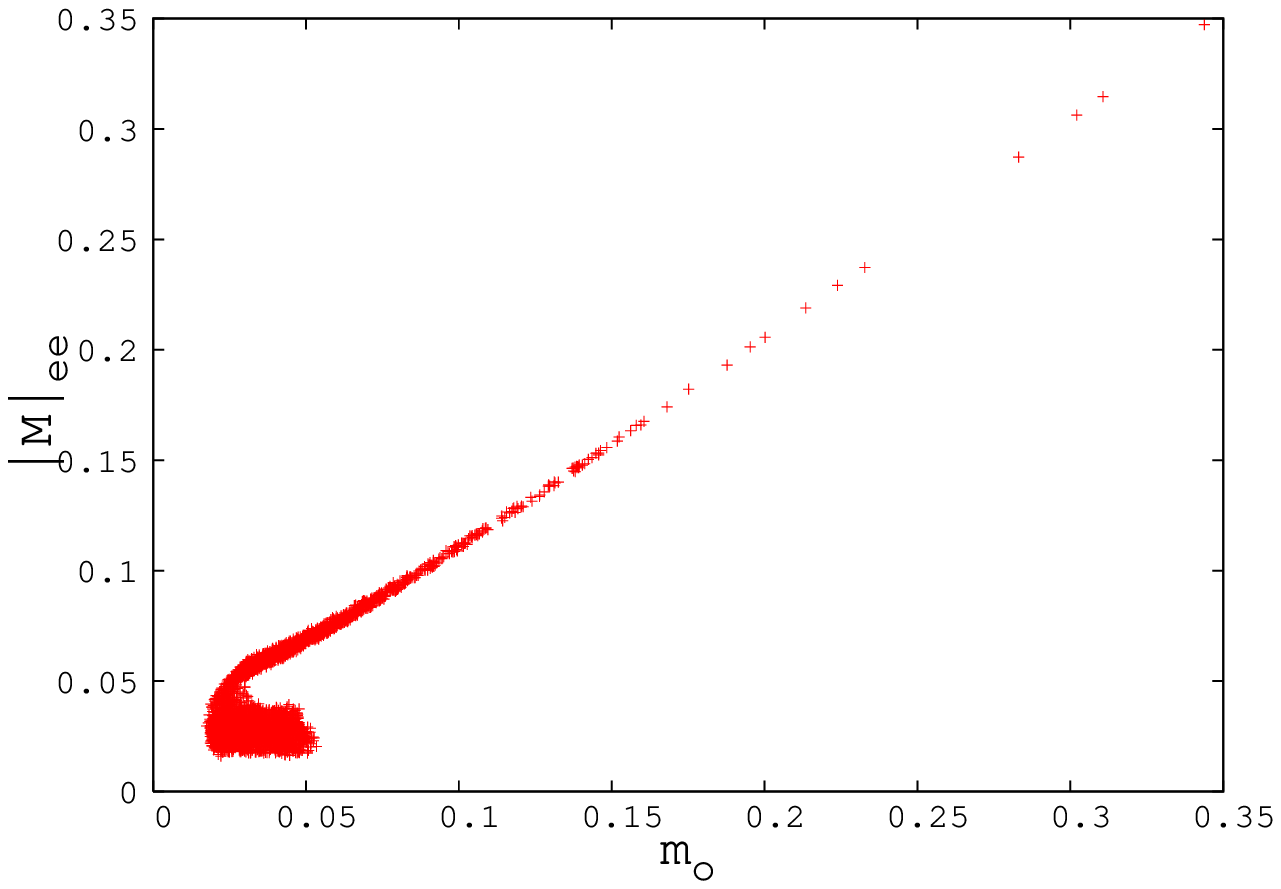}}
\caption{\label{fig10}Correlation plots for texture $D_{7}$ (IS) for type Y at 3 $\sigma$ CL. The symbols have their usual meaning. The  $\delta, \rho, \sigma$ are measured in degrees, while $|M|_{ee}$ and $m_{0}$ are in eV units.}
\end{center}
\end{figure}

\begin{figure}[h!]
\begin{center}
\subfigure[]{\includegraphics[width=0.45\columnwidth]{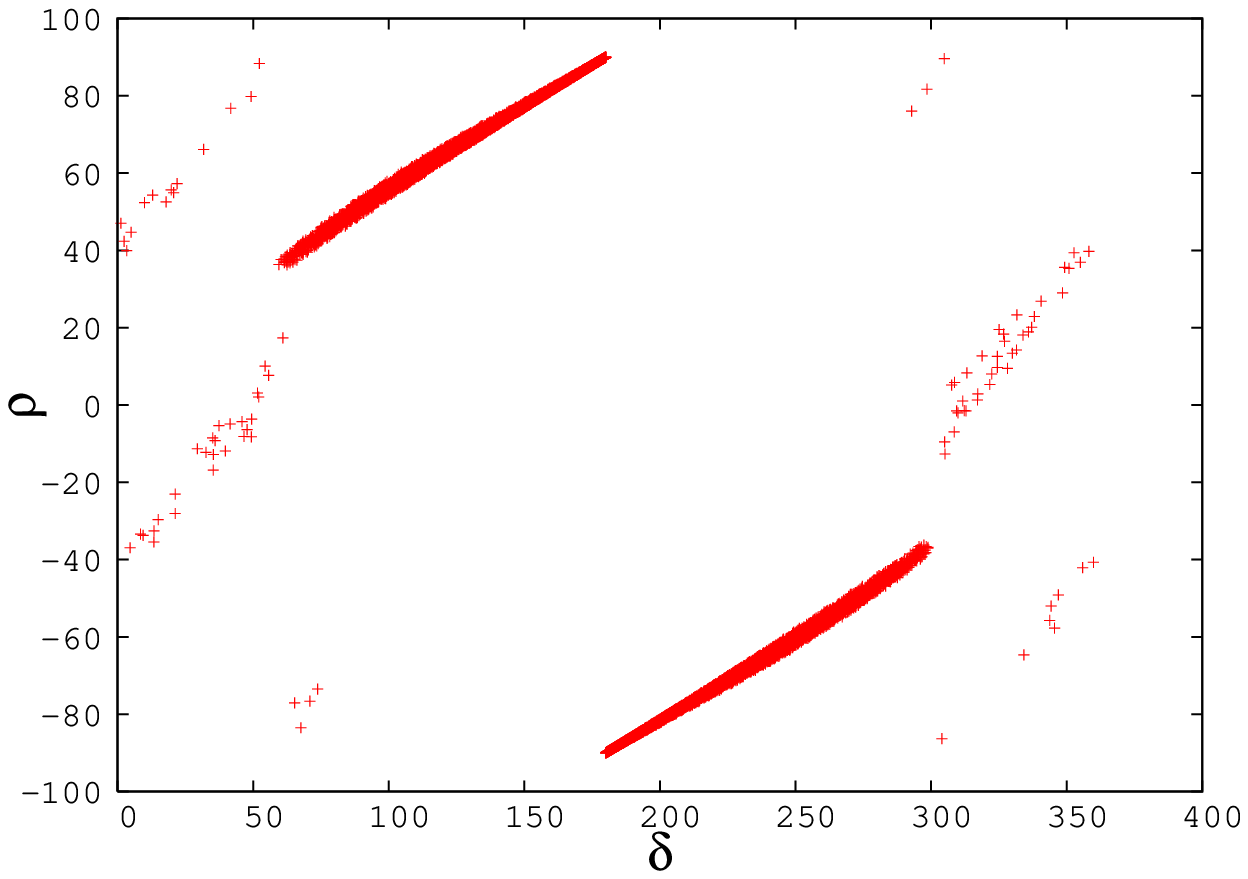}}
\subfigure[]{\includegraphics[width=0.45\columnwidth]{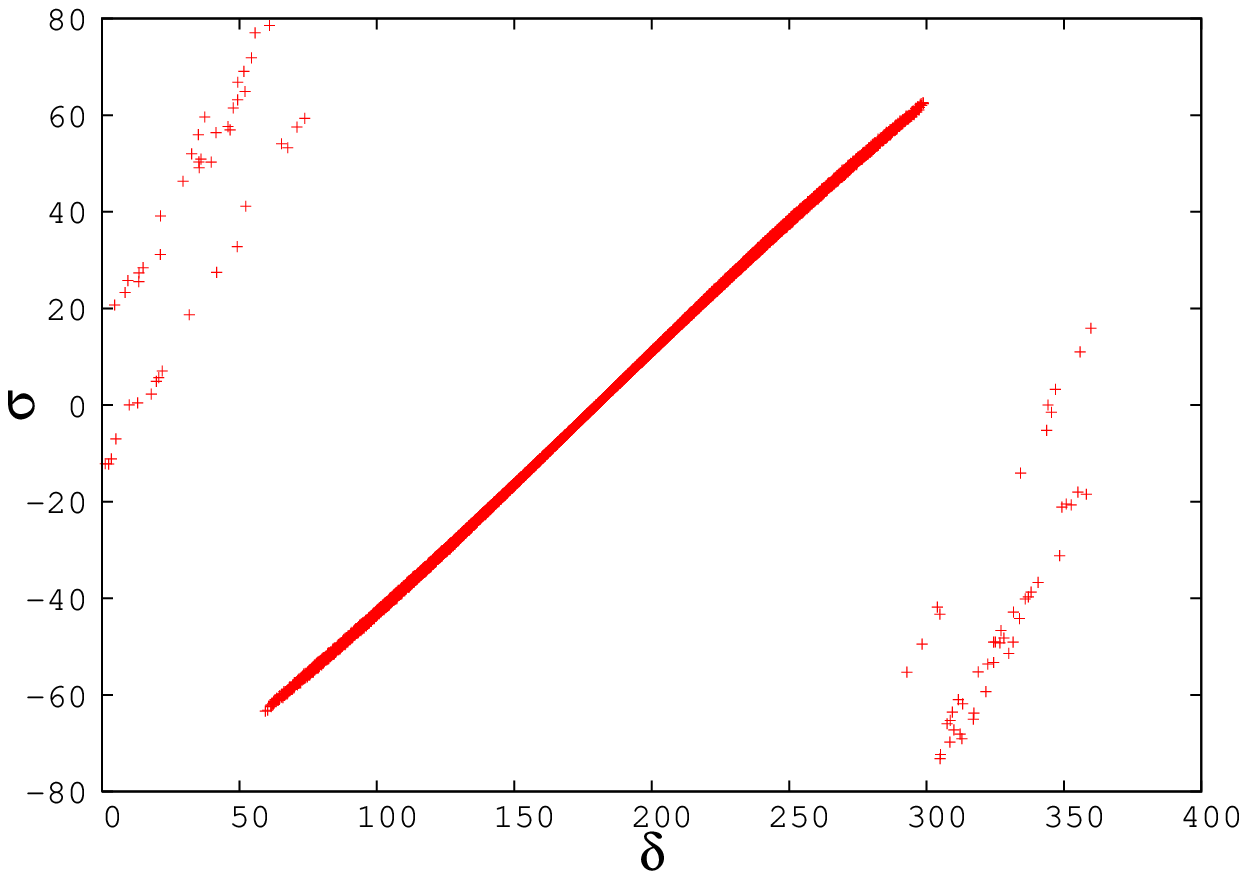}}
\subfigure[]{\includegraphics[width=0.45\columnwidth]{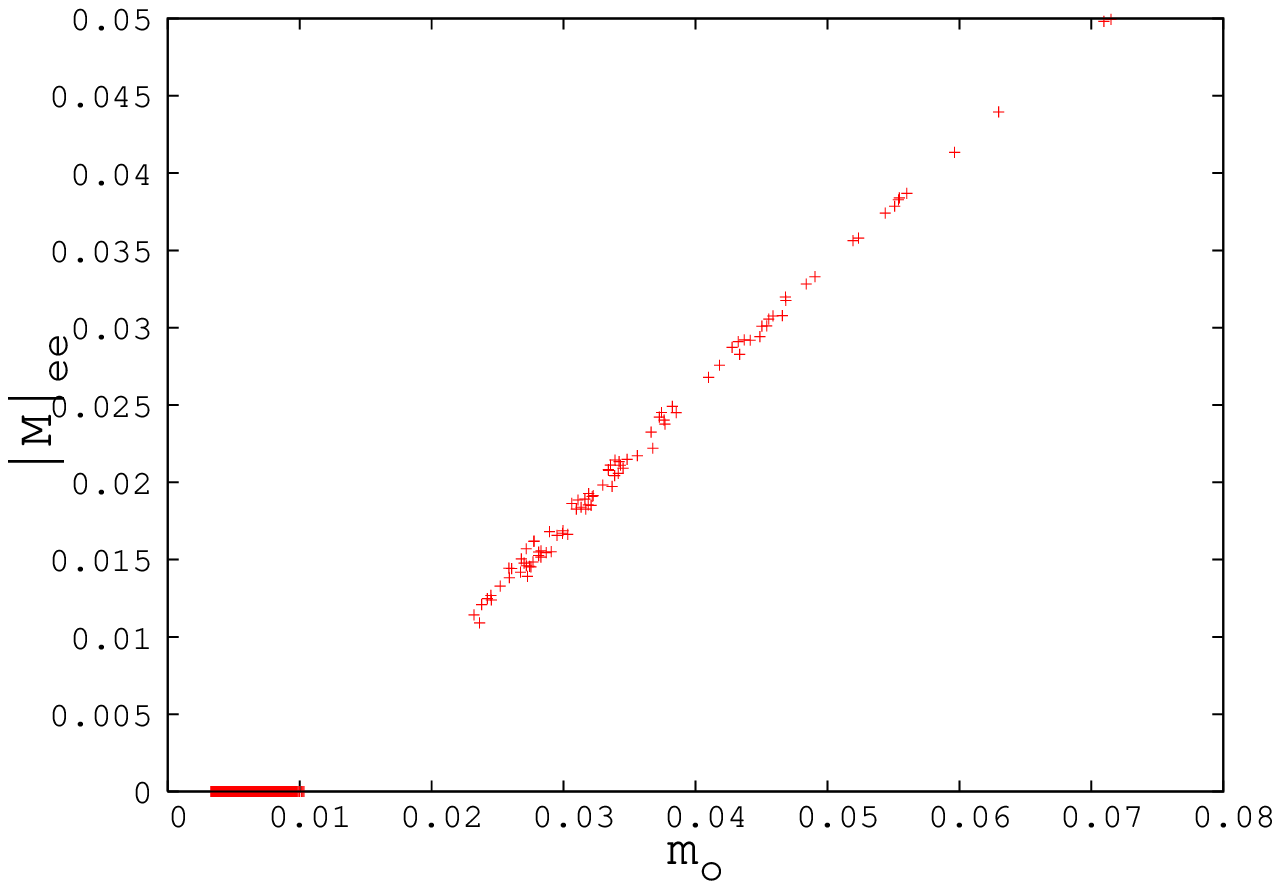}}
\caption{\label{fig11}Correlation plots for texture $E_{1}$ (NS) for type X at 3 $\sigma$ CL. The symbols have their usual meaning. The  $\delta, \rho, \sigma$ are measured in degrees, while $|M|_{ee}$ and $m_{0}$ are in eV units..}
\end{center}
\end{figure}

\begin{figure}[h!]
\begin{center}
\subfigure[]{\includegraphics[width=0.45\columnwidth]{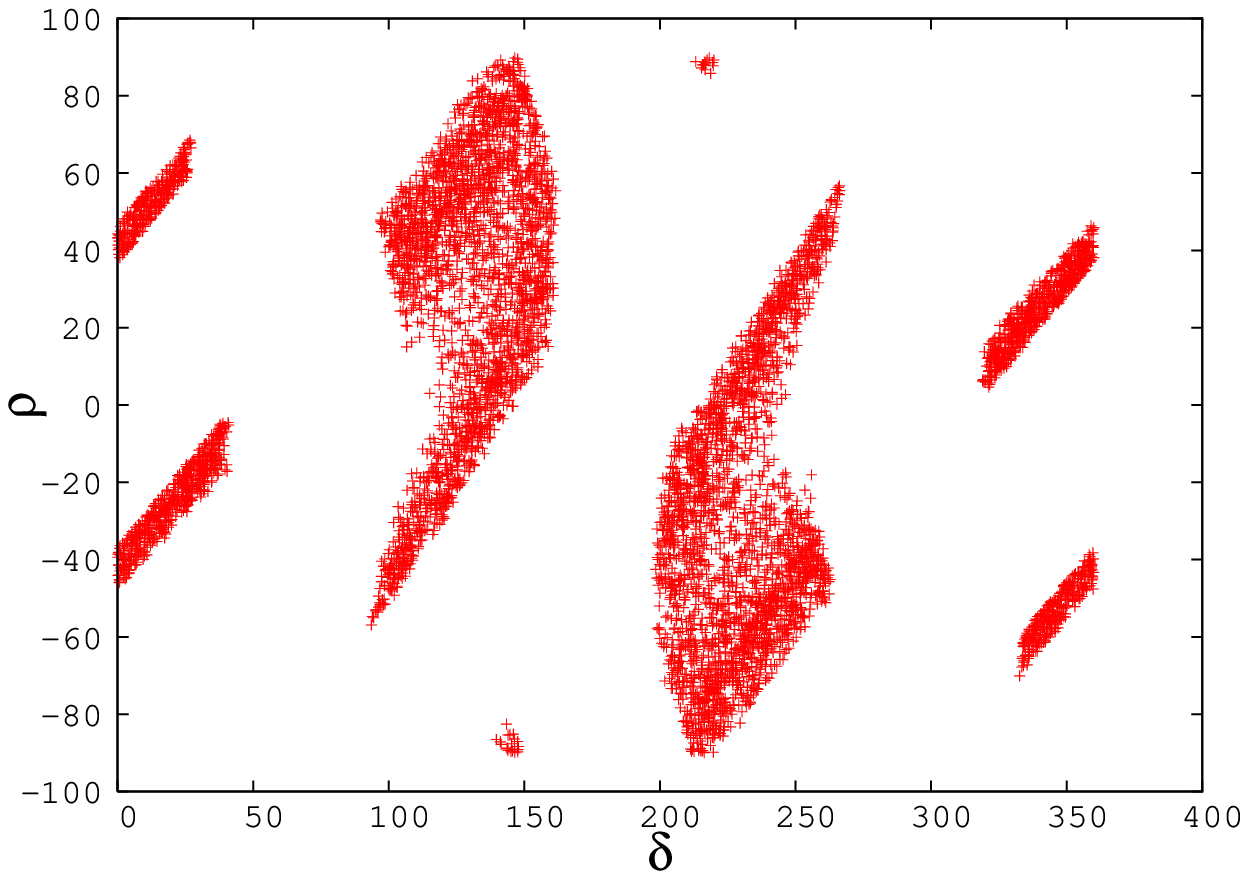}}
\subfigure[]{\includegraphics[width=0.45\columnwidth]{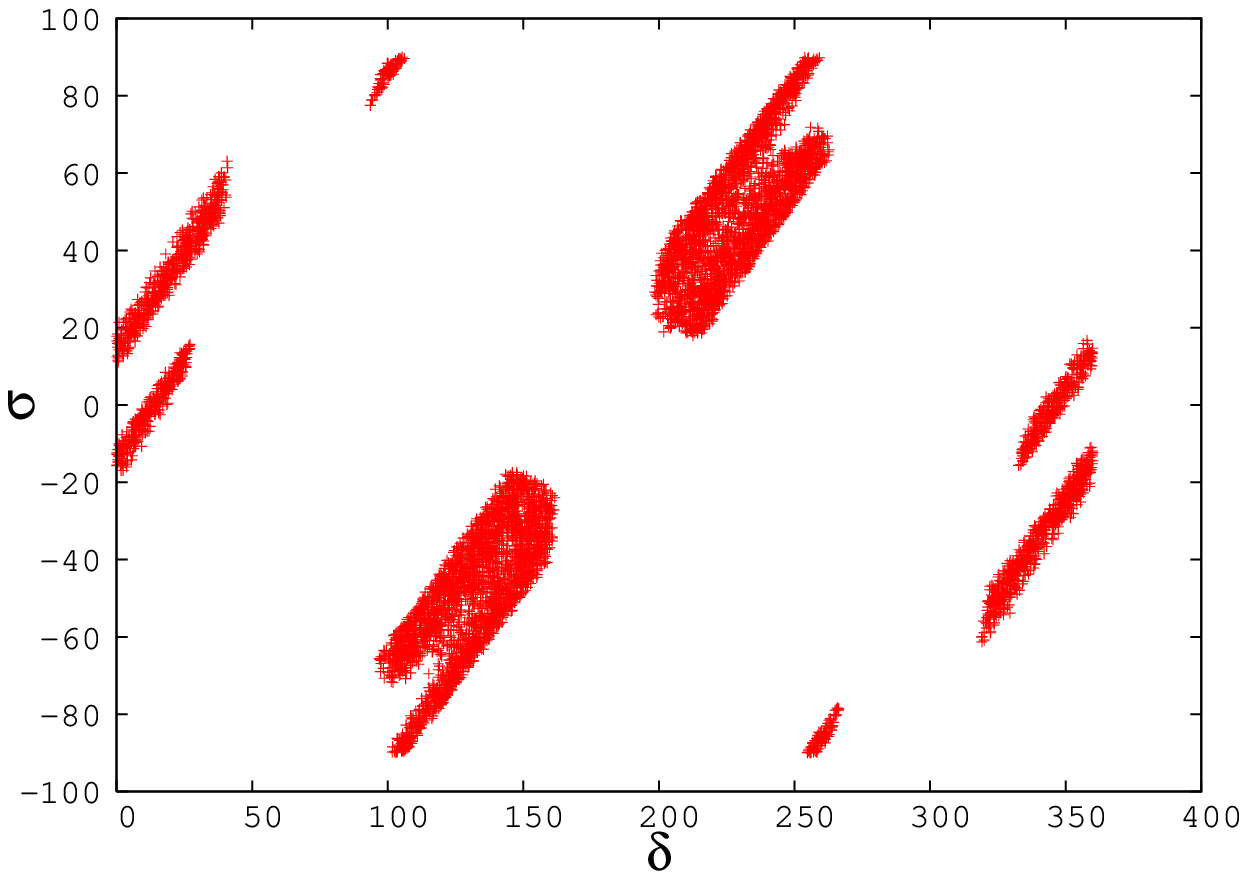}}
\subfigure[]{\includegraphics[width=0.45\columnwidth]{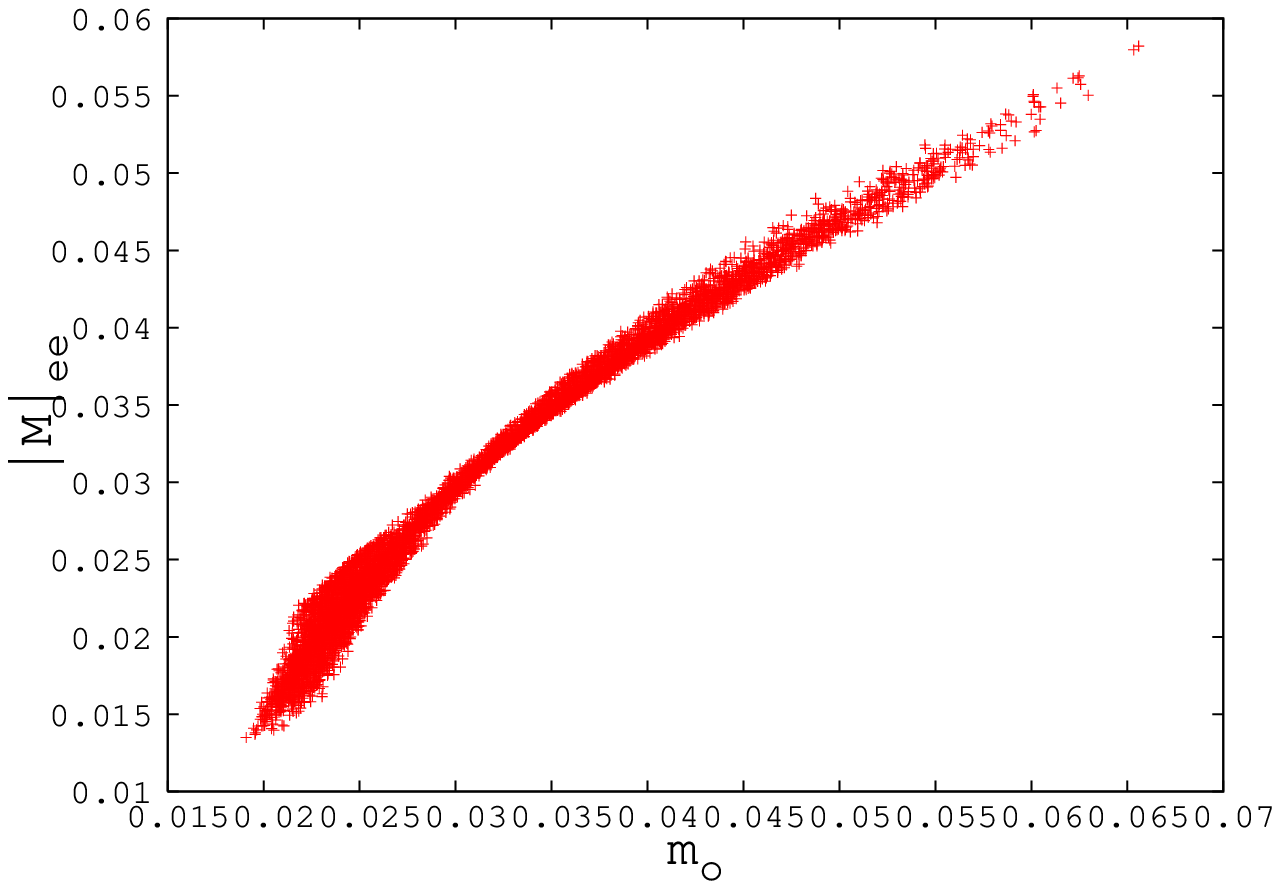}}
\caption{\label{fig12}Correlation plots for texture $E_{1}$ (IS) for type Y at 3 $\sigma$ CL. The symbols have their usual meaning. The  $\delta, \rho, \sigma$ are measured in degrees, while $|M|_{ee}$ and $m_{0}$ are in eV units.}
\end{center}
\end{figure}

 In Figs. \ref{fig1}, \ref{fig2}, \ref{fig3}, \ref{fig4}, \ref{fig5}, \ref{fig6}, \ref{fig7}, \ref{fig8}, \ref{fig9}, \ref{fig10}, \ref{fig11}, \ref{fig12} we demonstrate the correlations for $A_{1}$, $B_{2}$, $D_{7}$ and $E_{1}$ cases. Since there are large number of viable cases, therefore it is not practically possible to show all the plots. We have simply taken arbitrary independent cases from each category for the purpose of illustration of our results. The predictions regarding three CP-violating phases ($\rho, \sigma, \delta$), effective neutrino mass $|M|_{ee}$ and lowest neutrino mass  $m_{o}$ for all the allowed cases of type X and type Y textures have been encapsulated in Table \ref{tab3},\ref{tab4} ,\ref{tab5}, \ref{tab6}. Before proceeding further, it is worth pointing out that the phenomenological results for $\rho, \sigma, \delta$, $|M|_{ee}$ and $m_{o}$ have been obtained using the two possible solutions of $\lambda_{13}$ and $\lambda_{23}$ respectively[Eqs. \ref{eq20},\ref{eq21}, \ref{eq22}, \ref{eq23}]. All the sixty phenomenologically possible cases have been divided into six categories A, B, C, D, E, F. Among them large number of cases are found to overlap in their predictions regarding $\delta$, $\rho$, $\sigma$  $|M|_{ee}$ and $m_{0}$ and  are related via permutation symmetry as pointed out earlier.  The main results and the discussion are summarized as follows:\\

\textbf{Category A}: In Category A, all the ten cases  $A_{1}$, $A_{2}$, $A_{3}$, $A_{4}, A_{5}, A_{6}, A_{7}, A_{8}, A_{9}, A_{10}$ are found to be viable with the data at 3$\sigma$ CL for type X structure, and normal mass spectrum (NS) remain ruled out for all these cases [Table \ref{tab3}]. On the other hand, only four $A_{1}$, $A_{4}$, $A_{5}$, $A_{6}$ seem to be viable with current oscillation data for type Y, while inverted mass spectrum (IS) is ruled out for these cases.

For both type X and Y, no noticeable constraint has been found on the parameter space of  CP violating phases ($\rho, \sigma, \delta$). For type X , all the viable cases  predict the value of $|M|_{ee}$  in the range of 0.01eV to 0.05eV. This prediction lies well within the sensitivity limit of neutrinoless double beta decay experiments as mentioned above. On the other hand, for type Y, $|M|_{ee}$ is predicted to be zero implying that neutrinoless double beta decay is forbidden. Also the lower bound on lowest neutrino mass ($m_{o}$) is found to be extremely small ($\sim 10^{-3}$ or less) for all the viable cases of type X and type Y structure [Table \ref{tab3}]. For the purpose of illustration, we have presented the correlation plots for $A_{1}$ indicating the parameter space of $\rho, \sigma, \delta$, $|M|_{ee}$ and lowest neutrino mass ($m_{o})$ [Figs.\ref{fig1},\ref{fig2}]. \\

\textbf{Category B (C)}: In Category B, all the ten possible cases are allowed for both type X and type Y  structure, respectively at 3$\sigma$ CL [Table\ref{tab4}].  Cases $B_{2,3,4,5,8,9,10}$ allow both NS as well as IS for type X, while cases $B_{1, 2,3,4,5,8,9,10}$ allow both NS and IS for type Y. As mentioned earlier, cases of Category B are related to cases belonging to Category C  via permutation symmetry, therefore we can obtain the results for Category C from B by using Eq.\ref{eq41}.


Type X cases $B_{1}$ (IS), $B_{2}$ (IS), $B_{3}$ (NS, IS), $B_{4}$ (IS), $B_{5}$ (NS, IS), $B_{7}$ (NS)
, $B_{8}$ (IS), $B_{9}$ (NS, IS), $B_{10}$ (IS), $C_{1}$ (IS), $C_{2} $(IS), $C_{4} $(IS), $C_{5} $(NS, IS), $C_{6}$(NS, IS), $C_{7}$ (NS), $C_{8}$ (IS), $C_{8}$ (IS), $C_{9}$ (NS, IS), $C_{10} $(IS) cover literally the complete range of $\delta$. However, for $B_{2}$ (NS), $B_{4}$ (NS), $B_{6}$(NS), $B_{8}$(NS), $B_{10}$(NS), $C_{3}$(NS), $C_{5}$(NS), $C_{7}$(NS), $C_{8}$(NS) and $C_{10}$(NS) the parameter space of $\delta$ is found to be reduced to an appreciable extent [Table \ref{tab4}].

On the other hand, type Y cases  $B_{1}$ (NS), $B_{2} $(NS), $B_{3}$ (NS, IS), $B_{4}$ (NS), $B_{5}$(NS, IS), $B_{7} $(IS), $B_{8} $(NS), $B_{9}$ (NS, IS), $B_{10}$ (NS), $C_{1}$ (NS), $C_{2}$ (NS), $C_{4}$ (NS), $C_{5}$ (NS, IS), $C_{6}$(NS, IS), $C_{7}$ (IS), $C_{8}$ (NS), $C_{8}$ (NS), $C_{9}$ (NS, IS), $C_{10}$ (NS)  cover approximately the complete range of $\delta$. $B_{1}$(IS), $B_{2}$(IS),  $B_{4}$(IS), $B_{6}$(IS), $ B_{8}$(IS), $B_{10}$(IS), $C_{1}$(IS), $C_{3}$(IS), $C_{5}$(IS), $C_{7}$(IS), $C_{8}$(IS) and $C_{10}$(IS) the $\delta$ is found to be reduced appreciably [Table \ref{tab4}].

From the analysis, it is found that textures $B_{2}, B_{4}, C_{5}$ and $C_{7}$ belonging to type X predict near maximal Dirac type CP violation (i.e. $\delta \approx 90^{0} $ and $270^{0}$) for NS. In addition, the Majorana phases $\rho$ and $\sigma$ are found to be very close to $0^{0}$ for these cases. On the other hand, in case of type Y, $B_{1}, B_{2},  B_{4}, B_{6}, C_{1}, C_{4}, C_{5}$ and $C_{7}$ show almost similar constraints on the parameter space for $\delta$ however for opposite mass spectrum [Table \ref{tab4}].
In Figs. \ref{fig3},\ref{fig4}, \ref{fig5}, \ref{fig6}, we have complied the correlation plots for case $B_{2}$ for both type X and Y comprising the unknown parameters  $\rho, \sigma, \delta$, $|M|_{ee}$ and lowest neutrino mass ($m_{o})$. As explicitly shown in Figs. \ref{fig3}(a, b) and \ref{fig6}(a, b), $\delta\approx 90^{0}$ and $270^{0}$, while $\rho,    \sigma \approx 0^{0}$. The correlation plots between  $|M|_{ee}$ and $m_{o}$ have been encapsulated in Figs. \ref{fig3}(c), \ref{fig4}(c), \ref{fig5}(c), \ref{fig6}(c). The plots indicate the strong linear relation correlation  between these parameters and in addition, the lower bound on both the parameters is somewhere in the range from 0.001 to 0.01 eV. The prediction for the allowed space of $|M|_{ee}$ for all the cases of category B is given in Table \ref{tab4}.

\textbf{Category D (F)}:  In Category D, only nine cases  are acceptable with neutrino oscillation data at 3$\sigma$ CL for both type X and type Y structures respectively, while case $D_{8}$ is excluded for both of them [Table \ref{tab5}].  Cases $D_{1}$, $D_{2}$, $D_{4}$, $D_{5}$, $D_{6}$, $D_{7}$, $D_{9}$  show both NS and IS for type X and type Y respectively, while $D_{3}$ and $D_{10}$ are acceptable for IS (NS) and NS(IS) respectively in case of type X (type Y) structure. Similarly, the results for cases belonging to Category F can be obtained from Category D since both are related via permutation symmetry. It is found that only nine cases are allowed with data in category F, while $F_{7}$ is excluded at 3$\sigma$ CL.\\
Cases $D_{1}$(NS), $D_{2}$ (NS, IS), $D_{3}$(IS), $D_{4}$(NS), $D_{5}$(NS, IS), $D_{6}$(NS), $D_{7}$(NS), $D_{9}$(NS), $D_{10}$(NS), $F_{1}$(NS), $F_{2}$ (NS, IS), $F_{3}$(IS), $F_{4}$(NS), $F_{5}$(NS, IS), $F_{6}$(NS), $F_{7}$(NS), $F_{9}$(NS), $F_{10}$(NS) predict literally no constraints on $ \delta$ for type X texture. These cases give identical predictions for type Y as well, however for opposite mass ordering. On the other hand, for cases $D_{6}$(IS), $D_{4}$(IS),  $D_{7}$ (IS), $D_{9}$ (IS), $F_{3}$(IS),  $F_{6}$ (IS), $F_{8}$ (IS), $F_{9}$(IS) $\delta$ is notably constrained for type X, and similar observation have been found for these cases in type Y, however for opposite mass ordering [Table \ref{tab5}].

It is found that textures $D_{7}$ (IS), $D_{9}$ (IS), $F_{6}$ (IS) and $F_{8}$ (IS) belonging to type X predict near maximal Dirac CP violation (i.e. $\delta \approx 90^{0} $ and $270^{0}$). In addition, the Majorana phases $\rho$ and $\sigma$ are found to be very close to $0^{0}$ for these cases. The similar predictions hold for these cases belonging to type Y structure however for opposite mass spectrum.

The prediction on the allowed range of $|M|_{ee}$ for all the cases of category D is provided in Table \ref{tab5}.
As an illustration, in Figs. \ref{fig7}, \ref{fig8}, \ref{fig9}, \ref{fig10} we have complied the correlation plots for case $D_{7}$ for type X and type Y structures. Figs. \ref{fig7}(a, b)(\ref{fig10}(a, b)) indicate no constraint on $\delta, \rho, \sigma$  for NS(IS) corresponding to type X (type Y) structure at 3$\sigma$ CL.  On the other hand, $\delta \approx 90^{0}$ and $270^{0}$, while $\rho$ and $\sigma$ approach to $0^{0}$ for IS in case of type X structure [Fig. \ref{fig8}]. However, similar predictions for $\delta, \rho, \sigma$ have been observed for type Y, however for NS [Fig.\ref{fig9}].  In Figs. \ref{fig7} (c), \ref{fig8}(c), \ref{fig9}(c), \ref{fig10}(c), we have presented the correlation plots between $|M|_{ee}$ and $m_{0}$ indicating the linear correlation.

\textbf{Category E}: In Category E, only eight out of ten cases are allowed with experimental data for both type X and type Y structures at 3$\sigma$ CL [Table \ref{tab6}]. Cases $E_{7}$ and $E_{8}$  are ruled out for both type X and type Y structures. Only $E_{5}$ and $E_{10}$ favor both NS and IS, while rest of the cases favor either NS or IS  for type X and type Y structure [Table \ref{tab6}]. From table \ref{tab6}, it is clear that $E_{1}$, $E_{2}, E_{3}, E_{4}, E_{5}, E_{10}$  cover literally full range of $\delta$ for type X. Same cases show identical prediction for type Y, however for opposite mass spectrum.
 For NS, cases  $E_{1}, E_{2}, E_{5}, E_{9}, E_{10}$  belonging to type X predict the lower bound on effective mass $|M|_{ee}$ to be zero , while for IS, cases $E_{3}, E_{4}, E_{5}, E_{10}$ predict larger lower bound (greater than 0.01eV) on $|M|_{ee}$ [Table \ref{tab6}]. However for type Y, all these cases show larger lower bound on $|M|_{ee}$ ($\geq 0.01$eV) for both NS and IS.

For the purpose of illustration, we have presented the correlation plots for $E_{1}$  indicating the parameter space of $\rho, \sigma, \delta$, $|M|_{ee}$ and lowest neutrino mass ($m_{o})$ [Figs. \ref{fig11}, \ref{fig12}]. As shown in Figs.\ref{fig11}, \ref{fig12}, $\rho, \sigma, \delta$  remain literally unconstrained for both type X and type Y structures.  In addition, there is a linear correlation among $\rho, \sigma$ and $\delta$ at 3$\sigma$ CL for type X structure [Table \ref{tab6}]. Fig. \ref{fig11}(c)  indicates the strong linear correlation between $|M|_{ee}$ and $m_{o}$ and in addition, the lower bound on  $|M|_{ee}$ is predicted to be zero.

\section{Summary and Conclusion}
  To summarize, we have discussed the noval possibilities of hybrid textures in the flavor basis wherein  the assumption of either one zero minor and an equality between the elements  or one zero element and an equality between the cofactors in the Majorana neutrino mass matrix is considered. Out of sixty phenomenologically possible cases, only 56 are found to be viable for type X, while only 50 are viable with the present data for type Y at 3$\sigma$ CL. Therefore, out of 120 only 106 cases are found to be viable with the existing data. However only 38 seems to restrict the parameteric space of CP violating phases $\delta$, $\rho$, $\sigma$, while 16 out of these predict near maximal Dirac CP violation i.e. $\delta \simeq 90^{0}, 270^{0}$. The allowed parameter space for effective mass term $|M|_{ee}$ related to neutrinoless double beta decay as well as lowest neutrino mass term for all viable cases have been carefully studied. The present viable cases may be  derived from the discrete symmetry. However the symmetry realization for each case in a systematic and self consistent way deserve  fine-grained research.  The viability of these cases suggests that there are still rich unexplored structures of the neutrino mass matrix from both the phenomenological and theoretical points of view.

To conclude our discussion, we would like add that the hybrid textures comprising either one zero element and an equality between the elements or one zero minor and an equality between the cofactors lead to 106 viable cases , therefore there are now total 212 viable cases pertaining to the hybrid textures of $M_{\nu}$  in the flavor basis.  Since most of these cases overlap in their predictions regarding the experimentally undetermined parameters, therefore we expect that only the future longbaseline experiments, neutrinoless double beta decay experiments and cosmological observations could help us to select the appropriate structure of mass texture.

\section*{Acknowledgment}

The author would like to thank the Director, National Institute of Technology Kurukshetra, for providing the necessary facilities to work.  \\

 \begin{table}
 \begin{center}
\begin{footnotesize}
\resizebox{10cm}{!}{
\begin{tabular}{|c|c|c|c|c|}
\hline
  Cases& X &P&Y&Q\\
  \hline
   $A_{1}$& $C_{11}=0, M_{12}=M_{13}$& $e^{i(\phi_{\mu}-\phi_{\tau})}$& $M_{11}=0,C_{12}=C_{13}$ &$e^{i(\phi_{\tau}-\phi_{\mu})}$\\
   \hline
   $A_{2}$& $C_{11}=0, M_{12}=M_{22}$, &$e^{i(\phi_{e}-\phi_{\mu})} $& $M_{11}=0,C_{12}=C_{22}$&$ e^{i(\phi_{\mu}-\phi_{e})}$\\
   \hline
   $A_{3}$& $C_{11}=0, M_{13}=M_{23} $&$  e^{i(\phi_{e}-\phi_{\mu})}$& $M_{11}=0,C_{13}=C_{23}$&$ e^{i(\phi_{\mu}-\phi_{e})}$\\
   \hline
   $A_{4}$& $C_{11}=0, M_{22}=M_{23}$&$ e^{i(\phi_{\mu}-\phi_{\tau})}$& $M_{11}=0,C_{22}=C_{23}$&$ e^{i(\phi_{\tau}-\phi_{\mu})}$\\
   \hline
   $A_{5}$& $C_{11}=0,M_{22}=M_{33}$&$ e^{2i(\phi_{\mu}-\phi_{\tau})}$& $M_{11}=0,C_{22}=C_{33}$&$ e^{2i(\phi_{\tau}-\phi_{\mu})}$\\
   \hline
   $A_{6}$& $C_{11}=0,M_{23}=M_{33}$&$ e^{i(\phi_{\mu}-\phi_{\tau})}$& $M_{11}=0,C_{23}=C_{33}$&$ e^{i(\phi_{\tau}-\phi_{\mu})}$\\
   \hline
   $A_{7}$& $C_{11}=0,M_{12}=M_{23} $&$ e^{i(\phi_{e}-\phi_{\tau})}$& $M_{11}=0,C_{12}=C_{23}$&$ e^{i(\phi_{\tau}-\phi_{e})}$\\
   \hline
   $A_{8}$& $C_{11}=0,M_{13}=M_{33}$&$ e^{i(\phi_{e}-\phi_{\tau})}$& $M_{11}=0,C_{13}=C_{33} $&$ e^{i(\phi_{\tau}-\phi_{e})}$\\
   \hline
   $A_{9}$& $C_{11}=0,M_{13}=M_{22}$&$ e^{i(\phi_{e}+\phi_{\tau}-2\phi_{\mu})}$& $M_{11}=0,C_{13}=C_{22} $&$ e^{i(-\phi_{e}-\phi_{\tau}+2\phi_{\mu})}$\\
   \hline
   $A_{10}$& $C_{11}=0,M_{12}=M_{33}$&$ e^{i(\phi_{e}+\phi_{\mu}-2\phi_{\tau})}$& $M_{11}=0,C_{12}=C_{33}$&$ e^{i(-\phi_{e}-\phi_{\mu}+2\phi_{\tau})}$\\
   \hline

   $B_{1}$& $C_{12}=0,M_{11}=M_{13}$&$e^{i(\phi_{e}-\phi_{\tau})}$& $M_{12}=0,C_{11}=C_{13}$& $e^{i(\phi_{\tau}-\phi_{e})}$\\
   \hline
    $B_{2}$& $C_{12}=0,M_{12}=M_{22}$& $e^{i(\phi_{e}+\phi_{\tau}-2\phi_{\mu})}$& $M_{12}=0,C_{12}=C_{22}$& $e^{i(2\phi_{\mu}-\phi_{e}-\phi_{\tau})}$\\
   \hline
    $B_{3}$& $C_{12}=0,M_{13}=M_{23}$&$e^{i(\phi_{e}-\phi_{\mu})}$& $M_{12}=0,C_{13}=C_{23}$&$e^{i(\phi_{\mu}-\phi_{e})}$\\
   \hline
    $B_{4}$& $C_{12}=0,M_{13}=M_{33}$&$e^{i(\phi_{e}-\phi_{\tau})}$& $M_{12}=0,C_{13}=C_{33}$&$e^{i(\phi_{\tau}-\phi_{e})}$\\
   \hline
    $B_{5}$& $C_{12}=0,M_{11}=M_{22}$&$e^{2i(\phi_{e}-\phi_{\mu})}$& $M_{12}=0,C_{11}=C_{22}$&$e^{2i(\phi_{\mu}-\phi_{e})}$\\
   \hline
    $B_{6}$& $C_{12}=0,M_{11}=M_{23}$&$e^{i(2\phi_{e}-\phi_{\mu}-\phi_{\tau})}$& $M_{12}=0,C_{11}=C_{23}$&$e^{i(-2\phi_{e}+\phi_{\mu}+\phi_{\tau})}$\\
   \hline
    $B_{7}$& $C_{12}=0,M_{11}=M_{33}$&$e^{2i(\phi_{e}-\phi_{\tau})}$& $M_{12}=0,C_{11}=C_{33}$&$e^{2i(\phi_{\tau}-\phi_{e})}$\\
   \hline
    $B_{8}$& $C_{12}=0,M_{22}=M_{23}$&$e^{i(\phi_{\mu}-\phi_{\tau})}$& $M_{12}=0,C_{22}=C_{23}$&$e^{i(\phi_{\tau}-\phi_{\mu})}$\\
   \hline
    $B_{9}$& $C_{12}=0,M_{22}=M_{33}$&$e^{2i(\phi_{\mu}-\phi_{\tau})}$& $M_{12}=0,C_{22}=C_{33}$&$e^{2i(\phi_{\tau}-\phi_{\mu})}$\\
   \hline
    $B_{10}$& $C_{12}=0,M_{23}=M_{33}$&$e^{i(\phi_{\mu}-\phi_{\tau})}$& $M_{12}=0,C_{23}=C_{33}$&$e^{i(\phi_{\tau}-\phi_{\mu})}$\\
   \hline

    $C_{1}$& $C_{13}=0,M_{11}=M_{12}$&$e^{i(\phi_{e}-\phi_{\mu})}$& $M_{13}=0,C_{11}=C_{12}$&$e^{i(\phi_{\mu}-\phi_{e})}$\\
   \hline
    $C_{2}$& $C_{13}=0,M_{11}=M_{22}$&$e^{2i(\phi_{e}-\phi_{\mu})}$& $M_{13}=0,C_{11}=C_{22}$&$e^{2i(\phi_{\mu}-\phi_{e})}$\\
   \hline
    $C_{3}$& $C_{13}=0,M_{11}=M_{23}$&$e^{i(2\phi_{e}-\phi_{\mu}-\phi_{\tau})}$& $M_{13}=0,C_{11}=C_{23}$&$e^{i(-2\phi_{e}+\phi_{\mu}+\phi_{\tau})}$\\
   \hline
   $C_{4}$& $C_{13}=0,M_{11}=M_{33}$&$e^{2i(\phi_{e}-\phi_{\tau})}$& $M_{13}=0,C_{11}=C_{33}$&$e^{2i(\phi_{\tau}-\phi_{e})}$\\
   \hline
   $C_{5}$& $C_{13}=0,M_{12}=M_{22}$&$e^{i(\phi_{e}-\phi_{\mu})}$& $M_{13}=0,C_{12}=C_{22}$&$e^{i(\phi_{\mu}-\phi_{e})}$\\
   \hline
   $C_{6}$& $C_{13}=0,M_{12}=M_{23}$&$e^{i(\phi_{e}-\phi_{\tau})}$& $M_{13}=0,C_{12}=C_{23}$&$e^{i(\phi_{\tau}-\phi_{e})}$\\
   \hline
   $C_{7}$& $C_{13}=0,M_{12}=M_{33}$&$e^{i(\phi_{e}-\phi_{\mu}-2\phi_{\tau})}$& $M_{13}=0,C_{12}=C_{33}$&$e^{i(-\phi_{e}+\phi_{\mu}+2\phi_{\tau})}$\\
   \hline
   $C_{8}$& $C_{13}=0,M_{22}=M_{23}$&$e^{i(\phi_{\mu}-\phi_{\tau})}$& $M_{13}=0,C_{22}=C_{23}$&$e^{i(\phi_{\tau}-\phi_{\mu})}$\\
   \hline
   $C_{9}$& $C_{13}=0,M_{22}=M_{33}$&$e^{2i(\phi_{\mu}-\phi_{\tau})}$& $M_{13}=0,C_{22}=C_{33}$&$e^{2i(\phi_{\tau}-\phi_{\mu})}$\\
   \hline
   $C_{10}$& $C_{13}=0,M_{23}=M_{33}$&$e^{i(\phi_{\mu}-\phi_{\tau})}$& $M_{13}=0,C_{23}=C_{33}$&$e^{i(\phi_{\tau}-\phi_{\mu})}$\\
   \hline

    $D_{1}$& $C_{22}=0,M_{11}=M_{12}$&$e^{i(\phi_{e}-\phi_{\mu})}$& $M_{22}=0,C_{11}=C_{12}$&$e^{i(\phi_{\mu}-\phi_{e})}$\\
    \hline
    $D_{2}$& $C_{22}=0,M_{11}=M_{13}$&$e^{i(\phi_{e}-\phi_{\tau})}$& $M_{22}=0,C_{11}=C_{13}$&$e^{i(\phi_{\tau}-\phi_{e})}$\\
    \hline
    $D_{3}$& $C_{22}=0,M_{11}=M_{23}$&$e^{i(2\phi_{e}-\phi_{\mu}-\phi_{\tau})}$& $M_{22}=0,C_{11}=C_{23}$&$e^{i(-2\phi_{e}+\phi_{\mu}+\phi_{\tau})}$\\
    \hline
    $D_{4}$& $C_{22}=0,M_{11}=M_{33}$&$e^{2i(\phi_{e}-\phi_{\tau})}$& $M_{22}=0,C_{11}=C_{33}$&$e^{2_{•}i(\phi_{\tau}-\phi_{e})}$\\
    \hline
    $D_{5}$& $C_{22}=0,M_{12}=M_{13}$&$e^{i(\phi_{\mu}-\phi_{\tau})}$& $M_{22}=0,C_{12}=C_{13}$&$e^{i(\phi_{\tau}-\phi_{\mu})}$\\
    \hline
    $D_{6}$& $C_{22}=0,M_{12}=M_{23}$&$e^{i(\phi_{e}-\phi_{\tau})}$& $M_{22}=0,C_{12}=C_{23}$&$e^{i(\phi_{\tau}-\phi_{e})}$\\
    \hline
    $D_{7}$& $C_{22}=0,M_{12}=M_{33}$&$e^{i(\phi_{e}+\phi_{\mu}-2\phi_{\tau})}$& $M_{22}=0,C_{12}=C_{33}$&$e^{i(-\phi_{e}-\phi_{\mu}+2\phi_{\tau})}$\\
   \hline
    $D_{8}$& $C_{22}=0,M_{13}=M_{23}$&$e^{i(\phi_{e}-\phi_{\mu})}$& $M_{22}=0,C_{13}=C_{23}$&$e^{i(\phi_{\mu}-\phi_{e})}$\\
   \hline
    $D_{9}$& $C_{22}=0,M_{13}=M_{33}$&$e^{i(\phi_{e}-\phi_{\tau})}$& $M_{22}=0,C_{13}=C_{33}$&$e^{i(\phi_{\tau}-\phi_{e})}$\\
   \hline
    $D_{10}$& $C_{22}=0,M_{23}=M_{33}$&$e^{i(\phi_{\mu}-\phi_{\tau})}$& $M_{22}=0,C_{23}=C_{33}$&$e^{i(\phi_{\tau}-\phi_{\mu})}$\\
   \hline

   $E_{1}$& $C_{23}=0,M_{11}=M_{12}$&$e^{i(\phi_{e}-\phi_{\mu})}$& $M_{23}=0,C_{11}=C_{12}$&$e^{i(\phi_{\mu}-\phi_{e})}$\\
   \hline
    $E_{2}$& $C_{23}=0,M_{11}=M_{13}$&$e^{i(\phi_{e}-\phi_{\tau})}$& $M_{23}=0,C_{11}=C_{13}$&$e^{i(\phi_{\tau}-\phi_{e})}$\\
   \hline
    $E_{3}$& $C_{23}=0,M_{11}=M_{22}$&$e^{2i(\phi_{e}-\phi_{\mu})}$& $M_{23}=0,C_{11}=C_{22}$&$e^{2i(\phi_{\mu}-\phi_{e})}$\\
   \hline
   $E_{4}$& $C_{23}=0,M_{11}=M_{33}$&$e^{2i(\phi_{e}-\phi_{\tau})}$& $M_{23}=0,C_{11}=C_{33}$&$e^{2i(\phi_{\tau}-\phi_{e})}$\\
   \hline
   $E_{5}$& $C_{23}=0,M_{12}=M_{13}$&$e^{i(\phi_{\mu}-\phi_{\tau})}$& $M_{23}=0,C_{12}=C_{13}$&$e^{i(\phi_{\tau}-\phi_{\mu})}$\\
   \hline
   $E_{6}$& $C_{23}=0,M_{12}=M_{22}$&$e^{i(\phi_{e}-\phi_{\mu})}$& $M_{23}=0,C_{12}=C_{22}$&$e^{i(\phi_{\mu}-\phi_{e})}$\\
   \hline
   $E_{7}$& $C_{23}=0,M_{13}=M_{33}$&$e^{i(\phi_{e}-\phi_{\mu}-2\phi_{\tau})}$& $M_{23}=0,C_{13}=C_{33}$&$e^{i(-\phi_{e}+\phi_{\mu}+2\phi_{\tau})}$\\
   \hline
   $E_{8}$& $C_{23}=0,M_{13}=M_{22}$&$e^{i(\phi_{e}-\phi_{\tau}-2\phi_{\mu})}$& $M_{23}=0,C_{13}=C_{22}$&$e^{i(-\phi_{e}+\phi_{\tau}+2\phi_{\mu})}$\\
   \hline
   $E_{9}$& $C_{23}=0,M_{13}=M_{33}$&$e^{i(\phi_{e}-\phi_{\tau})}$& $M_{23}=0,C_{13}=C_{33}$&$e^{i(\phi_{\tau}-\phi_{e})}$\\
   \hline
   $E_{10}$& $C_{23}=0,M_{22}=M_{33}$&$e^{2i(\phi_{\mu}-\phi_{\tau})}$& $M_{23}=0,C_{22}=C_{33}$&$e^{2i(\phi_{\tau}-\phi_{\mu})}$\\
   \hline

   $F_{1}$& $C_{33}=0,M_{11}=M_{12}$&$e^{i(\phi_{e}-\phi_{\mu})}$& $M_{33}=0,C_{11}=C_{12}$&$e^{i(\phi_{\mu}-\phi_{e})}$\\
   \hline
    $F_{2}$& $C_{33}=0,M_{11}=M_{13}$&$e^{i(\phi_{e}-\phi_{\tau})}$& $M_{33}=0,C_{11}=C_{13}$&$e^{i(\phi_{\tau}-\phi_{e})}$\\
   \hline
    $F_{3}$& $C_{33}=0,M_{11}=M_{22}$&$e^{2i(\phi_{e}-\phi_{\mu})}$& $M_{33}=0,C_{11}=C_{22}$&$e^{2i(\phi_{\mu}-\phi_{e})}$\\
   \hline
   $F_{4}$& $C_{33}=0,M_{11}=M_{23}$& $e^{i(2\phi_{e}-\phi_{\mu}-\phi_{\tau})}$& $M_{33}=0,C_{11}=C_{23}$&$e^{i(-2\phi_{e}+\phi_{\mu}+\phi_{\tau})}$\\
   \hline
   $F_{5}$& $C_{33}=0,M_{12}=M_{13}$&$e^{i(\phi_{\mu}-\phi_{\tau})}$& $M_{33}=0,C_{12}=C_{13}$&$e^{i(\phi_{\tau}-\phi_{\mu})}$\\
   \hline
   $F_{6}$& $C_{33}=0,M_{12}=M_{22}$&$e^{i(\phi_{e}-\phi_{\mu})}$& $M_{33}=0,C_{12}=C_{22}$&$e^{i(\phi_{\mu}-\phi_{e})}$\\
   \hline
   $F_{7}$& $C_{33}=0,M_{12}=M_{23}$&$e^{i(\phi_{e}-\phi_{\tau})}$& $M_{33}=0,C_{12}=C_{23}$&$e^{i(\phi_{\tau}-\phi_{e})}$\\
   \hline
   $F_{8}$& $C_{33}=0,M_{13}=M_{22}$&$e^{i(\phi_{e}+\phi_{\tau}-2\phi_{\mu})}$& $M_{33}=0,C_{13}=C_{22}$&$e^{i(-\phi_{e}-\phi_{\tau}+2\phi_{\mu})}$\\
   \hline
   $F_{9}$& $C_{33}=0,M_{13}=M_{23}$&$e^{i(\phi_{e}-\phi_{\mu})}$& $M_{33}=0,C_{13}=C_{23}$&$e^{i(\phi_{\mu}-\phi_{e})}$\\
   \hline
   $F_{10}$& $C_{33}=0,M_{22}=M_{23}$&$e^{i(\phi_{\mu}-\phi_{\tau})}$& $M_{33}=0, C_{22}=C_{23}$&$e^{i(\phi_{\tau}-\phi_{\mu})}$\\
   \hline
    \end{tabular}}
\caption{\label{tab2}All the sixty phenomenological possible cases belonging to type X and Y respectively have been shown. P and Q are unobservable phases associated with type X and Y respectively. }
\end{footnotesize}
\end{center}
\end{table}

 \begin{table}
 \begin{center}
\begin{footnotesize}

\resizebox{14cm}{!}{
\begin{tabular}{|c|c|c|c|c|}
  \hline
&\multicolumn{2}{c|}{X} &\multicolumn{2}{c|}{Y} \\
\hline
Cases&NS&IS&NS&IS\\
\hline
$A_{1}$  &$\times$ & $\rho=
-90^{0}--72^{0}\oplus 72^{0}-90^{0}$ & $\rho=-90^{0}-90^{0}$
&$\times$  \\

&$\times$&$=(
-90^{0}--72^{0}\oplus 72^{0}-90^{0})$&$= (-90^{0}-90^{0})$
&$\times$\\
&$\times$ &
$\sigma= -90^{0}-90^{0}$ & $\sigma= -90^{0}-90^{0}$ & $\times$ \\

&$\times$ &$=(-90^{0}-90^{0})$ & $=(-90^{0}-90^{0})$ & $\times$ \\

&$\times$ & $\delta= 0^{0}-93.2^{0}\oplus 116.6^{0}-247.7^{0}\oplus 268.8^{0}-360^{0}$
& $\delta= 0^{0}-360^{0}$ & $\times$ \\

&$\times$ & $=(0^{0}-166.27^{0}\oplus 191.98^{0}-360^{0})$
& $=(0^{0}-360^{0})$ & $\times$ \\

&$\times$&$|M|_{ee}=0.0114-0.0540$&$|M|_{ee}=0.0$&$\times$\\
&$\times$&$m_{0}=0.000820-0.0470$&$m_{0}=0.00155-0.0103$&$\times$\\

\hline
$A_{2}(A_{8})$  &$\times$ & $\rho=
-85.6^{0}--76.3^{0}\oplus 76.4^{0}-85.6^{0}$ & $\times$
&$\times$  \\

&$\times$&$=(
-85.4^{0}--78.1^{0}\oplus 78.2^{0}-85.1^{0})$&$\times$
&$\times$\\

&$\times$ &$\sigma= -70.1^{0}--43.1^{0}\oplus 43.1^{0}-70^{0}$ & $\times $ & $\times $  \\
&$\times$ &
$=(-70.1^{0}--44.2^{0}\oplus 43.3^{0}-76^{0})$ & $\times $ & $\times $  \\

&$\times$ & $\delta= 0^{0}-360^{0}$
& $\times$ & $\times $  \\

&$\times$ & $=(0^{0}-360^{0})$
& $\times$ & $\times $  \\

&$\times$&$|M|_{ee}=0.0273-0.0489$&$\times$&$\times $ \\
&$\times$&$m_{0}=0.000820-0.0484$&$\times$&$\times $ \\
\hline
$A_{3}(A_{7})$ &$\times$ & $\rho=
-81.94^{0}--75.9^{0}\oplus 75.4^{0}-81.93^{0}$ & $\times$
&$\times $   \\

&$\times$&$=(
-82.4^{0}--77.1^{0}\oplus 76.9^{0}-82^{0})$&$\times$
&$\times $ \\
&$\times$ &
$\sigma= -70.1^{0}--52.1^{0}\oplus 50.47^{0}-66.41^{0}$ & $\times $ & $\times $  \\
&$\times$ &
$=(-72^{0}--54.2^{0}\oplus 53.3^{0}-73^{0})$ & $\times $ & $\times $  \\

&$\times$ & $\delta= 0^{0}-360^{0}$
& $\times$ & $\times $ \\

&$\times$ & $=(0^{0}-360^{0})$
& $\times$ & $\times $  \\

&$\times$&$|M|_{ee}=0.0308-0.0451$&$\times$& $\times$ \\
&$\times$&$m_{0}=0.000969-0.00187$&$\times$&$\times $   \\
\hline
$A_{4}(A_{6})$  &$\times$ & $\rho=
-89.8^{0}--71.2^{0}\oplus 71.1^{0}-89.9^{0}$ & $\rho=-90^{0}-90^{0}$
&$\times$  \\

&$\times$&$=(
-90^{0}--72.1^{0}\oplus 72^{0}-90^{0})$&$=(-90^{0}-90^{0}$)
&$\times $ \\
&$\times$ &
$\sigma= -90^{0}-90^{0}$ & $\sigma=-90^{0}-90^{0}$& $\times $  \\

&$\times$ &$=(-90^{0}-90^{0})$ & $=(-90^{0}-90^{0})$ &$\times $ \\

&$\times$ & $\delta= 29.78^{0}-89.64^{0}\oplus 148.6^{0}-208^{0}\oplus 269.8^{0}-329^{0}$
& $\delta= 0^{0}-360^{0}$ & $\times $ \\

&$\times$ & $=(0^{0}-29.78^{0} \oplus 90.98^{0}-151.64^{0}\oplus 208.6^{0}-271^{0}\oplus 329.8^{0}-360^{0})$
& $\delta= 0^{0}-360^{0}$ &$\times $  \\

&$\times$&$|M|_{ee}=0.0108-0.0501$&$|M|_{ee}=0.0$&$\times $ \\
&$\times$&$m_{0}=0.000904-0.00440$&$m_{0}=0.00143-0.0103$&$\times $ \\
\hline
$A_{5}(A_{5})$  &$\times$ & $\rho=
-90^{0}--71^{0}\oplus 71^{0}-90^{0}$ &  $\rho=
-90^{0}--42.3^{0}\oplus 41.5^{0}-90^{0}$
&$\times $   \\

&$\times$&$=(
-90^{0}--71.2^{0}\oplus 72^{0}-90^{0})$&$=(
-56.6^{0}-56.7^{0})$
&$\times $ \\
&$\times$ &
$\sigma= -90^{0}-90^{0}$ & $\sigma= -62.5^{0}-62.4^{0}$ & $\times $  \\
&$\times$ &
$=(-90^{0}-90^{0})$ & $=(
-90^{0}--41.2^{0}\oplus 41.4^{0}-90^{0})$ & $\times $  \\

&$\times$ & $\delta= 0^{0}-44^{0}\oplus 92^{0}-162^{0}\oplus 198^{0}-268^{0}\oplus 327^{0}-360^{0}$
& $\delta=0^{0}-360^{0}$ & $\times $  \\

&$\times$ & $=(27.36^{0}-109.3^{0}\oplus 147.9^{0}-209.37^{0}\oplus 250.7^{0}-331^{0})$
& $=0^{0}-360^{0}$ & $\times $  \\

&$\times$&$|M|_{ee}=0.0113-0.0500$&$|M|_{ee}=0.0$& $\times $ \\
&$\times$&$m_{0}=0.000822-0.00440$&$m_{0}=0.00306-0.0105$& $\times $ \\
\hline
$A_{9}(A_{10})$  &$\times$ & $\rho=
-84.2^{0}--74.53^{0}\oplus 73.52^{0}-84.2^{0}$ & $\times$
&$\times $  \\
&$\times$&$=(
-84.2^{0}--75^{0}\oplus 75^{0}-85.1^{0})$&$\times$
&$\times $ \\

&$\times$ &
$\sigma= -74.83^{0}--50.3^{0}\oplus 50.1^{0}-73.4^{0}$ & $\times $ & $\times $  \\
&$\times$ &
$=(-75^{0}--50.3^{0}\oplus 50.1^{0}-75^{0})$ & $\times $ & $\times $  \\

&$\times$ & $\delta= 0^{0}-360^{0}$
& $\times$ & $\times $  \\

&$\times$ & $=(0^{0}-360^{0})$
& $\times$ & $\times $  \\

&$\times$&$|M|_{ee}=0.0293-0.0469$&$\times$&$\times $ \\
&$\times$&$m_{0}=0.000940-0.00190$&$\times$&$\times $ \\

 \hline
 \end{tabular}}
\caption{\label{tab3} The allowed ranges of Dirac CP-violating phase $\delta$, the  Majorana phases $\rho, \sigma$,  effective neutrino mass $|M|_{ee}$, and lowest neutrino mass $m_{0}$ for the experimentally allowed cases of Category A at 3$\sigma$ CL. The predictions corresponding to $(\lambda_{13})_{-}$ and $(\lambda_{23})_{-}$ neutrino mass ratios have been put into brackets.}
\end{footnotesize}
\end{center}
\end{table}

 \begin{table}
 \begin{center}
\begin{footnotesize}
\resizebox{18cm}{!}{
\begin{tabular}{|c|c|c|c|c|}
  \hline
&\multicolumn{2}{c|}{X} &\multicolumn{2}{c|}{Y} \\
\hline
Cases&NS&IS&NS&IS\\
\hline
$B_{1}(C_{1})$  &$\times$ & $\rho=
-90^{0}--58.7^{0}\oplus -23.8^{0}-15^{0} \oplus 58.8^{0}-90^{0}$ & $\rho=
-90^{0}-90^{0}$
&$\rho=-87.04-86.7^{0}$  \\

&$\times$&$=(
-67.7^{0}--16.6^{0}\oplus 16.6^{0}-68.5^{0})$&$=(-90^{0}-90^{0})$
&$=(-90^{0}--3.17^{0} \oplus 2.38^{0}-90^{0})$\\

&$\times$ &
$\sigma= -87^{0}-87.9^{0}$ & $ \sigma=-90^{0}--64.8^{0} \oplus -24.98^{0}-24.62^{0} $ & $\sigma=-90^{0}--64.96^{0}\oplus -24.8^{0}-24.5^{0}\oplus 65^{0}-90^{0} $ \\
&$\times$ &
$=(-90^{0}-90^{0})$ & $=(
-72.3^{0}--20.2^{0}\oplus 20.44^{0}-72.44^{0})$ & $=(-90^{0}-90^{0})$ \\

&$\times$ & $\delta= 0^{0}-78.9^{0}\oplus 98.7^{0}-261^{0}\oplus 282^{0}-360^{0}$
& $\delta= 0^{0}-75.9^{0}\oplus 111.4^{0}-247^{0}\oplus 28^3{0}-360^{0}$ & $\delta=80.8^{0}-93.6^{0}\oplus 267.6^{0}-279.5^{0}$  \\

&$\times$ & $=(6.14^{0}-166.27^{0}\oplus 194.7^{0}-353.44^{0})$
& $=(19.7^{0}-157^{0}\oplus 205.9^{0}-340.5^{0})$ & $=(87.8^{0}-101.6^{0}\oplus 259.6^{0}-273.5^{0}$) \\

&$\times$&$|M|_{ee}=0.0221-0.0371$&$|M|_{ee}=0.0-0.00991$&$|M|_{ee}=0.0417-0.0540$\\
&$\times$&$m_{0}=0.00412-0.0139$&$m_{0}=0.002286-0.0165$&$m_{0}=0.0430-0.0540$\\

\hline
$B_{2}(C_{7})$  &$\rho=
-4.13^{0}--0.073^{0}\oplus -0.120^{0}-4.07^{0}$ & $\rho=
-86.6^{0}-87.08^{0}$ & $\rho=-90^{0}-90^{0}$
&$\rho=-2.98^{0}-2.97^{0}$  \\

&$=(-2.08^{0}--0.63^{0}\oplus 0.48^{0}-2.15^{0})$&$=(
-90^{0}-90^{0})$&$=(-77.08^{0}-76.8^{0})$
&$=(-2.14^{0}--0.96^{0} \oplus 0.89^{0}-2.15^{0})$\\

&$\sigma= -7.11^{0}--0.088^{0}\oplus 0.121^{0}-7.17^{0}$ &
$\sigma= -90^{0}-90^{0}$ & $\sigma=-68.7-67.9 $ & $\sigma=-1.01^{0}-1.09^{0}$ \\
&$=(-4.9^{0}--2.4^{0}\oplus 2.4^{0}-4.9^{0}$ &
$=(-90^{0}-90^{0})$ & $=(-90^{0}-90^{0})$ & $=(-1.38^{0}--0.131^{0}\oplus 0.140^{0}-1.28^{0})$ \\

&$\delta=89.13^{0}-97.15^{0}\oplus 262.4^{0}-271.0^{0}$ & $\delta= 20.1^{0}-89.64^{0}\oplus 127.8^{0}-229.0^{0}\oplus 268.4^{0}-340^{0}$
& $\delta=30.59-328.9^{0}$ & $\delta= 88.13^{0}-92.15^{0}\oplus 267.4^{0}-271.0^{0}$ \\

&$=(90.5^{0}-92.65^{0}\oplus 267.4^{0}-269.0^{0})$ & $=(0.0^{0}-53.7^{0}\oplus 86.8^{0}-160^{0}\oplus 213.5^{0}-271^{0}\oplus 306^{0}-360^{0})$
& $=(0^{0}-30.08^{0}\oplus 220.4^{0}-360^{0})$ & $=(88.29^{0}-97.02^{0}\oplus 267.8^{0}-271^{0}$) \\

&$|M|_{ee}=0.0298-0.263$&$|M|_{ee}=0.0213-0.303$&$|M|_{ee}=0.00231-0.303$&$|M|_{ee}=0.0535-0.445$\\
&$m_{0}=0.0297-0.262$&$m_{0}=0.00499-0.312$&$m_{0}=0.00203-0.304$&$m_{0}=0.0300-0.441$\\
\hline
$B_{3}(C_{6})$  &$\rho=
-80^{0}--0.088^{0}\oplus 0.088^{0}-80^{0}$ & $\rho=
-50^{0}-50^{0}$ & $\rho=
-53.4^{0}- 53.69^{0}$
&$\rho=-90^{0}--5.48^{0} \oplus 5.46^{0}-90^{0}$  \\

&$=(-90^{0}-90^{0})$&$=(
-62.5^{0}-62^{0})$&$=(-52.5^{0}--9.48^{0} \oplus 10.46^{0}-52.5^{0})$
&$=(-83.4^{0}--5.95^{0} \oplus 3.30^{0}-84.56^{0})$\\

&$\sigma= -90^{0}--0.09^{0}\oplus 0.09^{0}-90^{0}$ &
$\sigma= -63.2^{0}--38.6^{0}\oplus 38^{0}-63.8^{0}$ & $\sigma=-80^{0}--3.28^{0}\oplus 3.28^{0}-80^{0} $ & $\sigma=-90^{0}--2.28^{0}\oplus 2.54^{0}-90^{0}$ \\
&$=(-90^{0}-90^{0})$ &
$=(-56.7^{0}-57^{0})$ & $=(-77.7^{0}-79.1^{0}) $ & $=(-90^{0}--3.58^{0}\oplus 2.68^{0}-90^{0} )$ \\

&$\delta=31.49^{0}-95.98^{0}\oplus 146^{0}-176^{0}\oplus 184^{0}-213^{0}\oplus 263^{0}-329^{0}$ & $\delta= 0^{0}-40.5^{0}\oplus 87.5^{0}-156.7^{0}\oplus 158.9^{0}-272^{0}\oplus 321^{0}-360^{0}$
& $\delta=0^{0}-91.6^{0}\oplus 156.8^{0}-204^{0}\oplus 271^{0}-360^{0}$ & $\delta=82.39^{0}-173.02^{0}\oplus 192.8^{0}-277^{0}$ \\

&$=(5.5^{0}-32^{0}\oplus 93^{0}-150.8^{0}\oplus 208^{0}-267^{0})$ & $=(3^{0}-90^{0}\oplus 153^{0}-205^{0}\oplus 272^{0}-358^{0})$
& $=(4.79^{0}-41.6^{0}\oplus 82.5^{0}-277.9^{0}\oplus 319^{0}-354^{0})$ & $=(6.48^{0}-91.8^{0}\oplus 268.7^{0}-351^{0}$) \\

&$|M|_{ee}=0.0180-0.500$&$|M|_{ee}=0.0311-0.450$&$|M|_{ee}=0.00154-0.0490$&$|M|_{ee}=0.0441-0.497$\\
&$m_{0}=0.0180-0.494$&$m_{0}=0.00386-0.498$&$m_{0}=0.0-0.484$&$m_{0}=0.0231-0.498$\\

\hline
$B_{4}(C_{5})$  &$\rho=
-2.06^{0}-2.10^{0}$ & $\rho=
-72.4^{0}-73.2^{0}$ & $\rho=
-90^{0}--11.5^{0}\oplus 2.5^{0}-90^{0}$
&$\rho=-3.0^{0}--0.0329^{0} \oplus 0.0321^{0}-3.0^{0}$  \\

&$\times$&$=(
-90^{0}-90^{0})$&$=(-79.4^{0}-80.2^{0})$
&$=(-3.0^{0}--0.0329^{0} \oplus 0.0321^{0}-3.0^{0})$\\

&$\sigma= -7.45^{0}-7.45^{0}$ &
$\sigma= -90^{0}-90^{0}$ & $\sigma= -71.45^{0}-70.45^{0} $ & $\sigma=-3.22^{0}-3.29^{0}$ \\
&$\times$ &
$=(-90^{0}-90^{0})$ & $=(-90^{0}--17.0^{0}\oplus 16.2^{0}-90^{0}) $ & $=(-3.38^{0}--1.24^{0}\oplus 1.28^{0}-3.48^{0})$ \\

&$\delta=89.2^{0}-97^{0}\oplus 263.4^{0}-270.58^{0}$ & $\delta= 21.19^{0}-89.6^{0}\oplus 141^{0}-218^{0}\oplus 276^{0}-339^{0}$
& $\delta=85.6^{0}-274^{0}$ & $\delta= 88.69^{0}-92.25^{0}\oplus 267.8^{0}-270.1^{0}$ \\

&$\times$ & $=(0^{0}-33^{0}\oplus 85.6^{0}-158.5^{0}\oplus 214.4^{0}-274^{0}\oplus 322^{0}-360^{0})$
& $=(0^{0}-105^{0}\oplus 255.6^{0}-360^{0})$ & $=(88.9^{0}-92.57^{0}\oplus 267.8^{0}-270.9^{0})$ \\

&$|M|_{ee}=0.0226-0.275$&$|M|_{ee}=0.0344-0.384$&$|M|_{ee}=0.00532-0.405$&$|M|_{ee}=0.0512-0.0414$\\
&$m_{0}=0.0282-0.284$&$m_{0}=0.00386-0.468$&$m_{0}=0.0-0.404$&$m_{0}=0.0260-0.410$\\
\hline

$B_{5}(C_{4})$  &$\rho=
-61.6^{0}-60.23^{0}$ & $\rho=
-90^{0}--9^{0} \oplus 9^{0}-90^{0}$ & $\rho=
-90^{0}--13.6^{0} \oplus 11.8^{0}-90^{0}$
&$\rho=-16.6^{0}-16.4^{0}$  \\

&$=(-65^{0}--57^{0}\oplus -19^{0}-19^{0} \oplus 57^{0}-65^{0})$&$=(
-90^{0}--10^{0} \oplus 10^{0}-90^{0})$&$=(
-90^{0}--8.79^{0} \oplus 7.03^{0}-90^{0})$
&$=(-24.5^{0}--1.8^{0} \oplus 1.2^{0}-23.4^{0})$\\

&$\sigma= -64^{0}-64.08^{0}$ &
$\sigma= -90^{0}-90^{0}$ &$\sigma=-71.8^{0}--6.6^{0}\oplus 7.56^{0}-72^{0}$  & $\sigma=-21.5^{0}-21.5^{0}$ \\

&$=(-65^{0}--57^{0}\oplus -19^{0}-19^{0} \oplus 57^{0}-65^{0})$ &
$=(-87^{0}--0.4^{0}\oplus 0.06^{0}-88.1^{0})$ & $=(-90^{0}--13.6^{0}\oplus 9.4^{0}-90^{0})$ & $=(-18.6^{0}-18.9^{0})$ \\

&$\delta=52.5^{0}-175.6^{0}\oplus 184^{0}-307^{0}$ & $\delta= 28.17^{0}-87^{0}\oplus 120^{0}-239.5^{0}\oplus 272^{0}-331^{0}$
& $\delta=103.5^{0}-171.8^{0}\oplus 188.9^{0}-255^{0}$ & $\delta= 35.6^{0}-92.4^{0}\oplus 140.8^{0}-220.6^{0}\oplus 266^{0}-325^{0}$ \\

&$=(6.5^{0}-63.4^{0}\oplus 148^{0}-158^{0}\oplus 204^{0}-213^{0}\oplus 295^{0}-353^{0})$ & $=(7.3^{0}-29.8^{0}\oplus 56.3^{0}-165.4^{0}\oplus 196^{0}-348^{0})$
& $=(7.67^{0}-106.9^{0}\oplus 249.8^{0}-350^{0})$ & $=(5.4^{0}-37.7^{0}\oplus 87.8^{0}-144.5^{0}\oplus 215^{0}-273^{0}\oplus 328^{0}-353^{0})$ \\

&$|M|_{ee}=0.00710-0.465$&$|M|_{ee}=0.0114-0.550$&$|M|_{ee}=0.0441-0.469$&$|M|_{ee}=0.0417-0.498$\\
&$m_{0}=0.00855-0.478$&$m_{0}=0.00386-0.500$&$m_{0}=0.0540-0.473$&$m_{0}=0.00687-0.498$\\
\hline

$B_{6}(C_{3})$  &$\rho=
-80^{0}-80^{0}$ & $\times$ & $\times$
&$\rho=-90^{0}-90^{0} $  \\

&$=(-90^{0}--25^{0}\oplus -8.27^{0}-10.26^{0} \oplus 25^{0}-90^{0})$&$\times$&$\times$
&$=(-90^{0}-90^{0})$\\

&$\sigma= -90^{0}-90^{0}$ &
$\times$ & $\times $ & $\sigma=-90^{0}-90^{0}$ \\
&$=(-90^{0}--1.7^{0}\oplus -1.7^{0}-1.4^{0} \oplus 16.8^{0}-90^{0})$ &
$\times$ & $\times $ & $=(-90^{0}-90^{0})$ \\

&$\delta=64.29^{0}-175.6^{0}\oplus 189^{0}-295^{0}$ & $\times$
& $\times$ & $\delta=58.7^{0}-96.7^{0}\oplus 261.8^{0}-301.9^{0} $ \\

&$=(64.38^{0}-151^{0}\oplus 207.8^{0}-296^{0})$ & $\times$
& $\times$ & $=(80.69^{0}-121.9^{0}\oplus 238.7^{0}-273.5^{0} $) \\

&$|M|_{ee}=0.0121-0.255$&$\times$&$\times$&$|M|_{ee}=0.0414-0.334$\\
&$m_{0}=0.0121-0.280$&$\times$&$\times$&$m_{0}=0.0129-0.332$\\
\hline

$B_{7}(C_{2})$  &$\rho=
-43.9^{0}-45.7^{0}$ & $\times$ & $\times$
&$\rho=-18.8^{0}-18.4^{0}$  \\

&$=(-18.06^{0}--2.56^{0}\oplus 1.98^{0}-18.27^{0} )$&$\times$&$\times$
&$=(-25.6^{0}-25.7^{0})$\\

&$\sigma= -51.27^{0}-50.63^{0}$ &
$\times$ & $\times $ & $\sigma=-25.7^{0}-25.8^{0}$ \\

&$=(-25.85^{0}--3.45^{0}\oplus 2.54^{0}-26.29^{0})$ &
$\times$ & $\times $ & $=(-16.9^{0}-16.8^{0})$ \\

&$\delta=55.28^{0}-175.6^{0}\oplus 186.5^{0}-305^{0}$ & $\times$
& $\times$ & $\delta= 30^{0}-94^{0}\oplus 141.6^{0}-175.6^{0}\oplus 184.5^{0}-220.6^{0} \oplus 266^{0}-325^{0}$ \\

&$=(5.33^{0}-60.4^{0}\oplus 300^{0}-352.3^{0})$ & $\times$
& $\times$ & $=(13.4^{0}-35.9^{0}\oplus 88.8^{0}-140.8^{0}\oplus 218^{0}-271^{0} \oplus 325^{0}-360^{0}$) \\

&$|M|_{ee}=0.00687-0.439$&$\times$&$\times$&$|M|_{ee}=0.0417-0.479$\\
&$m_{0}=0.00687-0.439$&$\times$&$\times$&$m_{0}=0.00687-0.471$\\

\hline
$B_{8}(C_{10})$  &$\rho=
-90^{0}-90^{0}$ & $\rho=
-51.2^{0}-51^{0} $ & $\rho=
-72.8^{0}-72.7^{0} $
&$\rho=-89.8^{0}-88.7^{0}$  \\

&$=(-90^{0}-90^{0})$&$=(
-52.5^{0}--40^{0} \oplus 40^{0}-53^{0})$&$=(-90^{0}-90^{0})$
&$=(-89.8^{0}-89.7^{0})$\\

&$\sigma= -90^{0}-90^{0}$ &
$\sigma= -90^{0}-90^{0}$ & $\sigma= -90^{0}-90^{0} $ & $\sigma=-89^{0}-89^{0}$ \\
&$=(-90^{0}-90^{0})$ &
$=(-90^{0}-90^{0})$ & $=(-67.8^{0}-67.7^{0})$ & $=(-89^{0}-89^{0})$ \\

&$\delta=44.5^{0}-100^{0}\oplus 139^{0}-221^{0}\oplus 260.3^{0}-315^{0}$ & $\delta= 0^{0}-40.5^{0}\oplus 87.5^{0}-156.7^{0}\oplus 158.9^{0}-272^{0}\oplus 321^{0}-360^{0}$
& $\delta=0^{0}-360^{0}$ & $\delta= 45.6^{0}-133.6^{0}\oplus 226^{0}-313^{0}$ \\

&$=(93^{0}-157.8^{0}\oplus 201.3^{0}-267.5^{0})$ & $=(0^{0}-80.7^{0}\oplus 127^{0}-147.1^{0}\oplus 210^{0}-234.5^{0}\oplus 280^{0}-360^{0})$
& =($0^{0}-360^{0}) $& $=(48.6^{0}-134.6^{0}\oplus 226^{0}-313.1^{0})$ \\

&$|M|_{ee}=0.0-0.295$&$|M|_{ee}=0.00968-0.250$&$|M|_{ee}=0.000637-0.174$&$|M|_{ee}=0.0408-0.265$\\
&$m_{0}=0.00486-0.300$&$m_{0}=0.00412-0.269$&$m_{0}=0.00136-0.179$&$m_{0}=0.0-0.248$\\
\hline
$B_{9}(C_{9})$  &$\rho=
-80^{0}-80^{0}$ & $\rho=
-70^{0}-70^{0} $ & $\rho=
-90^{0}-90^{0}$
&$\rho=-90^{0}-3.7^{0} \oplus 16.7^{0}-90^{0}$  \\

&$=(-90^{0}-90^{0})$&$=(
-69.5^{0}-69.3^{0})$&$=(-68.6^{0}-68.9^{0})$
&$=(-84.7^{0}-82.3^{0})$\\

&$\sigma= -88^{0}-88^{0}$ &
$\sigma= -90^{0}-90^{0}$ & $\sigma=-72.3^{0}-72.6^{0} $ & $\sigma=-90^{0}-3.7\oplus 18.7^{0}-90^{0}$ \\

&$=(-90^{0}-90^{0})$ &
$=(-78.9^{0}-87^{0})$ & $=(-90^{0}-90^{0}) $ & $=(-90^{0}-90^{0})$ \\

&$\delta=22.96^{0}-92.3^{0}\oplus 141^{0}-241.5^{0}\oplus 267.3^{0}-336^{0}$ & $\delta= 0^{0}-81^{0}\oplus 92.5^{0}-269^{0}\oplus 278^{0}-360^{0}$
& $\delta=0^{0}-360^{0}$ & $\delta= 40.6^{0}-161.7^{0}\oplus 198.7^{0}-337^{0}$ \\

&$=(8.47^{0}-157.8^{0}\oplus 201.3^{0}-348.5^{0})$ & $=(0^{0}-135^{0}\oplus 158.7^{0}-205.4^{0}\oplus 225^{0}-360^{0})$
& $=(0^{0}-360^{0})$ & $=(9.68^{0}-151.1^{0}\oplus 200.3^{0}-352.5^{0}$) \\

&$|M|_{ee}=0.00761-0.250$&$|M|_{ee}=0.00688-0.450$&$|M|_{ee}=0.0-0.465$&$|M|_{ee}=0.0441-0.207$\\
&$m_{0}=0.00688-0.146$&$m_{0}=0.00109-0.476$&$m_{0}=0.00-0.463$&$m_{0}=0.00239-0.199$\\
\hline
$B_{10}(C_{8})$  &$\rho=
-90^{0}-90^{0}$ & $\rho=
 -75.19^{0}-75.1^{0}$ & $\rho=
-90^{0}-90^{0}$
&$\rho=
-81.5^{0}-82^{0}$  \\

&$=(-90^{0}-90^{0})$&$=(
-90^{0}--55^{0}\oplus -23.1^{0}-22.7^{0} \oplus 57^{0}-90^{0})$&$=(
-90^{0}-90^{0}$)
&$=(-90^{0}-90^{0})$\\

&$\sigma= -90^{0}-90^{0}$ &
$\sigma= -88^{0}-88^{0}$ & $\sigma=
-90^{0}-90^{0}$ & $\sigma=-90^{0}-90^{0}$ \\

&$=(-90^{0}-90^{0})$ &$=(-90^{0}-90^{0})$ & $=(-90^{0}-90^{0})$ & $=(-90^{0}-90^{0})$ \\

&$\delta=81.8^{0}-155.3^{0}\oplus 204.5^{0}-277.9^{0}$ & $\delta= 9.2^{0}-178^{0}\oplus 185^{0}-352.5^{0}$
& $\delta= 0^{0}-360^{0}$  & $\delta=46.6^{0}-91.2^{0}\oplus 271.8^{0}-314.9^{0}$\\

&$=(45.6^{0}-85.8^{0}\oplus 144.5^{0}-215.5^{0}\oplus 274^{0}-315^{0})$ & $=(0^{0}-80.7^{0}\oplus 124^{0}-249.1^{0}\oplus 280^{0}-360^{0})$
& $=(0^{0}-360^{0})$ & $=(89.9^{0}-132^{0}\oplus 228^{0}-270^{0}$) \\

&$|M|_{ee}=0.00581-0.155$&$|M|_{ee}=0.0138-0.280$&$|M|_{ee}=0.0-0.136$&$|M|_{ee}=0.0491-0.0540$\\
&$m_{0}=0.00706-0.153$&$m_{0}=0.00460-0.134$&$m_{0}=0.0-0.138$&$m_{0}=0.0-0.269$\\

\hline
\end{tabular}}
\caption{\label{tab4}The allowed ranges of Dirac CP-violating phase $\delta$, the  Majorana phases $\rho, \sigma$,  effective neutrino mass $|M|_{ee}$, and lowest neutrino mass $m_{0}$ for the experimentally allowed cases of Category B(C) at 3$\sigma$ CL. The predictions corresponding to $(\lambda_{13})_{-}$ and $(\lambda_{23})_{-}$ neutrino mass ratios have been put into brackets. }
\end{footnotesize}
\end{center}
\end{table}

 \begin{table}
 \begin{center}
\begin{footnotesize}
\resizebox{18cm}{!}{
\begin{tabular}{|c|c|c|c|c|}
  \hline
&\multicolumn{2}{c|}{X} &\multicolumn{2}{c|}{Y} \\
\hline
Cases&NS&IS&NS&IS\\
\hline
$D_{1}(F_{2})$ &$\rho=
-90^{0}-90^{0}$ & $\rho=
-87.9^{0}--30.97 \oplus 32.5-90^{0}$ & $\rho=-84.2^{0}-84.3^{0}$
&$\rho=-86.5^{0}-84.3^{0}$  \\

&$=(-90^{0}-90^{0})$&$=(
-63.8^{0}-46.16^{0})$&$=(-43.5^{0}-45.3^{0})$
&$=(-90^{0}-90^{0})$\\

&$\sigma= -90^{0}-90^{0}$ &
$\sigma= -44.5^{0}-39.55^{0}$ & $\sigma=-90^{0}-90^{0}$ & $\sigma=-0.0273^{0}-0.0271^{0}$ \\

&$=(-90^{0}--25.6^{0}\oplus 25.5^{0}-90^{0})$ &
$=(-90^{0}--24.6^{0}\oplus 23.5^{0}-90^{0})$ & $=(-74.4^{0}--15.6^{0}\oplus 11.5^{0}-73.5^{0})$ & $=(-90^{0}-90^{0})$ \\

&$\delta=0^{0}-360^{0}$ & $\delta= 31.7^{0}-98.4^{0}\oplus 260.8^{0}-329^{0}$
& $\delta=33.3^{0}-88.7^{0}\oplus 152.3^{0}-207.7^{0}\oplus 274.5^{0}-320^{0}$ & $\delta= 0^{0}-85.7^{0}\oplus 135.3^{0}-224.7^{0}\oplus 272.5^{0}-360^{0}$ \\

&$=(10^{0}-350^{0})$ & $=(100.0^{0}-260^{0})$
& $=(89.7^{0}-150.3^{0} \oplus 209.5^{0}-269.4^{0})$ & $=(0^{0}-155.3^{0} \oplus 206.5^{0}-360^{0}$) \\

&$|M|_{ee}=0.0406-0.173$&$|M|_{ee}=0.0443-0.167$&$|M|_{ee}=0.0376-0.267$&$|M|_{ee}=0.0114-0.0540$\\
&$m_{0}=0.00137-0.137$&$m_{0}=0.0629-0.159$&$m_{0}=0.0604-0.384$&$m_{0}=0.00205-0.258$\\

\hline
$D_{2}(F_{1})$  &$\rho=
-90^{0}-90^{0}$ & $\rho=
-40.9^{0}-40.1^{0}$ & $\rho=-46.9^{0}-45.9^{0}$
&$\rho=-90^{0}-90^{0}$  \\

&$=(-90^{0}-90^{0})$&$=(
-90^{0}-90^{0})$&$=(-64.5^{0}--46.1^{0} \oplus -46.8^{0}-65^{0} )$
&$=(-90^{0}-90^{0})$\\

&$\sigma= -90^{0}-90^{0}$ &
$\sigma= -67.5^{0}-27^{0} \oplus 27.0^{0}-68^{0} $ & $\sigma=-74.1^{0}--24^{0} \oplus 23.4^{0}-70.12^{0}$ & $\sigma=-90^{0}-90^{0}$ \\
&$=(-90^{0}-90^{0})$ &
$=(-90^{0}-90^{0})$ & $=(-90^{0}-70^{0} \oplus -21.0^{0}-21^{0} \oplus 70^{0}-90^{0}$) & $=(-90^{0}-90^{0})$ \\

&$\delta=0^{0}-360^{0}$ & $\delta= 95.78^{0}-258.5^{0}$
& $\delta= 97.78^{0}-150^{0}\oplus 209.78^{0}-254.6^{0} $ & $\delta= 0^{0}-31.2^{0}\oplus 93.2^{0}-150^{0}\oplus 208.2^{0}-270^{0}\oplus 328^{0}-360^{0} $ \\

&$=(0.0^{0}-360^{0})$ & $=(0.0^{0}-360^{0})$
& $=(70.5^{0}-90.87^{0}\oplus 152^{0}-209^{0}\oplus 271^{0}-282^{0})$ & $=(0^{0}-31.2^{0}\oplus 93.2^{0}-154^{0}\oplus 208.2^{0}-270^{0}\oplus 328.2^{0}-360^{0})$ \\

&$|M|_{ee}=0.00771-0.0793$&$|M|_{ee}=0.0545-0.142$&$|M|_{ee}=0.0505-0.132$&$|M|_{ee}=0.0245-0.143$\\
&$m_{0}=0.00281-0.156$&$m_{0}=0.101-0.298$&$m_{0}=0.101-0.268$&$m_{0}=0.0124-0.142$\\
\hline

$D_{3}(F_{4})$  &$\times$ & $\rho=-85.2^{0}-85.1^{0}$ & $\rho=-90^{0}-90^{0}$
&$\times$  \\

&$\times$&$=(-90^{0}-90^{0})$&$\rho=-64.45^{0}-65.6^{0}$
&$\times$\\

&$\times$ &
$\sigma= -55.2^{0}-46.8^{0}$ & $\sigma=-90^{0}-90^{0}$ & $\times$ \\

&$\times$ &$=(-90^{0}--44.8^{0}\oplus 43.6^{0}-90^{0})$ & $=(-77.7^{0}-78.9^{0})$ & $\times$ \\

&$\times$ & $\delta=26.97^{0}-137.87^{0}\oplus 209.3^{0}-326.3^{0}$
& $\delta=0^{0}-91.1^{0}\oplus 117.8^{0}-245.6^{0}\oplus 266.8^{0}-360^{0}$ & $\times$ \\

&$\times$ & $=(0.0^{0}-63^{0}\oplus 120.3^{0}-238.8^{0}\oplus 300.8^{0}-356.8^{0})$
& $=(31.5^{0}-146.7^{0}\oplus 208.9^{0}-333.6^{0})$ & $\times$ \\

&$\times$&$|M|_{ee}=0.0461-0.325$&$|M|_{ee}=0.0241-0.243$&$\times$\\
&$\times$&$m_{0}=0.0499-0.326$&$m_{0}=0.0371-0.278$&$\times$\\

\hline
$D_{4}(F_{3})$  &$\rho=
-90^{0}--17^{0} \oplus 18^{0}-90^{0}$ & $\rho=
-60^{0}--38.9^{0}  \oplus 36.75^{0}-66.66^{0}$ & $\rho=
-65.6^{0}-67.7^{0}$
&$\rho=-90^{0}-90^{0}$  \\

&$=(-67.7^{0} \oplus 67^{0})$&$=(-31.25^{0} \oplus 31.5^{0})$&$=(-59.8^{0}-51.6^{0})$&$=(-88.9^{0}-88^{0})$\\

&$\sigma= -62.5^{0} \oplus 64.5^{0}$ &
$\sigma= -31.29^{0}-30.89^{0}$ & $\sigma=-87.9^{0}-88.1^{0}$ & $\sigma=-90^{0}-90^{0}$ \\

&$=(-90^{0}--25.4^{0}\oplus 25.4^{0}-90^{0}$ &
$=(-70.38^{0}--38.1^{0}\oplus 40.8^{0}-73.45^{0})$ & $=(-77.06^{0}-16.7^{0} \oplus 13.7^{0}-76.8^{0})$ & $=(-90^{0}-90^{0})$ \\

&$\delta=0.0^{0}-360^{0}$ & $\delta= 71.92^{0}-104.9^{0}\oplus 253.8^{0}-296.3^{0}$
& $\delta=67.7^{0}-300^{0}$ & $\delta= 0^{0}-86.5^{0}\oplus 108.9^{0}-254.6^{0}\oplus 273.4^{0}-360^{0}$ \\

&$=(0^{0}-360^{0})$ & $=(117.1^{0}-154.4^{0}\oplus 211.8^{0}-251.4^{0})$& $=( 97.8^{0}-157^{0}\oplus 204.5^{0}-275.6^{0})$ & $=(0^{0}-161^{0}\oplus 198^{0}-360^{0}$) \\

&$|M|_{ee}=0.00735-0.158$&$|M|_{ee}=0.0608-0.220$&$|M|_{ee}=0.0491-0.153$ & $|M|_{ee}=0.0120-0.196$\\
&$m_{0}=0.00844-0.310$&$m_{0}=0.114-0.416$&$m_{0}=0.0981-0.308$ & $m_{0}=0.0249-0.387$\\
\hline

$D_{5}(F_{5})$  &$\rho=-90^{0}-90^{0}$ & $\rho=-90^{0}-90^{0}$ & $\rho=-90^{0}-90^{0}$
&$\rho=-90^{0}-90^{0}$ \\

&$=(-90^{0}-90^{0})$&$=(-90^{0}--8.74^{0} \oplus 8.40^{0}-90^{0})$&$=(-90^{0}-90^{0})$
&$=(-90^{0}-90^{0})$\\

&$\sigma= -90^{0}-90^{0}$ &
$\sigma= -90^{0}-90^{0}$ & $\sigma= -90^{0}-90^{0}$ & $\sigma= -90^{0}-90^{0}$ \\
&$=(-90^{0}-90^{0})$ &
$=(-90^{0}--8.75^{0}\oplus 8.40^{0}-90^{0})$ & $=(-90^{0}-90^{0})$ & $=(-90^{0}-90^{0})$ \\

&$\delta=0^{0}-360^{0}$ & $\delta= 76.72^{0}-284.5^{0}$
& $\delta= 0^{0}-175.6^{0}\oplus 185.6^{0}-360^{0}$ & $\delta= 0^{0}-159.8^{0}\oplus 200.2^{0}-360^{0}$ \\

&$=(0.0^{0}-152.5^{0}\oplus 215^{0}-360^{0})$ & $=(0.0^{0}-88.6^{0}\oplus 275^{0}-360^{0})$
& $=(0^{0}-360^{0}$) & $=(0^{0}-159.8^{0}\oplus 199.2^{0}-360^{0}$) \\

&$|M|_{ee}=0.0478-0.490$&$|M|_{ee}=0.0478-0.490$&$|M|_{ee}=0.0241-0.493$&$|M|_{ee}=0.00868-0.493$\\
&$m_{0}=0.00193-0.345$&$m_{0}=0.0325-0.495$&$m_{0}=0.0307-0.501$&$m_{0}=0.0-0.490$\\

\hline

$D_{6}(F_{9})$  &$\rho=
-80^{0}--3.17^{0} \oplus 6.54^{0}-80^{0}$ & $\rho=-62.97^{0}-62.98^{0}$ &$\rho=-43.6^{0}-43.1^{0}$
&$\rho=-57.8^{0}-54.6^{0}$  \\

&$=(-59.4^{0}-52.59^{0})$&$=(-60^{0}-60^{0})$&$=(-64.5^{0}--23.5^{0} \oplus 24.1^{0}-64^{0})$
&$=(-77.1^{0}--22.89^{0} \oplus 18.8^{0}-81.4^{0})$\\

&$\sigma= -66.30^{0}-66.30^{0}$ & $\sigma= -70.1^{0}-70^{0}$ & $\sigma=-88.9^{0}--36.7^{0} \oplus 38.9^{0}-89^{0}$ & $\sigma=-90^{0}--50.9^{0} \oplus 48.9^{0}-90^{0}$ \\

&$=(-90^{0}-90^{0})$ &$=(-88.4^{0}-88.0^{0})$ & $=(-43.5^{0}-42.4^{0})$& $=(-59.8^{0}-59.7^{0})$ \\

&$\delta=52.1^{0}-144.08^{0}\oplus 218.7^{0}-306.7^{0}$ & $\delta=70.79^{0}-149.5^{0}\oplus 210.5^{0}-290^{0}$
&$\delta=111.2^{0}-253^{0}$  & $\delta= 130^{0}-228.9^{0}$ \\

&$=(0.0^{0}-360^{0})$ & $=(67.9^{0}-292.08^{0})$
& $=(73.4^{0}-119.8^{0}\oplus 239.8^{0}-303.5^{0})$ & $=(52.1^{0}-140.08^{0}\oplus 220.7^{0}-306.7^{0}$) \\

&$|M|_{ee}=0.00712-0.101$&$|M|_{ee}=0.0467-0.134$&$|M|_{ee}=0.0371-0.217$&$|M|_{ee}=0.0144-0.108$\\
&$m_{0}=0.0341-0.209$&$m_{0}=0.0659-0.261$&$m_{0}=0.0610-0.428$&$m_{0}=0.0395-0.209$\\
\hline

$D_{7}(F_{8})$  &$\rho=-90^{0}-90^{0}$ & $\rho=-4.12^{0}-4.00^{0}$ & $\rho=-4.69^{0}-4.94^{0}$ &$\rho=-90^{0}-90^{0}$  \\

&$=(-69.6^{0}-69.4^{0})$&$=(-2.22^{0}--0.120^{0} \oplus 0.120^{0}-2.25^{0})$&$=(-3.98^{0}-4.65^{0})$
&$=(-90^{0}-90^{0})$\\

&$\sigma= -69.2^{0}-66^{0}$ &
$\sigma= -1.56^{0}-1.56^{0}$ & $\sigma= -7.26^{0}-7.28^{0}$ & $\sigma=-90^{0}-90^{0}$ \\

&$=(-90^{0}-90^{0})$ &$=(-2.45^{0}-2.34^{0})$ & $=(-7.03^{0}-7.93^{0})$ & $=(-90^{0}-90^{0})$ \\

&$\delta=0.0^{0}-165^{0}\oplus 197^{0}-360^{0}$ & $\delta=86.91^{0}-90.69^{0}\oplus 268.65^{0}-273^{0}$
& $\delta=89.80^{0}-97.17^{0}\oplus 263.3^{0}-270.4^{0}$& $\delta= 0^{0}-360^{0}$ \\

&$=(0^{0}-360^{0})$ & $=(88.91^{0}-92.69^{0}\oplus 267.65^{0}-271^{0})$
& $=(89.13^{0}-96.2^{0}\oplus 264.2^{0}-270.8^{0})$ & $=(0^{0}-152.8^{0} \oplus 201.6^{0}-360^{0}$) \\

&$|M|_{ee}=0.0481-0.311$&$|M|_{ee}=0.0387-0.311$&$|M|_{ee}=0.0357-0.216$&$|M|_{ee}=0.0169-0.0340$\\
&$m_{0}=0.0141-0.313$&$m_{0}=0.0312-0.465$&$m_{0}=0.0334-0.212$&$m_{0}=0.0153-0.308$\\

\hline
$D_{8}(F_{7})$  &$\times$ & $\times$ & $\times$
&$\times$  \\

&$\times$&$\times$&$\times$
&$\times$\\

&$\times$ &$\times$ & $\times $ & $\times$ \\
&$\times$ &$\times$ & $\times $ & $\times$ \\

&$\times$ & $\times$& $\times$ & $\times$\\

&$\times$& $\times$ & $\times$ &$\times$ \\

&$\times$&$\times$&$\times$&$\times$\\
&$\times$&$\times$&$\times$&$\times$\\
\hline
$D_{9}(F_{6})$  &$\rho=-90^{0}-90^{0}$ & $\rho=-0.82^{0}-2.99^{0} $ &$\rho=-1.51^{0}-0.548^{0} $
&$\rho=-90^{0}-90^{0}$  \\

&$=(-68^{0}-68^{0})$&$=(-0.39^{0}-2.53^{0})$&$=(-7.34^{0}-1.45^{0} )$
&$=(-90^{0}-90^{0})$\\

&$\sigma= -66^{0}-66^{0}$ &
$\sigma= -2.68^{0}-0.46^{0}$ & $\sigma=-7.34^{0}-1.45^{0}$ & $\sigma=-90^{0}-90^{0}$ \\

&$=(-90^{0}--24.97^{0} \oplus -9.6^{0}-1.46^{0} \oplus 20.91^{0}-90^{0})$ &
$=(-3.70^{0}-2.13^{0})$ & $=(-6.34^{0}-1.15^{0}) $ & $=(-90^{0}-90^{0})$ \\

&$\delta=0^{0}-158^{0}\oplus 200^{0}-360^{0}$ & $\delta= 87.87^{0}-91.83^{0}\oplus 268.4^{0}-272^{0}$
& $\delta= 85.87^{0}-90.83^{0}\oplus 268.4^{0}-274.5^{0}$ & $\delta= 0^{0}-360^{0}$ \\

&$=(0^{0}-360^{0})$ & $=(88.87^{0}-91.43^{0}\oplus 268.68^{0}-271.6^{0})$
& $=(85.57^{0}-91.23^{0}\oplus 269^{0}-275^{0})$ & $=(0^{0}-160^{0}\oplus 198.8^{0}-360^{0}$) \\

&$|M|_{ee}=0.00551-0.277$&$|M|_{ee}=0.0564-0.237$&$|M|_{ee}=0.0378-0.343$&$|M|_{ee}=0.0239-0.413$\\
&$m_{0}=0.00792-0.277$&$m_{0}=0.0351-0.238$&$m_{0}=0.0364-0.342$&$m_{0}=0.0124-0.412$\\
\hline

$D_{10}(F_{10})$  &$\rho=-90^{0}-90^{0}$ & $\times$ & $\times$
&$\rho=-90^{0}-90^{0}$  \\

&$=(-90^{0}-90^{0})$&$\times$&$\times$
&$=(-90^{0}-90^{0})$\\

&$\sigma= -90^{0}-90^{0}$ &
$\times$ & $\times $ & $\sigma=-90^{0}-90^{0}$ \\

&$=(-90^{0}-90^{0})$ &
$\times$ & $\times $ & $=(-90^{0}-90^{0})$ \\

&$\delta=11.70^{0}-348^{0}$ & $\times$
& $\times$ & $\delta= 0^{0}-160^{0} \oplus 209.2^{0}-360^{0}$ \\

&$=(0^{0}-74.56^{0}\oplus 89.17^{0}-275^{0}\oplus 285^{0}-360^{0})$ & $\times$
& $\times$ & $=(16.5^{0}-344^{0}$) \\

&$|M|_{ee}=0.0-0.0340$&$\times$&$\times$&$|M|_{ee}=0.0107-0.0268$\\
&$m_{0}=0.000594-0.0887$&$\times$&$\times$&$m_{0}=0.000649-0.0641$\\

\hline
\end{tabular}}
\caption{\label{tab5}The allowed ranges of Dirac CP-violating phase $\delta$, the  Majorana phases $\rho, \sigma$,  effective neutrino mass $|M|_{ee}$, and lowest neutrino mass $m_{0}$ for the experimentally allowed cases of Category D(F) at 3$\sigma$ CL. The predictions corresponding to $(\lambda_{13})_{-}$ and $(\lambda_{23})_{-}$ neutrino mass ratios have been put into brackets. }

\end{footnotesize}
\end{center}
\end{table}

 \begin{table}
 \begin{center}
\begin{footnotesize}
\resizebox{18cm}{!}{
\begin{tabular}{|c|c|c|c|c|}
  \hline
&\multicolumn{2}{c|}{X} &\multicolumn{2}{c|}{Y} \\ 
\hline 
Cases&NS&IS&NS&IS\\
 \hline
 
 $E_{1}(E_{2})$  &$\rho=
-90^{0}-90^{0}$ & $\times$ & $\times$
& $\rho= -90^{0}-90^{0}$  \\

&$=(-90^{0}-90^{0})$&$\times$&$\times$
&$=(-90^{0}-90^{0}$)\\ 
 
&$\sigma= -83.1^{0}-83^{0}$ &
$\times$ & $\times $ & $\sigma=-90^{0}-90^{0}$ \\
&$=(-90^{0}-90^{0})$ &
$\times$ & $\times $ & $=(-90^{0}-90^{0})$ \\

&$\delta=0.0^{0}-360^{0}$ & $\times$
& $\times$ & $\delta=0^{0}-42.29^{0} \oplus 97.8^{0}-162.1^{0} \oplus 198.7^{0}-266.8^{0} \oplus 319.2^{0}-360^{0} $ \\
 
&$=(0.0^{0}-360^{0})$ & $\times$
& $\times$ & $=(28.9^{0}-101.3^{0} \oplus 148.7^{0}-259.8^{0} \oplus 259.2^{0}-320^{0}$) \\ 

&$|M|_{ee}=0.0-0.0500$&$\times$&$\times$&$|M|_{ee}=0.0135-0.0580$\\
&$m_{0}=0.00287-0.0692$&$\times$&$\times$&$m_{0}=0.0189-0.0650$\\
\hline
$E_{3}(E_{4})$  & $\times$ & $\rho=-51.0^{0}--18.6 \oplus 18.0^{0}-51.5^{0}$& $\rho=-46.9^{0}--19.6 \oplus 17.0^{0}-53^{0}$ & $\times$  \\ 

&$\times$&$=(-90.0^{0}--42^{0} \oplus 48.24^{0}-90^{0})$&$=(-88.9^{0}--45.6^{0} \oplus 41.5^{0}-88.1^{0})$&$\times$\\ 

& $\times$&$\sigma=-51.6^{0}--19.6 \oplus 19.0^{0}-51.6^{0} $  & $\sigma=
-48.9^{0}--15.5^{0} \oplus 17.3^{0}-52.4^{0} $ & $\times$ \\

&$\times$&$=(-90.0^{0}--43^{0} \oplus 43.1^{0}-90^{0})$&$=(-90^{0}--40.7^{0} \oplus 40.7^{0}-90^{0})$
&$\times$\\ 

&$\times$ &$\delta= 15.6^{0}-41.3 \oplus 130^{0}-162^{0} \oplus 197^{0}-230^{0} \oplus 318^{0}-360^{0}$ & $\delta= 16.86^{0}-52.3^{0}\oplus 137.7^{0}-164.6^{0}\oplus 197.2^{0}-220^{0}\oplus 312.3^{0}-341.2^{0}$ & $\times$ \\ 

&$\times$ &$=(42.2^{0}-131.8 \oplus 230^{0}-316^{0})$ & $=(0^{0}-137.6^{0}\oplus 221.5^{0}-311.4^{0})$
& $\times$ \\ 
&$\times$&$|M|_{ee}=0.186-0.467$&$|M|_{ee}=0.178-0.494$&$\times$\\
&$\times$&$m_{0}=0.186-0.474$&$m_{0}=0.183-0.494$&$\times$\\

\hline
$E_{5}(E_{5})$  &$\rho=
-90^{0}-90^{0}$ & $\rho=
 -90^{0}--18^{0} \oplus 18^{0}-90^{0}$ & $\rho=-90^{0}--18.6^{0} \oplus 16.6^{0}-90^{0}$
&$\rho=-36.6^{0}-36.7^{0}$  \\

&$=(-81.49^{0}-82.5^{0})$&$=(
-86.1^{0}--36.9^{0}\oplus 33.89^{0}-86.7^{0})$&$=(-90^{0}--16.6^{0} \oplus 17.6^{0}-90^{0})$
&$=(-21.9^{0}-22.4^{0})$\\ 
 
&$\sigma= -70^{0}-70^{0}$ &
$\sigma= -90^{0}--19^{0} \oplus 19^{0}-90^{0}$ & $\sigma= -90^{0}--18.7^{0} \oplus 19^{0}-90^{0} $ & $\sigma=-76.8^{0}-75.6^{0} $\\

&$=(-90^{0}--8.7^{0} \oplus 8.6^{0}-90^{0})$ &
$=(-90^{0}--20.1^{0} \oplus 20^{0}-90^{0})$ & $=(-90^{0}--18.7^{0} \oplus 19^{0}-90^{0})$ & $=(-90^{0}--57.7^{0} \oplus 57.1^{0}-90^{0})$ \\

&$\delta=0.0^{0}-360^{0}$ & $\delta= 90.8^{0}-162.45^{0}\oplus 199.4^{0}-269.5^{0}$
& $\delta= 29.8^{0}-92.3^{0}\oplus 150.4^{0}-163^{0} \oplus 197^{0}-210^{0}\oplus 269.8^{0}-330.9^{0}$ & $\delta= 0^{0}-49.8^{0}\oplus 113.4^{0}-247.6^{0}\oplus 306.7^{0}-355.6^{0} $ \\
 
&$=(30.18^{0}-331.5^{0})$ & $=(33.3^{0}-85.7^{0}\oplus 270^{0}-325.1^{0})$
& $=(17.2^{0}-28.7^{0}\oplus 89.1^{0}-151.5^{0} \oplus 210.3^{0}-271.2^{0}\oplus 329^{0}-342.2^{0})$ & $=(52.3^{0}-117.6^{0}\oplus 247.4^{0}-307.8^{0} $) \\ 

&$|M|_{ee}=0.0-0.140$&$|M|_{ee}=0.138-0.430$&$|M|_{ee}=0.125-0.493$&$|M|_{ee}=0.0114-0.369$\\
&$m_{0}=0.00127-0.276$&$m_{0}=0.133-0.459$&$m_{0}=0.133-0.499$&$m_{0}=0.0120-0.368$\\
\hline
$E_{6}(E_{9})$  &$\rho=
-90^{0}- -65.2 \oplus 65.2 -90^{0}$ & $\times$ & $\times$
&$\rho=-90^{0}--64.8^{0} \oplus 64.1^{0}-90^{0}$  \\

&$=(-90^{0}--44.88 \oplus 55.95-90^{0})$&$\times$&$\times$
&$=(-85.3^{0}--44.8^{0} \oplus 44.08^{0}-84.5^{0})$\\ 
 
&$\sigma= -39.1^{0}-36.5^{0}$ &
$\times$ & $\times $ & $\sigma=-44.91^{0}-41.8^{0}$ \\

&$=(-67.41^{0}--22.19 \oplus 25.62-69^{0})$ &
$\times$ & $\times $ & $=(-71.8^{0}--24.8^{0}\oplus 26.7^{0}-69.7^{0})$ \\

&$\delta=0.0^{0}-42.29 \oplus 306^{0}-360^{0}$ & $\times$
& $\times$ & $\delta=0^{0}-53.5^{0}\oplus 304.5^{0}-360^{0}$ \\
 
&$=(40.0^{0}-90.86 \oplus 275^{0}-322.2^{0})$ & $\times$
& $\times$ & $=(39.4^{0}-89.9^{0}\oplus 271.2^{0}-324.5^{0}$) \\ 

&$|M|_{ee}=0.00722-0.019$&$\times$&$\times$&$|M|_{ee}=0.0114-0.0540$\\
&$m_{0}=0.00220-0.0417$&$\times$&$\times$&$m_{0}=0.0190-0.0520$\\
\hline
$E_{7}(E_{8})$  &$\times$ & $\times$ & $\times$
&$\times$  \\

&$\times$&$\times$&$\times$
&$\times$\\ 
 
&$\times$ &

$\times$ & $\times $ &$\times$ \\
&$\times$ &
$\times$ & $\times $ &$\times$\\

&$\times$ & $\times$
& $\times$ & $\times$\\
 
&$\times$ & $\times$
& $\times$ &$\times$ \\ 

&$\times$&$\times$&$\times$&$\times$\\
&$\times$&$\times$&$\times$&$\times$\\
\hline
$E_{10}(E_{10})$  &$\rho=
-90^{0}-90^{0}$ & $\rho=
 -90^{0}--28^{0} \oplus 28^{0}-90^{0}$ & $\rho=
 -87.8^{0}--31.0^{0} \oplus 31.1^{0}-88.1^{0}$
&$\rho=-89.8^{0}-89.7^{0}$  \\

&$=(-90^{0}-90^{0})$&$=(-81.3^{0}--20.79^{0}\oplus 25.08^{0}-83.1^{0})$&$=( -85.8^{0}--38.0^{0} \oplus 42.1^{0}-88.1^{0})$
&$=(-90^{0}-90^{0})$\\ 
 
&$\sigma= -90^{0}-90^{0}$ &
$\sigma= -87^{0}--30^{0} \oplus 30^{0}-71^{0}$ & $\sigma= -81.4^{0}--30.79^{0}\oplus 30.08^{0}-82.1^{0}$ & $\sigma= -88.7^{0}-87.9^{0}$ \\
&$=(-90^{0}-90^{0})$ &
$=(-84.2^{0}--19.8^{0} \oplus 23.1^{0}-88.1^{0})$ & $=(-81.4^{0}--30.79^{0}\oplus 30.08^{0}-82.1^{0})$ & $=(-89.8^{0}-89^{0})$ \\

&$\delta=0.0^{0}-20^{0} \oplus 89.59^{0}-281^{0} \oplus 331^{0}-360^{0}$ & $\delta= 21.35^{0}-71.9^{0}\oplus 272.4^{0}-331.5^{0}$
& $\delta= 25.6^{0}-28^{0}\oplus 90.4^{0}-149.1^{0}\oplus 212.4^{0}-268.5^{0}\oplus 338^{0}-339^{0}$ & $\delta= 0^{0}-101.05^{0}\oplus 149.2^{0}-210.2^{0}\oplus 259.7^{0}-360^{0}$ \\
 
&$=(0.0^{0}-103^{0} \oplus 158^{0}-207.8^{0} \oplus 254^{0}-360^{0})$ & $=(98.16^{0}-160.3^{0}\oplus 204.5^{0}-267.8^{0})$
& $=(41.84^{0}-89.65^{0}\oplus 272.5^{0}-321.4^{0})$ & $=(0^{0}-28.05^{0}\oplus 77^{0}-279.1^{0}\oplus 332^{0}-360^{0}$) \\ 

&$|M|_{ee}=0.0-0.261$&$|M|_{ee}=0.150-0.430$&$|M|_{ee}=0.135-0.352$&$|M|_{ee}=0.01088-0.432$\\
&$m_{0}=0.00098-0.276$&$m_{0}=0.147-0.420$&$m_{0}=0.142-0.352$&$m_{0}=0.00920-0.438$\\

\hline 
\end{tabular}}
\caption{\label{tab6} The allowed ranges of Dirac
CP-violating phase $\delta$, the  Majorana phases
$\rho, \sigma$,  effective neutrino mass $|M|_{ee}$,
and lowest neutrino mass $m_{0}$ for the experimentally allowed
cases of Category E at 3$\sigma$ CL. The predictions corresponding to $(\lambda_{13})_{-}$ and $(\lambda_{23})_{-}$ neutrino mass ratios have been put into brackets. }
\end{footnotesize}
\end{center}
\end{table}

\end{document}